\documentclass[prb,twocolumn,showpacs,floatfix,amssymb]{revtex4}

\usepackage{amsmath}
\usepackage{amsfonts}
\usepackage{graphicx}
\usepackage{color}
\usepackage{hyperref}

\newcommand{\tr}{\mathop{\rm tr}\nolimits}

\newcounter{addeq}

\begin{document}

\title{$\theta$ renormalization, electron-electron interactions and
super universality in the quantum Hall regime}

\author{A.\,M.\,M.\,Pruisken$^{1}$ and I.\,S.\,Burmistrov$^{1,2}$}

\affiliation{$^{1}$ Institute for Theoretical Physics, University
of Amsterdam, Valckenierstraat 65, 1018 XE Amsterdam, The
Netherlands} \affiliation{$^{2}$ L.\,D.\,Landau Institute for
Theoretical Physics, Kosygina street 2, 117940 Moscow, Russia}

\begin{abstract}
The renormalization theory of the quantum Hall effect relies
primarily on the non-perturbative concept of $\theta$
renormalization by instantons. Within the generalized non-linear
$\sigma$ model approach initiated by Finkelstein we obtain the
physical observables of the interacting electron gas, formulate
the general (topological) principles by which the Hall conductance
is robustly quantized and derive - for the first time - explicit
expressions for the non-perturbative (instanton) contributions to
the renormalization group $\beta$ and $\gamma$ functions. Our
results are in complete agreement with the recently proposed idea
of {\em super universality} which says that the fundamental
aspects of the quantum Hall effect are all generic features the
{\em instanton vacuum} concept in asymptotically free field
theory.
\end{abstract}

 \maketitle


\section{\label{Intro}Introduction}

One of the long standing mysteries in the theory of the plateau
transitions in the quantum Hall regime is the apparently
insignificant or subdominant role that is played by the long
ranged Coulomb interaction between the electrons. The pioneering
experiments on quantum criticality in the quantum Hall regimes by
H.\,P.\,Wei {\em et al.},~\cite{experiments} for example, are in
many ways a carbon copy of the scaling predictions based on the
field theory of Anderson localization in strong magnetic
fields.~\cite{freepart} The initial success of the free electron
theory has primarily led to a widely spread believe in Fermi
liquid type of
ideas~\cite{believe1,believe2,believe3,believe4,believe5,believe6}
as well as an extended literature on scaling and critical exponent
phenomenology.~\cite{Exponents1,Exponents2,Exponents3,Exponents4,
Exponents5,Exponents6,Exponents7,Exponents8,Exponents9,Exponents10,
Exponents11,Exponents12,Exponents13,Exponents14}

Except for experimental considerations, however, there exists
absolutely no valid (microscopic) argument that would even
remotely justify any of the different kinds of free (or {\em
nearly} free) electron scenarios that have frequently been
proposed over the years. In fact, Fermi liquid principles are
fundamentally in conflict with the novel insights that have more
recently emerged from the development of a microscopic theory on
interaction effects.~\cite{Unify1,Unify2,Unify5} These
developments are naturally based on the topological concept of an
{\em instanton vacuum}~\cite{LevineLibbyPruisken} which is very
well known to be the fundamental mechanism by which the free
electron gas {\em de-localizes} in two spatial dimensions and in
strong magnetic fields.~\cite{freepart} The outstanding and
difficult problem that one is faced with is whether or not the
topological concepts in quantum field theory retain their
significance also when the electron-electron interactions are
taken into account.~\cite{PruiskenBaranov}

For a variety of reasons, however, it has taken a very long time
before the subject matter gained the physical clarity that it now
has.~\cite{PruiskenBaranovVoropaev,PruiskenBaranovBurmistrov}
Perhaps the most awkward obstacles were provided by the historical
controversies~\cite{Coleman} in QCD where the idea of an instanton
parameter $\theta$ arose first but its exact meaning remained
rather obscure.~\cite{Polyakov} These controversies have mainly
set the stage for the wrong physical ideas and the wrong
mathematical objectives. For example, in sharp contrast to the
general expectations in the
field~\cite{Zirnbauer1,Zirnbauer2,Zirnbauer3,Affleck2,Affleck3,Affleck4}
the fundamental problems do not reside in the conventional aspects
of disordered systems such as the replica method or ``exact''
critical exponent values. A more fundamental issue has emerged,
the {\em massless chiral edge excitations},~\cite{Unify3} that
dramatically change the way in which the $\theta$ parameter is
generally being perceived.~\cite{PruiskenBaranovVoropaev} A
detailed understanding of the physics of the edge has resolved,
amongst many other things, the long standing controversies that
historically have spanned the subject such as the {\em
quantization of topological charge},~\cite{Witten} the meaning of
{\em instantons} and {\em instanton
gases}~\cite{Witten,Affleck1,Affleck5} etc. As a result of all
this we can now state that the instanton angle $\theta$ {\em
generically} displays all the basic features of the quantum Hall
effect, independent of the details such as the replica limit. This
includes not only the appearance of {\em gapless} excitations at
$\theta =\pi$ but also the most fundamental and much sought after
aspect of the theory, the existence of {\em robust topological
quantum numbers} that explain the precision and observability of
the quantum Hall
effect.~\cite{PruiskenBaranovVoropaev,PruiskenBurmistrov}

A second major complication in dealing with interaction effects is
the notorious complexity of the underlying
theory.~\cite{Finkelstein1,Finkelstein2,Finkelstein3} Although
Finkelstein's original ideas in the field have been very
illuminating, it has nevertheless taken herculean efforts to
understand how the generalized non-linear $\sigma$ model approach
can be studied as a field theory. This includes not only the
theory of perturbative expansions~\cite{Unify2,Unify5} but also
such basic aspects like the global symmetries of the problem
($\mathcal F$ invariance), electrodynamic $U(1)$ gauge invariance
as well as the physical observables of the theory.~\cite{Unify1}
Advances along these lines are absolutely necessary if one wants
to extend the perturbative theory of localization and interaction
effects to include the highly non-trivial consequences of the
$\theta$ vacuum.

It obviously makes an enormously big difference to know that the
instanton vacuum theory of the quantum Hall effect is NOT merely
an isolated critical exponent problem that exists in replica field
theory or ``super symmetric" extensions of free electron
approximations alone. Contrary to this widely spread misconception
in the literature the fundamental features of the quantum Hall
effect actually reveal themselves as a {\em super universal}
consequence of topological principles in quantum field theory that
until to date have not been well understood. The concept of {\em
super universality} makes it easier and more natural to comprehend
why the basic phenomena of scaling are retained by the electron
gas also when the Coulomb interaction between the electrons is
taken into account. Moreover, it facilitates the development of a
unifying theory that includes completely different phenomena such
as the fractional quantum Hall regime. Unlike Fermi liquid ideas,
however, {\em super universality} does not necessarily imply that
the quantum critical details at $\theta = \pi$ remain the same.
The various different applications of the $\theta$ vacuum concept
do in general have different exponent values at $\theta = \pi$
and, hence, they belong to different {\em universality
classes}.~\cite{PruiskenBurmistrov}

In this paper we revisit the problem of topological excitations
(instantons) and $\theta$ renormalization~\cite{Pruisken2,Pruisken3}
in the theory of the interacting
electron gas. The results of an early analysis of instanton
effects have been reported in a short paper by Pruisken and
Baranov.~\cite{PruiskenBaranov} However, much of the conceptual
structure of the theory was not known at that time, in particular
the principle of $\mathcal{F}$ {\em invariance} and the appearance
of the {\em massless edge excitations} that together elucidate the
fundamental aspects of the $\theta$ vacuum on the strong coupling
side.~\cite{PruiskenBaranovVoropaev} These novel insights
unequivocally define the physical observables (i.e. the
conductance parameters $\sigma_{xx}$ and $\sigma_{xy}$) that
control the dynamics of the $\theta$ vacuum at low energies. These
physical observables should therefore quite generally be regarded
as some of the most fundamental quantities of the theory.

A detailed knowledge of instanton effects on the physical
observables of the theory has fundamental significance since it
bridges the gap that exists between the {\em weak coupling}
Goldstone singularities at short distances, and the {\em super
universal} features of the quantum Hall effect that generally
appear at much larger distances only. The theory of observable
parameters, as it now stands, provides the general answer to the
``arena of bloody controversies" that historically arose because
of a complete lack of any physical objectives of the theory. A
prominent and exactly solvable example of these statements is
given by the large $N$ expansion of the $CP^{N-1}$ model that,
unlike the previous expectations, sets the stage for all the
non-perturbative features of the $\theta$ parameter that one is
interested in Ref.~[\onlinecite{PruiskenBaranovVoropaev}].

The main objective of the present work is to review the instanton
methodology, provide the technical details of the computation and
extend the analysis in several ways. Our study of the interacting
electron gas primarily relies on the procedure of {\em spatially
varying masses} that very recently has been applied, with great
success, in the context of the ordinary $U(M+N)/U(M) \times U(N)$
non-linear $\sigma$ model.~\cite{PruiskenBurmistrov} The important
advantage of this procedure is that it facilitates
non-perturbative computations of the renormalization group $\beta$
and $\gamma$ functions of the theory. These computations, together
with the new insights into the strong coupling features and
symmetries of the problem, lay out the complete phase and
singularity structure of the disordered electron gas. The results
of this paper, which include the non-Fermi liquid behavior of the
Coulomb interaction problem, obviously cannot be obtained in any
different manner.

This paper is organized as follows. We start out in
Section~\ref{Form} with a brief introduction to the formalism and
recall the effective action procedure for massless chiral edge
excitations. In Section~\ref{qHe} we briefly elaborate on the
general topological principles that explain the {\em robust}
quantization of the Hall conductance. The general argument is
deeply rooted in the methods of quantum field theory and relies on
the relation that exists between the conductances on the one hand,
and the sensitivity of the interacting electron gas to
infinitesimal changes in the boundary conditions on the other. The
argument is furthermore based on the relation between Kubo
formalism, the background field methodology and the effective
action for chiral edge excitations which is described in
Appendix~\ref{AppA}.

In Section~\ref{PO} we give the complete list of physical
observables which then serves as the basic starting point for the
remainder of this paper. We show the general relationship between
the physical observables and the renormalization group $\beta$ and
$\gamma$ functions and briefly discuss the results of the theory
in $2+\epsilon$ spatial dimensions.

In Section~\ref{Form.Instan} we recall the various different
aspects associated with instanton matrix field configurations and
embark on the problem of quantum fluctuations. We introduce the
method of spatially varying masses and end the Section with the
complete action for the quantum fluctuations in Table~\ref{Tf}.

In Section~\ref{EC} together with Appendices~\ref{ME} and
\ref{AppC} we present the results of detailed computations that
deal, amongst many other things, with the technical difficulties
associated with the theory in Pauli-Villars regularization, the
{\em replica method} as well as the {\em infinite sums} over
Matsubara frequency indices that are inherent to the problem of
electron-electron interactions.

In Section~\ref{TCFS} we address the various different aspects
associated with the integration over zero modes and embark on the
general problem of transforming the Pauli-Villars masses in curved
space back into flat space following the methodology introduced by
't Hooft.~\cite{tHooft} This finally leads to the most important
advances of this paper, the renormalization-group $\beta$ and
$\gamma$ functions which are evaluated at a non-perturbative
level. These final results provide a unified theory of the
disordered electron gas that includes the effects of both {\em
finite} range electron-electron interactions and {\em infinite}
range interactions such as the Coulomb potential. We end this
paper with a discussion in Section~\ref{SUM}.

\section{\label{Form}Formalism}
%

%
\subsection{\label{Form.FNLSM}The action}
%

The generalized replica non-linear sigma model involves unitary
matrix field variables $Q^{\alpha \beta}_{n m}(\textbf{r})$ that
obey the following constraints
\begin{equation}  \label{Const}
Q=Q^{\dagger}, \quad \tr Q=0,\quad Q^{2} = 1.
\end{equation}
The superscripts $\alpha, \beta = 1,\dots, N_{r}$ represent the
\textit{replica} indices and the subscripts $n, m$ are the indices
of the \textit{Matsubara} frequencies $\omega_{k} = \pi T (2 k +
1)$ with $k=n, m$. A convenient representation in terms of unitary
matrices $\mathcal{T}(\textbf{r})$ is obtained by writing
\begin{equation}\label{QrepU}
Q(\textbf{r}) = \mathcal{T}^{-1}(\textbf{r}) \Lambda
\mathcal{T}(\textbf{r}), \quad \Lambda_{nm}^{\alpha
\beta}=\mathop{\rm sign}\nolimits(\omega_{n})\delta ^{\alpha
\beta}\delta_{nm}.
\end{equation}
The effective action for the two-dimensional interacting electron
gas in the presence of disorder and a perpendicular magnetic field
can be written as follows~\cite{Unify1}
\begin{equation}
\mathcal{Z} = \int \mathcal{D}[Q] \exp S, \qquad S = S_{\sigma} +
S_{F}. \label{S}
\end{equation}
Here, $S_{\sigma}$ is the \textit{free electron}
action~\cite{Pruisken1}
\begin{equation}
S_{\sigma } = -\frac{\sigma _{xx}}{8} \int d \mathbf{r}
\mathop{\rm tr} \nolimits (\nabla Q)^{2}+\frac{\sigma _{xy}}{8}
\int d \mathbf{r} \mathop{\rm tr}\nolimits \varepsilon_{jk}
Q\nabla_{j} Q \nabla_{k} Q. \label{Ssigma}
\end{equation}
The quantities $\sigma_{xx}$ and $\sigma_{xy}$ represent the
\textit{mean field} values for the \textit{longitudinal} and
\textit{Hall} conductances in units $e^{2}/h$ respectively. The
symbol $\varepsilon_{jk}=-\varepsilon_{kj}$ stands for the
antisymmetric tensor. Next, $S_{F}$ contains the \textit{singlet
interaction} term~\cite{Finkelstein3,Unify1}
\begin{equation}
S_{F}=\pi Tz\int d \mathbf{r} \mathcal{O}_F(Q) \label{SF}
\end{equation}
where
\begin{equation}
\mathcal{O}_F(Q) = c\sum\limits_{\alpha n}\mathop{\rm tr}
\nolimits I_{n}^{\alpha }Q\mathop{\rm tr}\nolimits I_{-n}^{\alpha
}Q+4 \mathop{\rm tr}\nolimits\eta Q- 6 \mathop{\rm
tr}\nolimits\eta \Lambda .\label{OF}
\end{equation}
Here, $z$ is the so-called \textit{singlet interaction amplitude},
$T$ the temperature and $c$ the \textit{crossover} parameter which
allows the theory be interpolated between the case of electrons
with Coulomb interaction ($c=1$) and the free electron case
($c=0$). The singlet interaction term involves a matrix
\begin{equation}
\left ( I_{n}^{\alpha }\right ) _{km}^{\beta \gamma }=\delta
^{\alpha \beta }\delta ^{\alpha \gamma }\delta _{k,n+m} \label{In}
\end{equation}
which is the Matsubara representation of the $U(1)$ generator
$\exp (-i\omega _{n}\tau )$ with $\tau$ being imaginary time.
Matrix
\begin{equation}
\eta _{nm}^{\alpha\beta}=n \delta ^{\alpha\beta}\delta_{nm}
\label{eta}
\end{equation}
is used to represent the set of the Matsubara frequencies
$\omega_n$.

%
\subsection{\label{Form.CutOff} $\mathcal{F}$ invariance and $\mathcal{F}$ algebra}
%

Unlike the free particle problem ($c=0$), the theory with
electron-electron interactions ($0<c \leq 1$) is mainly
complicated by the fact that the range of Matsubara frequency
indices $m,n$ must be taken from $-\infty $ to $+\infty $, along
with the replica limit $N_r \rightarrow 0$. Under these
circumstances one can show that the singlet interaction term
fundamentally affects the ultra violet singularity structure of
the theory (the renormalization group $\beta$ and $\gamma$
functions) which is one of the peculiar features of the theory of
electron-electron
interactions.~\cite{Finkelstein1,Finkelstein2,Finkelstein3,Unify2,Unify5}
Moreover, the problem with {\em infinite ranged interactions}
($c=1$) such as the Coulomb interaction displays an exact global
symmetry named $\mathcal{F}$ invariance.~\cite{Unify1} This means
that $S_F$ is invariant under electrodynamic $U(1)$ gauge
transformations which are spanned by the matrices $I^\alpha_n$.
This symmetry is broken by the problem with {\em finite ranged
interactions} ($0<c<1$). In order to retain the $U(1)$ algebra in
truncated frequency space with a cut-off $N_m$ a set of algebraic
rules has been developed named $\mathcal{F}$
algebra.~\cite{Unify1} These rules permit one to proceed in finite
frequency space where the index $n$ runs from $-N_{m}$ to
$N_{m}-1$, i.e the matrix field variables $Q$ have a finite size
\begin{equation}\label{QrepU}
Q(\textbf{r}) = \mathcal{T}^{-1}(\textbf{r}) \Lambda
\mathcal{T}(\textbf{r}), \quad \mathcal{T}(\textbf{r})\in U(2 N)
\end{equation}
where $N=N_r N_m$. The two limits of the theory, $N_r \to 0$ and
then $N_m\to \infty$ respectively, are taken at the end of all
computations. The main purpose of $\mathcal{F}$ algebra is to
ensure that electrodynamic $U(1)$ gauge invariance as well as
$\mathcal{F}$ invariance are preserved by the renormalization
group, both perturbatively and at a non-perturbative level.
%
%
\subsection{\label{qHe} Quantization of the Hall conductance}
%
The robust quantization of the Hall conductance can be
demonstrated on the basis of very general principles such as mass
generation and the fact that the conductances can be expressed in
terms of the response of the system to changes in the boundary
conditions. The subtleties of the argument involve a novel and
previously unexpected ingredient of the instanton vacuum concept,
however, which has been recognized very recently only. The main
problem resides in the $\sigma_{xy}$ term in Eq.~\eqref{Ssigma}
which is formally identified as the {\em topological charge}
$\mathcal{C}[Q]$ associated with the matrix field configuration
$Q$. Assuming for simplicity the geometry of a square of size $L
\times L$ then we can express the topological charge in terms of
both a {\em bulk} integral and an {\em edge} integral as follows
\begin{eqnarray}
\mathcal{C}[Q] &=& \frac{1}{16 \pi i} \int d \mathbf{r} \tr
\varepsilon _{ab} Q\nabla_{a}Q \nabla_{b} Q   \notag \\
&=& \frac{1}{4 \pi i}\oint \limits d x \tr \mathcal{T} \nabla_x
\mathcal{T}^{-1} \Lambda.\label{edgecharge}
\end{eqnarray}
The remarkable thing that is usually overlooked is that the matrix
field $Q$ generally splits up into distinctly different
components, each with a distinctly different topological
significance and very different physical properties. For this
purpose we introduce a change of variables
\begin{equation}
 Q = t^{-1} Q_0 t \label{changeofvar}.
\end{equation}
Here, the $Q_0$ is an arbitrary matrix field with boundary
conditions $Q_0 =\Lambda$ at the edge (or, equivalently,
$\mathcal{T}_0$ equals an arbitrary $U(N) \times U(N)$ gauge at
the edge). The unitary matrix field $t$ generally represents the
fluctuations about the special boundary conditions. This change of
variables is just a formal way of splitting the topological charge
$\mathcal{C}[Q]$ of an arbitrary matrix field configuration $Q$
into an {\em integral} piece $\mathcal{C}[Q_0]$ and a {\em
fractional} piece $\mathcal{C}[q]$,
\begin{equation}
\mathcal{C}[Q] = \mathcal{C}[Q_0] + \mathcal{C}[q],\qquad q=t^{-1}
\Lambda t.\label{addcharge}
\end{equation}
Without a loss in generality we can write
\begin{equation}\label{bulkedgeQ}
\mathcal{C}[Q_0] \in \mathbb{Z},\qquad
 - \frac{1}{2} < \mathcal{C}[q] \leq \frac{1}{2}.
\end{equation}
The main new idea is that the matrix field $t$ or $q$ should be
taken as a dynamical variable in the problem, rather than being a
fixed quantity that one can choose freely. The reason is that one
can generally associate {\em massless chiral edge excitations}
with the fluctuating matrix fields $q$. These so-called {\em edge
modes} $q$ are distinctly different from the {\em bulk modes}
$Q_0$ which usually (i.e. for arbitrary values of $\sigma_{xy}$)
generate dynamically a {\em mass gap} in the bulk of the system.
These various statements immediately suggest that the low energy
dynamics of the strong coupling phase is described by an effective
action of the matrix field variable $q$ obtained by formally
eliminating the {\em bulk modes} $Q_0$. This effective action
procedure is furthermore based on the fact that the mean field
quantity $\sigma_{xy}$ (which is equal to the filling fraction
$\nu$ of the Landau levels) can in general be split into an {\em
integral edge} part $k(\nu)$ and a {\em fractional bulk} piece
$\theta(\nu)$ as follows
\begin{equation}
 \sigma_{xy} = \nu = k(\nu) + \frac{\theta(\nu)}{2\pi}
\end{equation}
where
\begin{equation}\label{bulkedgesigmaxy}
 k(\nu) \in \mathbb{Z},\qquad  - \pi < \theta(\nu) \leq \pi.
\end{equation}
In what follows we shall separately consider the theory with $c=0$
(free particles) and $c=1$ (Coulomb interactions) both of which
are invariant under the action of renormalization group.

\begin{figure}[tbp]
\includegraphics[height=80mm]{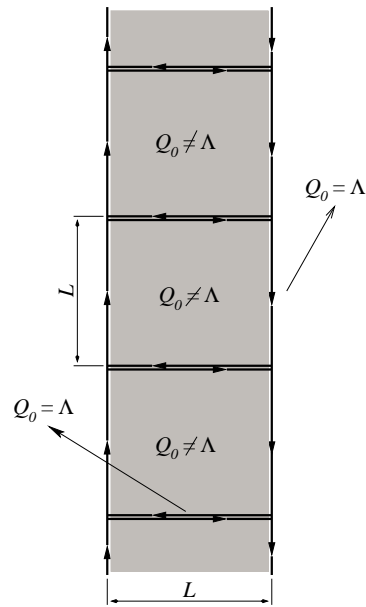}
\caption{Geometry of an infinite strip, see text.} \label{STRIP}
\end{figure}

\subsubsection{Free particles ($c=0$)}

In the absence of external frequencies and at $T=0$ we can write
the action for the free electron gas as follows
\begin{equation}
 S = S_\sigma^\textrm{edge}[q] + S_\sigma^\textrm{bulk}[Q]
\end{equation}
where
\begin{eqnarray}
 S_\sigma^\textrm{edge} [q] &=&
 2\pi i k(\nu) {\mathcal C}[q] \label{edgepiece}\\
 S_{\sigma }^\textrm{bulk} [Q] &=&   -\frac{\sigma _{xx}}{8} \int
 d\mathbf{r}
 \tr (\nabla Q)^{2}+ i{\theta(\nu)} {\mathcal C}[Q].
\end{eqnarray}
Provided the matrix field variable $t$ satisfies the classical
equations of motion we can obtain an effective action for $q$ by
eliminating the bulk matrix field $Q_0$
\begin{equation}
 S_\textrm{eff} [q] = S_\sigma^\textrm{edge} [q]
 + S_\textrm{eff}^\textrm{bulk}[q]
\end{equation}
where
\begin{equation}
 \exp S_\textrm{eff}^\textrm{bulk}[q] =
 \int_{\partial V} {\mathcal D}[Q_0] \exp
 S_{\sigma}^\textrm{bulk} [t^{-1} Q_0 t].\label{bulkpiece}
\end{equation}
Here the subscript $\partial V$ indicates that the functional
integral has to be performed with $Q_0=\Lambda$ at the edge. The
effective action for the bulk can be written as
\begin{equation}\label{Response1}
 S_\textrm{eff}^\textrm{bulk} [q] =   -\frac{\sigma^{\prime}_{xx}}{8}
 \int d \mathbf{r}
 \tr (\nabla q)^{2}+ i{\theta^\prime} {\mathcal C}[q].
\end{equation}
Here, $\sigma^{\prime}_{xx} =\sigma_{xx} (L)$ and $\theta^\prime =
\theta (L)$ play the role of response parameters that measure the
sensitivity of the system to an infinitesimal change in the
boundary conditions. For exponentially localized states these
parameters are expected to vanish for large enough $L$ and the
effective action is now given by the edge piece
(Eq.\eqref{edgepiece}) alone. This one dimensional action is known
to describe {\em massless chiral edge excitations}.~\cite{Unify3}
To obtain a suitably regulated action for the edge we may proceed
by stacking many blocks of size $L \times L$ on top of one another
to form an infinite strip (see Fig.~\ref{STRIP}). The action for
the quantum Hall state is then defined along infinite edges and
can be written as~\cite{Unify1}
\begin{equation}
 S_\textrm{eff} [q] = \frac{k(\nu)}{2} \oint dx \tr t \nabla_x t^{-1} \Lambda
 + \pi T \rho_\textrm{edge} \oint dx \tr \eta q .
\end{equation}
Here we have introduced a frequency term to regulate the infrared.
The quantity $\rho_{edge}$ stands for the density of edge states
and the integer $k(\nu) = \sigma_{xy}^\prime$ indicates that the
Hall conductance is robustly quantized.

At this stage several remarks are in order. First of all, from an
explicit (non-perturbative) computation of the response parameters
$\sigma_{xx}^\prime$ and $\theta^\prime$ we know that the argument
generally fails for $\theta(\nu)=\theta^\prime=\pi$ where the mass
gap vanishes and the system is quantum critical. This happens at
the center of the Landau bands where a transition takes place
between adjacent quantum Hall plateaus.

Secondly, it is important to keep in mind that the aforementioned
argument for an exact quantization of the Hall conductance is
entirely based on the fact that the edge modes $q$ are {\em
massless}. The beauty of the effective action procedure is that it
unequivocally demonstrates that the so-called {\em spherical
boundary conditions} (i.e. $Q_0 = \Lambda$ at the edge) are
dynamically generated by the system itself, independent of any
weak coupling arguments such as finite action requirements and
independent of $N_r$ and $N_m$. The bulk components $Q_0$ have
mathematically very interesting properties in that they are a
realization of the formal homotopy theory result
\begin{equation}
\pi_2 (G/H) = \pi_1 (H) = \mathbb{Z} .
\end{equation}
The integer $\mathbb{Z}$ is equal to the topological charge
$\mathcal{C} [Q_0]$ which is identified as the jacobian for the
mapping of the manifold $U(2N)/U(N) \times U(N)$ onto the
two-dimensional plane. Physically the quantization of
$\mathcal{C}[Q_0]$ represents the {\em quantization of flux} and
the integer $k(\nu) \mathcal{C} [Q_0]$ can be interpreted in terms
of a {\em discrete number} of electrons that have crossed the
Fermi energy at the edge of the system.

Notice that except for the massless chiral edge modes there exists
no compelling reason to believe why the topological charge
$\mathcal{C}[Q_0]$ and, hence, the Hall conductance is {\em
robustly quantized}. In fact, the quantization of topological
charge has been one of the longstanding and controversial issues
in quantum field theory~\cite{Witten} that have fundamentally
complicated the development of a microscopic theory of the quantum
Hall effect.

\subsubsection{Coulomb interaction ($c=1$)}

An extension of the effective action procedure to the problem with
the long ranged Coulomb interaction is by no means obvious. The
argument relies, to a major extend, on the detailed knowledge
obtained from an explicit analysis of the Finkelstein approach
which shows that the theory undergoes structural changes in the
limit where $N_r \rightarrow 0$ and $N_m \rightarrow \infty$. The
action is more complicated and now given by
\begin{equation}
 S = S_\sigma^\textrm{edge} [q] + S_\sigma^\textrm{bulk} [Q] + S_F [Q]
\label{Scoul}
\end{equation}
where $c=1$ is inserted in the expression for $S_F [Q]$.
Elimination of the matrix field variable $Q_0$ leads to the
definition of the effective action
\begin{equation}
 e^{S_\textrm{eff}^\textrm{bulk}[q]} = \int_{\partial V} {\mathcal D}[Q_0]
 e^{S_{\sigma}^\textrm{bulk} [t^{-1} Q_0 t] + S_F [t^{-1} Q_0 t]}.
 \label{Scoul1}
\end{equation}
On the basis of symmetries one can write down the following
explicit result
\begin{equation}
 S_\textrm{eff}^\textrm{bulk} [q] = -\frac{\sigma^{\prime}_{xx}}{8} \int d \mathbf{r}
 \tr (\nabla q)^{2}+ i{\theta^\prime} {\mathcal C}[q].\label{Scoul2}
\end{equation}
Here, the response parameters $\sigma^{\prime}_{xx} =\sigma_{xx}
(L)$ and $\theta^\prime = \theta (L)$ are evaluated in the limit
where $T \rightarrow 0$. It is important to emphasize that $S_F$
cannot be omitted from Eqs.~\eqref{Scoul}-\eqref{Scoul2}. The
reason is, as we already mentioned before, that this term
fundamentally affects the ultra violet singularity structure of
the
theory.~\cite{Finkelstein1,Finkelstein2,Finkelstein3,Unify2,Unify5}

The remaining part of the argument proceeds along similar lines as
before. Provided the system with Coulomb interactions generates a
mass gap, both parameters $\sigma^{\prime}_{xx}$ and
$\theta^\prime$ should vanish for $L$ large enough. A suitably
regulated action for the quantum Hall state has been obtained
previously and the result is as follows~\cite{Unify3}
\begin{eqnarray}
 S_\textrm{eff}[q] &=& \frac{k(\nu)}{2} \oint dx \tr t \nabla_x t^{-1}
 \Lambda \notag\\
 &+& \frac{\pi^2}{2} T \rho_\textrm{edge} \oint dx
 {\mathcal O}_F [q]\notag\\
 &-&\frac{\pi}{4} T k(\nu) \oint dx \oint dy  \tr I^\alpha_{-n}
 q(x)  v_\textrm{eff}^{-1} (x-y) \notag\\
 &\times & \tr I^\alpha_{n} q(y).
\end{eqnarray}
As before we have $\sigma_{xy}^\prime = k(\nu)$. Here, the
quantity $v_\textrm{eff}(x-y)$ contains the Coulomb interaction
$U_0(x-y)=1/|x-y|$. The Fourier transform is given by
\begin{equation}
v_\textrm{eff}(p) = \frac{k(\nu)}{2\pi\rho_\textrm{edge}}\left(1+
\rho_\textrm{edge} U_0 (p)\right).
\end{equation}
%
\subsection{\label{PO} Physical observables}
%

Next, for a detailed understanding of interaction effects it is
clearly necessary to develop a quantum theory for the {\em
observable} parameters $\sigma_{xx}^\prime$, $\sigma_{xy}^\prime$
or $\theta^\prime$, $z^\prime$ and $c^\prime$. At the same time it
is extremely important to show that the response quantities
defined by the effective action procedure are precisely the same
as those obtained from ordinary linear response theory. This will
be done in Appendix~\ref{AppA} where we embark on some of the
principal results of $\mathcal F$ algebra.

In this Section we recollect the $\mathcal F$ invariant
expressions for the observable parameters that will be used in the
remainder of this paper. As pointed out in the original
papers,\cite{Unify1,Unify2,Unify3,Unify5} the main advantage of
working with $\mathcal F$ invariant quantities is that they
facilitate renormalization group computations at finite
temperatures and frequencies. They are furthermore valid in the
entire range $0\leq c \leq 1$ and simpler to work with in general.
In the second part of this Section we briefly recall the results
of the theory in $2+\epsilon$ spatial dimensions.
\begin{widetext}
%
\subsubsection{\label{Form.PhysObser.Kubo}Kubo formula}
%

The response quantities $\sigma_{xx}^\prime$ and $\theta^\prime$
for arbitrary values of $c$ can be expressed in terms of
current-current correlations according to ~\cite{Pruisken3,Unify2}
\begin{eqnarray}
 \sigma_{xx}^\prime &=& - \frac{\sigma_{xx}}{4 n L^2}
 \int d \mathbf{r}\left \langle \tr
 [I_{n}^{\alpha },Q(\mathbf{r})][I_{-n}^{\alpha },Q(\mathbf{r})]
 \right \rangle
 + \frac{\sigma _{xx}^{2}}{8 n L^2} \int d \mathbf{r}
 \int d{\mathbf{r}^\prime}\left \langle \tr
 I_{n}^{\alpha }Q(\mathbf{r})\nabla Q(\mathbf{r})\tr I_{-n}^{\alpha
 }Q(\textbf{r}^{\prime })\nabla Q(\textbf{r}^{\prime})\right
 \rangle \label{resp1} \\
 \sigma_{xy}^\prime &=&  \sigma_{xy}
 + \frac{\sigma_{xx}^{2}}{8 n L^2} \int d \mathbf{r}\int d{\mathbf{r}^\prime}
 \varepsilon_{jk}\langle \tr
 I_{n}^{\alpha }Q(\mathbf{r})\nabla_j Q(\mathbf{r})
 \tr I_{-n}^{\alpha }Q(\textbf{r}^{\prime })\nabla_k
 Q(\textbf{r}^{\prime})\rangle . \label{resp2}
\end{eqnarray}
\end{widetext}
Here and from now onward the expectations are defined with the
respect to the theory of Eq.~\eqref{S} - ~\eqref{OF} and we assume
spherical boundary conditions.

\subsubsection{\label{Form.PhysObser.SH} Specific heat}
%

A natural definition of the observable quantity $z^\prime$ is
obtained through the derivative of the thermodynamic potential
with respect to temperature which is directly related to the
specific heat of the electron gas.~\cite{Unify2,Unify5} Write
\begin{eqnarray}
 \frac{\partial \ln \Omega}{\partial \ln T} &=& \pi T z
 \int d\mathbf{r} {\langle \mathcal{O}_F (Q)\rangle}
 \notag \\ \label{zcren1}
 &=& \pi T z^{\prime} \int d\mathbf{r} \mathcal{O}_F (\Lambda)
\end{eqnarray}
then the expression for $z^\prime$ becomes
\begin{equation}\label{zcren}
z^{\prime} = z \frac{\langle \mathcal{O}_F (Q)\rangle}{
\mathcal{O}_F (\Lambda)}.
\end{equation}
The expression for remaining observable $c^\prime$ is determined
by the general condition imposed on the {\em static response} of
the system which says that the quantity $z\alpha =z(1-c)$ remains
unaffected by the quantum
fluctuations.~\cite{Finkelstein1,Finkelstein2,Finkelstein3,Unify1,Unify2}
The second equation therefore reads as follows
\begin{equation}\label{zalpha}
z^\prime (1-c^\prime) = z (1-c)
\end{equation}
or $z^\prime \alpha^\prime = z \alpha$. Eq.~\eqref{zalpha} has
been explicitly verified in the theory of perturbative expansions.
In what follows we proceed and employ Eqs.~\eqref{zcren} and
\eqref{zalpha} for non-perturbative computational purposes as
well. A justification of this procedure is given in
Section~\ref{Omega} where we embark on the various different
subtleties associated with instanton calculus.
%
\subsubsection{\label{Form.PhysObser.BGF} $\beta$ and $\gamma$ functions}
%
The expressions of the previous Sections facilitate
renormalization group studies that include not only ordinary
perturbative expansions but also the non-perturbative effects of
instantons. Since much of the analysis is based on the theory in
$2+\epsilon$ spatial dimensions~\cite{Pruisken5} we shall first
recapitulate some of the results of the perturbative
renormalization group in two
dimensions.~\cite{Finkelstein3,Unify2,Unify5} Let $\mu^\prime$
denote the momentum scale associated with the observable theory
then the quantities $\sigma^\prime_{xx} =
\sigma_{xx}(\mu^\prime)$, $z^\prime = z(\mu^\prime)$ and $c^\prime
= c(\mu^\prime)$ can be expressed in terms of the renormalization
group $\beta$ and $\gamma$ functions according to (see
Ref.~[\onlinecite{PruiskenBurmistrov}])
\begin{eqnarray}
\sigma^\prime_{xx} &=& \sigma_{xx} +
 \int_{\mu_0}^{\mu^\prime} \frac{d\mu}{\mu} \beta_{\sigma} (\sigma_{xx},c)
 \label{rgsigma}\\
z^\prime &=& z~~~
 -\int_{\mu_0}^{\mu^\prime} \frac{d\mu}{\mu}\gamma_z(\sigma_{xx},c)
 z\label{rgz1} \\
 z^\prime \alpha^\prime &=& z \alpha\label{rgz2}
\end{eqnarray}
where
\begin{eqnarray}
\beta_{\sigma} (\sigma_{xx},c) &=& \beta_0(c) +
\frac{\beta_1(c)}{\sigma_{xx}} + O(\sigma_{xx}^{-2})\label{pb1}
\\
\gamma_z (\sigma_{xx},c) &=& \frac{c \gamma_0}{\sigma_{xx}} +
\frac{c \gamma_1(c)}{\sigma^2_{xx}} +
O(\sigma_{xx}^{-3}).\label{pb2}
\end{eqnarray}
The one-loop results are known for arbitrary value of the
crossover parameter $c$ and are given
by~\cite{Finkelstein3,Unify2}
\begin{equation}\label{bg0}
\beta_0(c) = \frac{2}{\pi}\left (1+\frac{\alpha}{c}\ln\alpha\right
), \qquad \gamma_0 = - \frac{1}{\pi}
\end{equation}
whereas the two-loop results were obtained for $c=1$ and $c=0$
only. In the case of electrons with the Coulomb interaction
($c=1$) the results are as follows~\cite{Unify2,Unify5}
\begin{eqnarray}
\beta_1(1) &=& \frac{4}{\pi^2} \Biggl [ 50+\frac{1}{6}-3\pi
^{2}+\frac{\pi ^{4}}{12}+\frac{19}{2}\zeta (3)+16G \notag \\ & & \hspace{0.7cm} -8\textrm{li}_{4}\left(\frac{1}{2}\right)+
\left (\frac{\pi ^{2}}{2}-44-7\zeta (3)\right )\ln 2 \notag \\
&& \hspace{0.7cm} +\left (16+\frac{\pi ^{2}}{3}\right )\ln ^{2}2
-\frac{1}{3}\ln
^{4}2-8\textrm{li}_{4}\left(\frac{1}{2}\right)\Biggr ] \notag \\
&\approx & 0.66
\label{bg1_0}\\
\gamma_1(1) &=& - \frac{3}{\pi^2} - \frac{1}{6}\approx
0.47\label{bg1_1}
\end{eqnarray}
where $\zeta(n)$ denotes the Riemann zeta function, $G=0.915\dots$
the Catalan constant, and $\textrm{li}_n(x) = \sum^\infty_{k=1}
x^k k^{-n}$ the polylogarithmic function. For free electrons
($c=0$) we have ~\cite{BHZ}
\begin{eqnarray}\label{bg1_2}
\beta_1(0) &=& \frac{1}{2\pi^2} \\
\gamma_1(0) &=& 0.\label{bg1_3}
\end{eqnarray}
The main objective of the present paper is to extend the results
for the observable theory Eqs.~\eqref{rgz1}-\eqref{pb1} to include
the effect of instantons and $\theta$ renormalization.

%
\subsubsection{\label{Flows} RG flows in $2+\epsilon$ dimensions}
%

\begin{figure}[tbp]
\includegraphics[width=80mm]{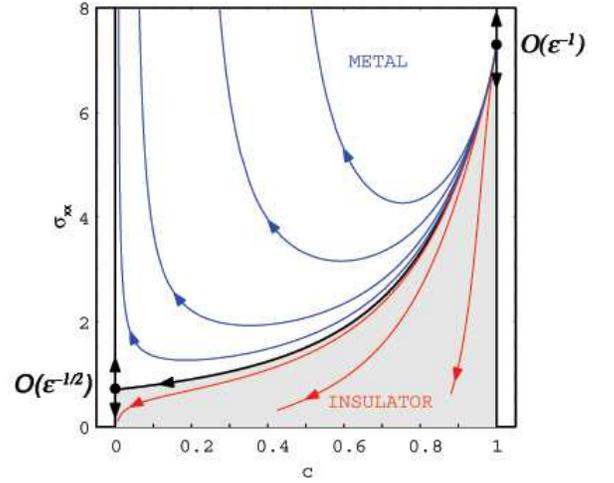}
\caption{Flow diagram in the $\sigma_{xx} ,c$ plane. Here
$\epsilon=0.1$, see text} \label{flowlines}
\end{figure}

For a general understanding of the problem it is important to
spell out the consequences of the theory in $2+\epsilon$
dimension. In this case the $\beta$ function is given by
\begin{equation}
\frac{d\sigma_{xx}}{d\ln\mu} = -\epsilon \sigma_{xx} +\beta_\sigma
(\sigma_{xx}, c).
\end{equation}
Following Eq.~\eqref{rgz2} we can express the renormalization of
$c$ as follows
\begin{equation}\label{bc1}
\frac{d c}{d\ln\mu} = \beta_c (\sigma_{xx}, c) = (1-c) \gamma_z
(\sigma_{xx}, c) .
\end{equation}
The renormalization group flow lines in the $(\sigma_{xx}, c)$
plane are sketched in Fig.~\ref{flowlines}. We see that there are
two critical fixed points describing a quantum phase transition
between a metal and an insulator. Along the Coulomb line ($c=1$)
the fixed point value is $\sigma_{xx}^{*} =O(\epsilon^{-1})$ and along the Fermi liquid line ($c=0$) we have
$\sigma_{xx}^{*} =O(\epsilon^{-1/2})$. The results show
that the problem with finite range interactions $0<c<1$ and the
Coulomb interaction problem $c=1$ belong to different universality
classes.

In two spatial dimensions the metallic phases ($\sigma_{xx} >
\sigma_{xx}^{*}$) disappear altogether indicating that all the
states of the (spin polarized or spinless) electron gas are now
Anderson localized, independent of the presence of
electron-electron interactions. This means that as far as the
quantum Hall effect is concerned one is generally faced with
identically the same fundamental difficulties as those previously
encountered in the scaling theory of the free electron
gas.~\cite{Pruisken1}

Let us see how the perturbative theory of localization and
interaction effects manages to describe an insensitivity of the
system to changes in the boundary conditions as outlined in the
previous Section. We consider for simplicity the problem with the
Coulomb interaction problem ($c=1$) in two dimensions
($\epsilon=0$). Since the response parameter $\sigma_{xx}^\prime$
is independent of the arbitrary momentum scale $\mu_0$ that
defines the ``renormalized'' theory $\sigma_{xx}(\mu_0)$ we
immediately obtain from Eq.~\eqref{rgsigma} the general scaling
result
\begin{equation}
\sigma_{xx}^\prime = \sigma_{xx} (\mu^\prime) = f_\sigma
(\mu^\prime \xi) \label{obs1}
\end{equation}
where $\mu^\prime$ is related to the linear dimension $L$ of the
system according to $\mu^\prime = L^{-1}$. The $\xi$ obeys the
differential equation
\begin{equation}
 \left( \frac{\partial}{\partial \ln \mu_0} +\beta_{\sigma}
 \frac{\partial}{\partial \sigma_{xx}} \right) \xi =0
\end{equation}
and can be identified with a dynamically generated correlation
length (localization length) of the system
\begin{equation}
 \xi = \mu_0^{-1} \sigma_{xx}^{-\beta_1 (1)/\beta_0^2 (1)}
 e^{\sigma_{xx}/\beta_0 (1)} . \label{obs2}
\end{equation}
Next, comparison of Eqs.~\eqref{obs1} and~\eqref{obs2} with the
expression of Eq.~\eqref{rgsigma} leads to the following explicit
(weak coupling) result for the scaling function $f_\sigma(X)$ with
$X = (\mu^\prime \xi)^{\beta_0 (1)}$
\begin{equation}
 f_\sigma (X)  \approx   \ln X +\frac{\beta_1 (1)}{\beta_0 (1)} \ln \ln X +
 \frac{\beta_1^2 (1)}{\beta_0^2 (1)} \frac{\ln \ln X}{\ln X}, \quad X\gg 1.
 \label{obs3}
\end{equation}
The statement of exponential localization can now be formulated by
saying that in the regime of strong coupling the scaling function
$f(X)$ vanishes according to
\begin{equation}
f_\sigma (X) \approx \exp \left ( -X^{-1/\beta_0(1)} \right ) =
\exp\left ( -1/(\mu^\prime \xi) \right ),\quad X\ll 1.
\end{equation}
These naive expectations are fundamentally modified by the
$\theta(\nu)$ dependence of the theory which is invisible in
perturbation theory. In anticipation of the results of the present
paper we can say that the fixed point structure of the theory in
$2+\epsilon$ dimension (Fig.~\ref{flowlines}) is reminiscent of
what happens in the theory in two dimensions at $\theta
(\nu)=\pi$. Although the physics is very different in both cases,
it is nevertheless important to keep in mind that the
renormalization is determined, to a major extend, by the global
symmetries of the problem. In particular, since $\mathcal{F}$
invariance is retained along the Coulomb line $c=1$ only and
broken otherwise one generally expects, like the theory in
$2+\epsilon$ dimensions, that the problem with finite range
interactions $0<c<1$ lies in the domain of attraction of the Fermi
liquid line $c=0$ whereas the Coulomb interaction problem $c=1$
describes a distinctly different, non-Fermi liquid universality
class. Armed with the insights obtained from the perturbative
renormalization group we next embark - for the remainder of this
paper - on the problem of instantons.

%
%
\section{\label{Form.Instan}Instantons}
%
%
In this Section we recapitulate the instanton analysis for the
Grassmannian non-linear $\sigma$ model
(Sections~\ref{Form.Instan1} and \ref{QFAI}). We introduce the
methodology of spatially varying masses which essentially adapts
the interaction part of the action $S_F$ to the metric of a sphere
(Section~\ref{P.Metric}). In Section~\ref{QFAI.GL1} we derive the
complete action for the small oscillator problem that will be used
as a starting point for the remainder of this paper.
%
%
\subsection{\label{Form.Instan1}Introduction}
%
%
%
%
\subsubsection{\label{Form.Instan12}The action $S_\sigma$}
%
%
On the basis of the Polyakov-Schwartz inequality~\cite{Pruisken2}
\begin{equation}
\frac{1}{8}\int d \mathbf{r}\mathop{\rm tr}\nolimits(\nabla
Q)^{2}\geq 2\pi |\mathcal{C}[Q]| \label{ineqS}
\end{equation}
one can construct stable matrix field configurations (instantons)
for each of the discrete topological sectors labelled by the
integer $\mathcal{C}[Q]$. The classical action $S_\sigma$ is
finite
\begin{equation}
S_{\sigma }^{\mathrm{inst}}=-2\pi \sigma _{xx}|\mathcal{C}[Q]|+\,
i\theta \mathcal{C}[Q]. \label{SsigmaInst}
\end{equation}
The single instanton configuration with the topological charge
$\mathcal{C}[Q]=\pm 1$ which is of interest to us can be
represented as follows ~\cite{Pruisken2,PruiskenBaranov}
\begin{equation}
Q_{\textrm{inst}}(\textbf{r})= \mathcal{T}_0^{-1}
\Lambda_{\textrm{inst}}(\textbf{r}) \mathcal{T}_0, \qquad
\Lambda_{\textrm{inst}}(\textbf{r}) = \Lambda + \rho(\textbf{r}).
\label{InstSol}
\end{equation}
Here, the matrix $\rho_{nm}^{\alpha\beta}(\textbf{r})$ has four
non-zero matrix elements only
\begin{eqnarray}
\rho _{00}^{11} &=&-\rho _{-1-1}^{11}=-\frac{2\lambda ^{2}}{
|z-z_{0}|^{2}+\lambda ^{2}}  \label{rho1} \\
\rho _{0-1}^{11} &=&\bar{\rho}_{-10}^{11}=\frac{2\lambda
(z-z_{0})}{|z-z_{0}|^{2}+\lambda ^{2}}  \notag
\end{eqnarray}
with $z= x+\,i y$. The manifold of instanton parameters consists
the quantity $z_{0}$ denoting the {\em position} of the instanton,
the parameter $\lambda$ which equals the {\em scale size} as well
as the global unitary rotation $\mathcal{T}_0$ which describes the
{\em orientation} in the coset space $U(2N)/U(N) \times U(N)$.
These parameters do not change the value of the classical action
$S_{\sigma }^{\mathrm{inst}}$. The {\em anti-instanton} with
$\mathcal{C}[Q]= -1$ is simply obtained by complex conjugation.
%
%
\subsubsection{\label{Form.Instan13}The action $S_F$}
%
%

In the presence of mass terms like the \emph{singlet interaction}
term $S_F$ the idea of stable topologically non-trivial field
configurations becomes generally more complicated. The minimum
action requirement, for example, immediately tells us that the
global matrix $\mathcal{T}_0$ is now restricted to run over the
subgroup $U(N)\times U(N)$ only.~\cite{PruiskenBurmistrov} Instead
of Eq.~\eqref{InstSol} we therefore write
\begin{equation}\label{InstSol1a}
Q_{\rm inst}(\textbf{r}) = U^{-1} {\Lambda}_{\rm inst}
(\textbf{r}) U = \Lambda + U^{-1} \rho(\textbf{r}) U
\end{equation}
with $U \in U(N) \times U(N)$. Next, by substituting
Eq.~\eqref{InstSol1a} into Eq.\eqref{SF} one can split $S_F$ into
a topologically trivial piece and an instanton peace as follows
\begin{equation}\label{SFInst1}
S_F[Q_{\textrm{inst}}] = S_F[\Lambda] + S_F^{\textrm{inst}}
\end{equation}
where
\begin{equation}\label{SFInst2}
S_F[\Lambda] = -2 \pi T z \int d \mathbf{r} \tr \eta \Lambda
\end{equation}
and
\begin{eqnarray}
S_F^{\textrm{inst}}[U] &=& \pi T z \int d \mathbf{r} \Bigl [
c\sum\limits_{\alpha n}\mathop{\rm tr} \nolimits I_{n}^{\alpha
}U^{-1}\rho U\mathop{\rm tr}\nolimits I_{-n}^{\alpha }U^{-1}\rho U
\notag \\ && \hspace{1.2cm} +4 \mathop{\rm tr}\nolimits \eta
U^{-1}\rho U\Bigr ].\label{SFInst3}
\end{eqnarray}
Similarly we can write the classical contribution to the
thermodynamic potential as the sum of two pieces
\begin{equation}\label{TPInst1}
\Omega^\textrm{class} = \Omega_0^\textrm{class}
+\Omega_{\textrm{inst}}^\textrm{class}
\end{equation}
where $\Omega_0^\textrm{class}$ is the contribution of the trivial
vacuum
\begin{equation}\label{TPInst2}
\Omega_0^\textrm{class} =  S_F[\Lambda] = -2 \pi T z \int d
\mathbf{r} \tr \eta \Lambda
\end{equation}
and $\Omega_{\textrm{inst}}$ is the instanton piece
\begin{equation}\label{TPInst3}
\Omega_{\textrm{inst}}^\textrm{class} = \int_\textrm{inst} \exp
\left ( -2\pi\sigma_{xx} \pm \, i \theta + S_F^{\textrm{inst}}[U]
\right ) .
\end{equation}
The subscript ``$\rm inst$'' indicates that the integral is over
the manifold of instanton parameters $z_0$, $\lambda$ and $U$.

%
\begin{figure}[tbp]
\includegraphics[width=80mm]{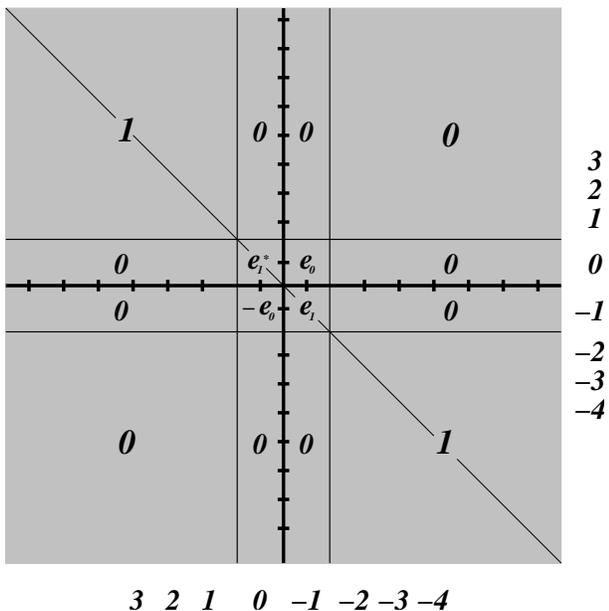}
\caption{The instanton matrix $R$, see text.} \label{matrixR}
\end{figure}
%

One of the main complications next is that the action
$S_F^{\textrm{inst}}[U]$ is not finite but, rather, it diverges
logarithmically in the size of the system. Although these and
other complications associated with mass terms are quite well
known, the resolution that has been proposed is formal at best and
useless for practical purposes.~\cite{ConstInst} There is in this
respect a true advantage to be gained if one follows up on the
idea of {\em spatially varying masses} which has recently been
introduced and analyzed in great detail by the
authors.~\cite{PruiskenBurmistrov} This methodology not only
extends the formalism developed for the massless theory in a
natural fashion, but also lends itself to a non-perturbative
analysis of the renormalization group $\beta$ and $\gamma$
functions of the theory. Before embarking on the specific problem
of the interacting electron gas it is necessary to first
recapitulate some of the main results obtained for the ordinary
Grassmannian manifold.~\cite{PruiskenBurmistrov} This will be done
in the Sections below where we generalize the harmonic oscillator
problem to include an arbitrary range of Matsubara frequencies.
The most important results are written in Section ~\ref{QFAI.GL1}
Table~\ref{Tf} which contains the complete action of quantum
fluctuations about the single instanton.
%
%
\subsection{\label{QFAI}Quantum fluctuations}
%
%
%
\subsubsection{\label{QFAI.P}Preliminaries}
%
To obtain the most general matrix field variable $Q$ with
topological charge equal to unity we first rewrite the instanton
solution $\Lambda_{\rm inst}$ in Eqs.~\eqref{rho1} and
\eqref{InstSol1a} as a unitary rotation $R$ about the trivial
vacuum $\Lambda$
\begin{equation}\label{RlambdaR}
\Lambda_{\rm inst} = R^{-1} \Lambda R .
\end{equation}
From now onward we use the following notation for an arbitrary
matrix $A$
\begin{eqnarray}
 A^{\alpha\beta}_{mn} =
\begin{pmatrix}
 A^{\alpha\beta}_{n_1 n_3} & ~~~ A^{\alpha\beta}_{n_1 n_2}\\ \\
 A^{\alpha\beta}_{n_2 n_1} & ~~~ A^{\alpha\beta}_{n_2n_4}
\end{pmatrix} .\\\nonumber
\end{eqnarray}
Here, the $n_i$ with {\em odd} subscripts $i$ denote the indices
for {\em positive} Matsubara frequencies. Similarly, the {\em
even} subscripts $i$ refer to the {\em negative} Matsubara
frequencies. Hence, the indices $n_1$ and $n_3$ run over the set
of {\em non-negative} integers $\{ 0, 1, 2,\dots\}$. The indices
$n_2$ and $n_4$ run over the set of {\em negative} integers $\{
-1, -2, -3,\dots\}$. Fully written out the different frequency
blocks of the unitary matrix $R^{\alpha\beta}_{mn}$ now become
\begin{eqnarray}
 R^{\alpha\beta}_{n_1 n_3} &=& {\delta}^{\alpha\beta} \delta_{n_1 n_3} \left[ 1+
 (\bar{e}_{1}-1) \delta^{\alpha 1} \delta _{n_1 ,0} \right] \\
 R^{\alpha\beta}_{n_2 n_4} &=& {\delta}^{\alpha\beta} \delta_{n_2 n_4} \left[ 1+
 ({e}_{1}-1) \delta^{\alpha 1} \delta _{n_2 ,-1} \right] \\
 R^{\alpha\beta}_{n_1 n_2} &=&  {\delta}^{\alpha\beta} \delta^{\alpha 1}
 \delta_{n_1 ,0} \delta_{n_2,-1} \left[ e_0 \right]
 \\
 R^{\alpha\beta}_{n_2 n_1} &=&  {\delta}^{\alpha\beta} \delta^{\alpha 1}
 \delta_{n_1 ,0} \delta_{n_2,-1} \left[ - e_0 \right] = - R^{\alpha\beta}_{n_1 n_2}
\end{eqnarray}
where the quantities $e_0$ and $e_1$ are defined by
\begin{eqnarray}\label{e0}
 e_0 &=& \frac{\lambda}{\sqrt{|z-z_0 |^2 + \lambda^2}}
 \\\label{e1}
 e_1 &=& \frac{z-z_0}{\sqrt{|z-z_0 |^2 + \lambda^2}} .
\end{eqnarray}
The structure of the matrix $R^{\alpha\beta}_{mn}$ is illustrated
in Fig.~\ref{matrixR}. It is a simple matter next to generalize
Eq.~\eqref{RlambdaR} and the result is
\begin{equation}\label{top1}
Q=\mathcal{T}^{-1}_0 R^{-1} {\mathcal V} \, R\, \mathcal{T}_0 .
\end{equation}
Here, $\mathcal{T}_0$ denotes a global $U(2N)$ rotation and the
matrix ${\mathcal V}$ with ${\mathcal V}^2 = {\bf 1}$  represents
the small fluctuations about the one instanton. Write
\begin{equation}\label{qpar}
{\mathcal V} = w +\Lambda \sqrt{1 - w^2}
\end{equation}
with
\begin{equation}\label{vfield}
w=\begin{pmatrix}
0 & v \\
v^\dag & 0
\end{pmatrix}
\end{equation}
then the matrix ${\mathcal V}$ can formally be written as a series
expansion in powers of the $N \times N$ complex matrices $v$,
$v^{\dag}$ which are taken as the independent field variables in
the problem.
%
\subsubsection{\label{P.Curved} Stereographic projection}
%
Eq. ~\eqref{top1} lends itself to an exact analysis of the small
oscillator problem. First we recall the results obtained for the
free electron theory,~\cite{Pruisken2}
\begin{equation}\label{Ai}
\frac{\sigma_{xx}}{8} \int d \textbf{r}\tr (\nabla_j Q)^2 =
\frac{\sigma_{xx}}{8} \int d \textbf{r}\tr [\nabla_j + {A}_j ,
{\mathcal V}]^2
\end{equation}
where the matrix $A_j$ contains the instanton degrees of freedom
\begin{equation}
{A}_j = R \mathcal{T}_0 \nabla_j \mathcal{T}^{-1}_0 R^{-1} =R
\nabla_j R^{-1} .
\end{equation}
By expanding the ${\mathcal V}$ in Eq.~\eqref{Ai} to quadratic
order in the quantum fluctuations $v,v^\dag$ we obtain the
following results
\begin{widetext}
\begin{eqnarray}
&&\frac{\sigma_{xx}}{8} \int d \textbf{r} \tr [\nabla_j + {A}_j ,
{\mathcal V}]^2  = \notag \\ \notag \\
&&=\frac{\sigma_{xx}}{4} \int d
\textbf{r} \mu^2(\textbf{r}) \Biggl [
\sum\limits_{\alpha=2}^{N_r}\sum\limits_{\beta=2}^{N_r}\sum\limits_{n_1n_2}
v^{\alpha\beta}_{n_1n_2}O^{(0)}v^{\dag\beta\alpha}_{n_2n_1}  +
\sum\limits_{\alpha=2}^{N_r} \Biggl
(\sum\limits_{n_1n_2}{}^{^{\prime\prime }}
v_{n_1n_2}^{1\alpha}O^{(0)} v^{\dag \alpha 1}_{n_2n_1} +
\sum\limits_{n_1n_2}{}^{''} v_{n_1n_2}^{\alpha 1 }O^{(0)} v^{\dag
1 \alpha}_{n_2n_1}  \notag \\
&& \hspace{2.3cm} + \sum\limits_{n_1}{}^{^{\prime }}
v_{n_1,-1}^{1\alpha}O^{(0)} v^{\dag \alpha 1}_{-1,n_1} + \sum\limits_{n_2}{}^{^{\prime }}
v_{0,n_2}^{\alpha 1}O^{(0)} v^{\dag 1\alpha}_{n_2,0} +
\sum\limits_{n_1} v_{n_1,-1}^{\alpha 1}O^{(1)} v^{\dag
1\alpha}_{-1,n_1} +
\sum\limits_{n_2} v_{0,n_2}^{1\alpha}O^{(1)} v^{\dag \alpha
1}_{n_2,0} \Biggr )   \notag \\
&& \hspace{2.3cm}+ \sum\limits_{n_1n_2}{}^{^{\prime\prime }} v_{n_1n_2}^{11}O^{(0)}
v^{\dag 11}_{n_2n_1} +
\sum\limits_{n_1}{}^{^{\prime }} v_{n_1,-1}^{11}O^{(1)} v^{\dag
11}_{-1,n_1} + \sum\limits_{n_2}{}^{^{\prime }}
v_{0,n_2}^{11}O^{(1)} v^{\dag 11}_{n_2,0}+
v_{0,-1}^{11}O^{(2)} v^{\dag 11}_{-1,0}
\Biggr
] \label{fluct11}
\end{eqnarray}
\end{widetext}
The ``prime'' on the summation signs are defined as follows
\begin{equation}
\sum\limits_{n_{1}}{}^{^{\prime
}}=\sum\limits_{n_{1}=1}^{N_{m}-1}~,~~
\sum\limits_{n_{2}}{}^{^{\prime}}
=\sum\limits_{n_{2}=-2}^{-N_{m}}. \label{sumsign}
\end{equation}
The three different operators $O^{(a)}$ with $a=0,1,2$ are given
as
\begin{equation}\label{Oa0}
O^{(a)} = \frac{(r^2 + \lambda^2 )^2}{4\lambda^2} \left[ \nabla_j
+ \frac{i a }{r^2 + \lambda^2} \varepsilon_{jk} r_k \right]^2
+\frac{a}{2}.
\end{equation}
The introduction of a measure $\mu^2(\textbf{r})$ for the spatial
integration in Eq.~\eqref{fluct11},
\begin{equation}\label{measure}
\mu({r}) = \frac{2\lambda}{r^2 + \lambda^2}
\end{equation}
indicates that the quantum fluctuation problem is naturally
defined on a sphere with radius $\lambda$. It is convenient to
employ the stereographic projection
\begin{eqnarray}\label{eta}
\eta &=& \frac{r^{2}-\lambda^{2}}{r^{2}+\lambda^{2}},\qquad -1 <
\eta < 1 \\
\theta &=& \tan^{-1} \frac{y}{x},\qquad 0 \leq \theta < 2 \pi
.
\end{eqnarray}
In terms of $\eta, \theta$ the integration can be written as
\begin{equation}
\int d \textbf{r}\mu^2({r})  = \int d \eta d\theta .
\end{equation}
Moreover,
\begin{equation}\label{e0e1}
e_{0} = \sqrt{\frac{1-\eta}{2}},\qquad \qquad e_{1} =
\sqrt{\frac{1+\eta}{2}} e^{i \theta}
\end{equation}
and the operators become
\begin{eqnarray}
O^{(a)} &=& \frac{\partial}{\partial \eta} \left [ (1-\eta^{2})
\frac{\partial}{\partial \eta} \right ] + \frac{1}{1-\eta^{2}}
\frac{\partial^{2}}{\partial^{2} \theta} - \frac{i a}{1-\eta}
\frac{\partial}{\partial \theta} \notag \\
 &-& \frac{a^{2}}{4}
\frac{1+\eta}{1-\eta} + \frac{a}{2} \label{Oa}
\end{eqnarray}
with $a=0,1,2$. Finally, using Eq.\eqref{fluct11} we can count the
total number of fields $v^{\alpha\beta}$ on which each of the
operators $O^{(a)}$ act. The results are listed in
Table~\ref{TZM}.

%

\begin{table}[tbp]
\begin{ruledtabular}
\caption{Number of zero modes, see text.}
\begin{tabular}{||c|c|c||}\hline
Operator & Number of fields $v_{n_{1}n_{2}}^{\alpha\beta}$
& Degeneracy \\ & & \\
\hline
$O^{(0)}$ & $(N-1)^{2}$ & $1$ \\
& &  \\
$O^{(1)}$ & $2(N-1)$ & $2$ \\
& &  \\
 $O^{(2)}$ & $1$ & $3$\\\hline
\end{tabular}
\label{TZM}
\end{ruledtabular}
\end{table}

%
\subsubsection{\label{ES}Energy spectrum}
%

We are interested in the eigenvalue problem
\begin{equation}\label{EigEq} O^{(a)} \Phi^{(a)} (\eta, \theta) =
E^{(a)} \Phi^{(a)} (\eta, \theta)
\end{equation}
where the set of eigenfunctions $\Phi^{(a)}$ are taken to be
orthonormal with respect to the scalar product
\begin{equation}
(\bar{\Phi}^{(a)}_1 , \Phi^{(a)}_2) = \int d\eta d\theta\,
\bar{\Phi}^{(a)}_1(\eta, \theta) \Phi^{(a)}_2(\eta, \theta) .
\end{equation}
The Hilbert space of square integrable eigenfunctions is given in
terms of Jacobi polynomials,
\begin{equation}
P_{n}^{\alpha ,\beta }(\eta
)=\frac{(-1)^{n}}{2^{n}n!}\frac{(1-\eta
)^{-\alpha }}{(1+\eta )^{\beta }}\frac{d^{n}}{d\eta ^{n}}%
\frac{(1-\eta )^{n+\alpha }}{(1+\eta )^{-n-\beta }}.
\label{Jacobi}
\end{equation}
Introducing the quantum number $J$ to denote the discrete energy
levels
\begin{equation}\label{Ev}
\begin{array}{cclrcl}
E^{(0)}_J &=& J(J+1),\qquad & J &=& 0,1, \dots\\
E^{(1)}_J &=& (J-1)(J+1),\qquad & J &=& 1,2, \dots\\
E^{(2)}_J &=& (J-1)(J+2), \qquad & J &=& 1,2, \dots
\end{array}
\end{equation}
then the eigenfunctions are labelled by $(J,M)$ and can be written
as follows
\begin{eqnarray}
\Phi^{(0)}_{J,M} &=& C_{J,M}^{(0)}
e^{i M\theta}(1-\eta^2)^{M/2}P_{J-M}^{M,M}(\eta)\notag \\
C_{J,M}^{(0)} &=&
\frac{\sqrt{\Gamma(J-M+1)\Gamma(J+M+1)(2J+1)}}{2^{M+1}\sqrt\pi\Gamma(J+1)}\notag\\
&&\hspace{3.5cm} M = -J,\dots,J\label{Ef1}
\end{eqnarray}
\begin{eqnarray}
\Phi^{(1)}_{J,M}&=& C_{J,M}^{(1)} e^{i
M\theta}(1-\eta^2)^{M/2}(1-\eta)^{1/2}
P_{J-M-1}^{M+1,M}(\eta)\notag\\
C_{J,M}^{(1)}&=&
\frac{\sqrt{\Gamma(J-M)\Gamma(J+M+1)}}{2^{M+1}\sqrt\pi\Gamma(J)}\notag
\\
&&\hspace{3cm} M  = -J,\dots,J-1 \label{Ef2}
\end{eqnarray}
\begin{eqnarray}
\Phi^{(2)}_{J,M}&=& C_{J,M}^{(2)} e^{i
M\theta}(1-\eta^2)^{M/2}(1-\eta)P_{J-M-1}^{M+2,M}(\eta)\notag\\
C_{J,M}^{(2)}&=&
\frac{\sqrt{\Gamma(J-M)\Gamma(J+M+2)(2J+1)}}{2^{M+2}\sqrt\pi\Gamma(J)\sqrt{J(J+1)}}
\notag\\
&&\hspace{2cm} M = -J-1,\dots,J-1.\label{Ef3}
\end{eqnarray}

\begin{table*}[tbp]
\begin{ruledtabular}
\caption{Zero energy modes $v_{n_1 n_2}^{\alpha\beta}$ expressed
in terms of $t_{mn}^{\alpha \beta }$, $\delta\lambda$, $\delta
z_0$ and $\Phi^{(a)}_{JM}$, see text.}
\begin{tabular}{||c|c||c|c|c||}\hline
 & & & & \\
$\alpha$ $\beta$ & $n_1$ $n_2$ & $O^{(0)}$ & $O^{(1)}$& $O^{(2)}$
\\
 & & & & \\
\hline \hline
 & & & & \\
  $\alpha > 1, \, \beta > 1$ & $n_1 \geq 0,\, n_2
\leq
-1$ & $2 i t^{\alpha\beta}_{n_1n_2} \Phi^{(0)}_{0,0}$ & & \\
 & & & & \\
\hline
 & & & & \\
$\alpha > 1, \, \beta = 1$ & $n_1 \geq 0,\, n_2 = -1$  &
 & $2 i \sqrt{2\pi} (t^{\alpha 1}_{n_1,-1} \Phi^{(1)}_{1,-1} - t^{\alpha 1}_{n_1,0}
 \Phi^{(1)}_{1,0})$&
 \\
 & & & & \\
& $n_1 \geq 0,\, n_2 < -1$
 & $2 i t^{\alpha 1}_{n_1 n_2} \Phi^{(0)}_{0,0}$ & &  \\
 & & & & \\
\hline
 & & & & \\
$\alpha = 1, \, \beta > 1$ & $n_1 > 0,\, n_2 \leq -1$ & $2 i
t^{1\beta}_{n_1 n_2} \Phi^{(0)}_{0,0}$  & &  \\
 & & & & \\
& $n_1 = 0,\, n_2 \leq -1$ & & $2 i \sqrt{2\pi} (t^{1
\beta}_{0,n_2} \Phi^{(1)}_{1,-1} + t^{1\beta}_{-1,n_2}
 \Phi^{(1)}_{1,0})$ & \\
 & & & & \\
\hline
& & & & \\
$\alpha = 1, \, \beta = 1$ & $n_1 > 0,\, n_2 < -1$ & $2 i
t^{\alpha\beta}_{n_1n_2} \Phi^{(0)}_{0,0}$ &  &\\
 & & & & \\
& $n_1 = 0,\, n_2 < -1$ &  & $2 i \sqrt{2\pi} (t^{11}_{0,n_2}
\Phi^{(1)}_{1,-1} + t^{11}_{-1,n_2}
 \Phi^{(1)}_{1,0})$ & \\
 & & & & \\
& $n_1 > 0,\, n_2 = -1$ & &$2 i \sqrt{2\pi} (t^{11}_{n_1,-1}
\Phi^{(1)}_{1,-1} - t^{11}_{n_1,0}
 \Phi^{(1)}_{1,0})$ & \\
 & & & & \\
& $n_1 = 0,\, n_2 = -1$ & & & $4 i\sqrt{\frac{\pi}{3}}\Bigl [
t_{-1,0}^{11} \Phi^{(2)}_{1,-2} + \frac{1}{\sqrt2}
(t_{-1,-1}^{11}-t_{0,0}^{11}+ i
\frac{\delta\lambda}{\lambda})\Phi^{(2)}_{1,-1}$ \\
& & & & $- (t_{0,-1}^{11} -
\frac{\delta \bar{z}_0}{\lambda})\Phi^{(2)}_{1,0}\Bigr ] $\\
 & & & & \\
\hline
\end{tabular}
\label{TZMROT}
\end{ruledtabular}
\end{table*}

%
\subsubsection{\label{ZM}Zero modes}
%

From Eq.~\eqref{Ev} we see that the operators $O^{(0)}$ has a zero
frequency mode $E^{(0)}_J = 0$ for $J=0$. Similarly, we have
$E^{(1)}_J = E^{(2)}_J = 0$ for $J=1$. The corresponding
eigenfunctions can be written as follows
\begin{equation}\label{ZM.0}
\begin{array}{cclcllcl}
O^{(0)} & \Longrightarrow & \Phi_{0,0}^{(0)} & = & 1 \quad & & &
\\
O^{(1)} & \Longrightarrow & \Phi_{1,-1}^{(1)} & = &
\displaystyle\frac{1}{\sqrt{2 \pi}}\bar{e}_{1} ,\quad  &
\Phi_{1,0}^{(1)}&= &\displaystyle\frac{1}{\sqrt{2
\pi}}e_{0} \\
 O^{(2)}& \Longrightarrow & \Phi_{1,-2}^{(2)} & = & \displaystyle
\sqrt{\frac{3}{4\pi}} \bar{e}_{1}^{2}, \quad
&\displaystyle\Phi_{1,-1}^{(2)} & = &\displaystyle\sqrt{\frac{3}{2
\pi}} e_{0} \bar{e}_{1} \\
& & \Phi_{1,0}^{(2)} & = & \displaystyle\sqrt{\frac{3}{4\pi}}
e_{0}^{2}. & & &
\end{array}
\end{equation}
Here, the quantities $e_0$ and $e_1$ are defined in
Eqs.~\eqref{e0} and \eqref{e1} (see also Eq.~\eqref{e0e1}). The
number of the zero modes of each $O^{(a)}$ is listed in
Table~\ref{TZM}. The total we find $2 (N^2+2N)$ zero modes in the
problem.

Next, it is important to show that these zero modes precisely
correspond to all the instanton degrees of freedom contained in
the matrices $R$ and $\mathcal{T}_0$ of Eq.~\eqref{top1}. For this
purpose we write the instanton solution as follows
\begin{equation}
Q_{\rm inst} (\xi_i) = U_\textrm{inst}^{-1} (\xi_i) \Lambda
U_\textrm{inst} (\xi_i).
\end{equation}
Here, $U_\textrm{inst} = R\, \mathcal{T}_0$ and the $\xi_i$ stand
for the parameters $z_0$, $\lambda$ and the generators of
$\mathcal{T}_0$. An infinitesimal change in the instanton
parameters $\xi_i \rightarrow \xi_i + \varepsilon_i$ can be
written in the form of Eq.~\eqref{top1} as follows
\begin{equation}
Q_{\rm inst} (\xi_i + \varepsilon_i) = U_\textrm{inst}^{-1}
(\xi_i) {\mathcal V} _\varepsilon U_\textrm{inst} (\xi_i)
\end{equation}
where to linear order in $\varepsilon_i$ we can write
\begin{eqnarray}
{\mathcal V}_\varepsilon &=& \Lambda - \varepsilon_i \left[
U_\textrm{inst}
\partial_{i} U_\textrm{inst}^{-1} , \Lambda \right] \notag \\
&=&
\begin{pmatrix}
  {\bf 1} & 2\varepsilon_i \left[ U_\textrm{inst} \partial_{i} U_\textrm{inst}^{-1}
  \right]_{n_1n_2}^{\alpha\beta} \\
  - 2\varepsilon_i \left[ U_\textrm{inst} \partial_{i} U_\textrm{inst}^{-1}
  \right]_{n_2n_1}^{\alpha\beta} &
  -{\bf 1}
\end{pmatrix}.\notag\\
\end{eqnarray}
We have written $\partial_{i} = \partial / \partial {\xi_i}$. By
comparing this expression with Eq.~\eqref{top1} we see that the
fluctuations tangential to the instanton manifold can be expressed
in terms of the matrix field variables $v, v^\dag$ according to
\begin{eqnarray}
v^{\alpha\beta}_{n_1n_2} &=& ~~ 2\varepsilon_i \left[
U_\textrm{inst}
\partial_{i} U_\textrm{inst}^{-1} \right]_{n_1n_2}^{\alpha\beta} \\
\left[ v^{\dag} \right]^{\alpha\beta}_{n_2n_1} &=& -2\varepsilon_i
\left[ U_\textrm{inst}
\partial_{i} U_\textrm{inst}^{-1}
\right]_{n_2n_1}^{\alpha\beta}.
\end{eqnarray}
To obtain explicit expressions it suffices to expand
$\mathcal{T}_0$ about unity
\begin{eqnarray}
\mathcal{T}_0 &=& 1 +i\, t
\end{eqnarray}
and write
\begin{equation}
 R(\lambda+\delta\lambda, z_0 + \delta z_0) = R(\lambda, z_0 )
 +\delta\lambda ~ \partial_\lambda R + \delta z_0 ~
 \partial_{z_0} R .
\end{equation}
The expression for $v$ now becomes
\begin{eqnarray}
v^{\alpha\beta}_{n_1n_2} &=& ~~ 2i \left[ R t R^{-1}
\right]_{n_1n_2}^{\alpha\beta} + 2 \delta\lambda \left[ R
\partial_{\lambda} R^{-1}
\right]_{n_1n_2}^{\alpha\beta} \nonumber \\
&& + 2 \delta z_0 \left[ R
\partial_{z_0} R^{-1}
\right]_{n_1n_2}^{\alpha\beta} .
\end{eqnarray}
Notice that $ v^\dag $ is just the hermitian conjugate of $ v $ as
it should be. In Table~\ref{TZMROT} we present the complete list
of zero energy modes $v_{n_{1}n_{2}}^{\alpha \beta }$ written in
terms of $t_{mn}^{\alpha \beta }$, $\delta\lambda$ and $\delta
z_0$ as well as the eigenfunctions $\Phi^{(a)}_{JM}$.

\begin{figure}[tbp]
\includegraphics[width=80mm]{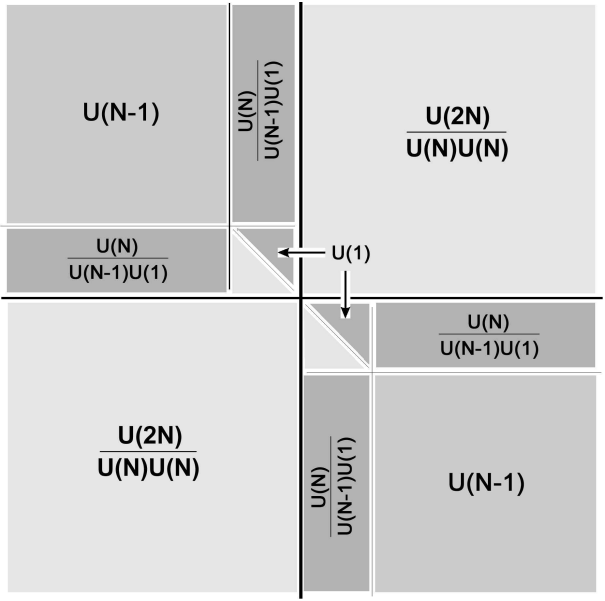}
\caption{Symmetry breaking by the one instanton, see text}
\label{FIG7}
\end{figure}

In these expressions $t^{\alpha\beta}_{n_1n_2}$ and
$t^{\alpha\beta}_{n_2n_1}$ denote the generators of $U(2N)/U(N)
\times U(N)$. The $t^{\alpha 1}_{n_1,0}$ and $t^{1\alpha}_{0,n_1}$
with $n_1\neq 0$ and $\alpha=1$ are the generators of a
$U(N)/U(N-1) \times U(1)$ rotation. The same holds for $t^{\alpha
1}_{n_2,-1}$ and $t^{1\alpha}_{-1,n_2}$ with $n_2\neq -1$ and
$\alpha=1$. Finally, $t_{0,0}^{11} -t_{-1,-1}^{11}$ denotes the
$U(1)$ generator corresponding to rotations of the $O(3)$
instanton in the $xy$ plane. The number of instanton degrees of
freedom adds up to $2(N^2+2N)$ which is that same as the number of
zero modes in the problem. The various different generators $t$ of
the instanton manifold is illustrated in Fig.~\ref{FIG7}.

%
\subsection{\label{P.Metric} Spatially varying masses}
%

In the previous Section we have seen that the instanton problem
naturally acquires the geometry of a {\em sphere}. This clearly
complicates the problem of mass terms in the theory which are
usually written in {\em flat} space. To deal with this problem we
shall modify the definition of the singlet interaction term and
introduce a spatially varying momentum scale $\mu (\textbf{r})$ as
follows
\begin{equation}\label{zRen}
z\to z \mu^2(\textbf{r}),\qquad zc\to zc \mu^2(\textbf{r})
\end{equation}
such that the action $S_F$ is now {\em finite} and can be written
as
\begin{eqnarray}\notag
S_{F}[Q] &\to& \pi T z \int d \mathbf{r} \mu^2(\textbf{r})\Bigl (
c\sum\limits_{\alpha n}\mathop{\rm tr} \nolimits I_{n}^{\alpha
}Q\mathop{\rm tr}\nolimits I_{-n}^{\alpha }Q \\ && \hspace{2.1cm}
+ 4 \mathop{\rm tr}\nolimits\eta Q - 6 \mathop{\rm
tr}\nolimits\eta \Lambda \Bigr ).\label{Ohmu}
\end{eqnarray}
As we will show below, in Sections~\ref{EC} and \ref{Imani}, the
introduction of a spatially varying momentum scale
$\mu(\textbf{r})$ permits the development of a complete quantum
theory of the interacting electron gas that is defined on a
sphere. Although the philosophy sofar proceeds along similar lines
as those employed in the ordinary Grassmannian
model,~\cite{PruiskenBurmistrov} it is important to keep in mind
that the presence of $S_F$ is itself affecting the ultraviolet
singularity structure of the theory. This means that both the
physics and the conceptual structure of the problem with
interactions are fundamentally different from what one is used to.
Moreover, in view of the mathematical peculiarities of the theory,
in particular those associated with the limits $N_r \rightarrow 0$
and $N_m \rightarrow \infty$, it must be shown explicitly that
instantons are well defined at a quantum level and that the
aforementioned ultraviolet behavior of the interacting electron
gas does not depend on the specific geometry that one chooses,
i.e. the introduction of $\mu (\textbf{r})$ in Eq.~\eqref{Ohmu}.
In this respect, we shall in what follows greatly benefit from our
theory of {\em observable parameters} since it provides the
appropriate framework for a general understanding of the theory at
short distances. To study the {\em ultraviolet} we first address
the problem of quantum fluctuations for the special case where
unitary matrix $\mathcal{T}_0$ in Eq.~\eqref{top1} is equal to
unity. We will come back to the general case not until
Section~\ref{TCFS} where embark on the {\em infrared} of the
system, in particular the various different steps that are needed
in order to change the geometry of the system from {\em curved}
space to {\em flat} space.
%
\subsection{\label{QFAI.GL1}Action for the quantum fluctuations}
%
Keeping the remarks of the previous Section in mind we obtain the
complete action as the sum of a classical part $S^{\rm inst}$ and
a quantum part $\delta S$ as follows
\begin{equation}
S=S_F[\Lambda]+ S^\textrm{inst}+\delta S \label{Sfluc1}
\end{equation}
where
\begin{equation}\label{Sfluc1_0}
S^\textrm{inst} = -2\pi\sigma_{xx} + i \theta + S_F^\textrm{inst}
\end{equation}
and
\begin{equation}
\delta S=\delta S^{(0)}+\delta S^{(1)}+\delta S^{(2)}+\delta
S_{\mathrm{linear}}^{(2)}.  \label{Sfluc2}
\end{equation}
Here $S_F^{\rm inst}$ stands for the classical action of the
modified \emph{singlet interaction} term, Eq.~\eqref{SFInst3},
with $U=1$ and is given by
\begin{eqnarray}\notag
S_F^{\rm inst} &=& \pi T z \int d \mathbf{r}
\mu^2(\textbf{r})\left ( c\sum\limits_{\alpha n}\mathop{\rm tr}
\nolimits I_{n}^{\alpha }\rho\mathop{\rm tr}\nolimits
I_{-n}^{\alpha }\rho + 4 \mathop{\rm
tr}\nolimits\eta \rho\right ) \\
&=& 16\pi^{2}Tz\Bigl(\frac{c}{3}-1\Bigr).\label{OCSM}
\end{eqnarray}
Next, the results for $\delta S$ in Eq.~\eqref{Sfluc2} are
classified in four different parts. The complete list of
contributions is presented in Table~\ref{Tf}. We use the following
notations $n_{12}=n_{1}-n_{2}$ and $\kappa ^{2}=8\pi T/\sigma
_{xx}$ from now onward. We will first briefly comment on the
different parts of $\delta S$.
%
\subsubsection{\label{Szero} $\delta S^{(0)}$}
%
This term contains all the fluctuations $v_{mn}^{\alpha \beta}$
with replica indices $\alpha, \beta >1$ that do not couple to the
instanton. $\delta S^{(0)}$ has therefore the same form as the
fluctuations about the trivial vacuum.
%
\subsubsection{\label{S12} $\delta S^{(1)}$, $\delta S^{(2)}$}
%
The terms $\delta S^{(1)}$ and $\delta S^{(2)}$ contain all the
fluctuations $v_{mn}^{\alpha \beta}$ with either $\alpha=1$ or
$\beta=1$. $\delta S^{(2)}$ only contains the fluctuations in the
first replica channel $v_{mn}^{11}$ and the remaining
contributions are collected in $\delta S^{(1)}$. In both $\delta
S^{(1)}$ and $\delta S^{(2)}$ we distinguish between the
``diagonal'' contributions that mainly originate from $S_\sigma$
(first four lines in Table~\ref{Tf}) and the ``off-diagonal'' ones
originating from $S_F$ (fifth and subsequent lines).

\begin{table*}[tbp]
\begin{ruledtabular}
\caption{Complete list of quantum fluctuations about the one
instanton, see text.}
\label{Tf}
\begin{eqnarray*}
\hline\hline
\end{eqnarray*}
\begin{eqnarray*}
\color{red}\delta S^{(0)}=\color{blue}- &&\color{blue}\frac{\sigma _{xx}}{4}%
\int d\eta d\theta \sum\limits_{\alpha ,\beta
=2}^{N_{r}}\sum\limits_{n_{1}\cdots n_{4}}\delta
_{n_{12},n_{34}}v_{n_{1}n_{2}}^{\alpha \beta
}\Bigl[(O^{(0)}+\kappa
^{2}zn_{12})\delta _{n_{1}n_{3}}-\kappa ^{2}zc\delta ^{\alpha \beta }\Bigr]%
v_{n_{4}n_{3}}^{\dag\beta \alpha } \\
&& \\
\hline
&& \\
\color{red}\delta S^{(1)}=\color{blue}- &&\color{blue}\frac{\sigma _{xx}}{4}%
\int d\eta d\theta \sum\limits_{\alpha =2}^{N_{r}}\Biggl \{%
\sum\limits_{n_{1}n_{2}}{}^{^{\prime \prime
}}v_{n_{1}n_{2}}^{1\alpha }(O^{(0)}+\kappa
^{2}zn_{12})v_{n_{2}n_{1}}^{\dag\alpha
1}+\sum\limits_{n_{1}n_{2}}{}^{^{\prime \prime
}}v_{n_{1}n_{2}}^{\alpha
1}(O^{(0)}+\kappa ^{2}zn_{12})v_{n_{2}n_{1}}^{\dag 1\alpha } \\
&\color{blue}+&\color{blue}\sum\limits_{n_{1}}{}^{^{\prime
}}v_{n_{1},-1}^{1\alpha }(O^{(0)}+\kappa
^{2}z(n_{1}+1))v_{-1,n_{1}}^{\dag\alpha
1}+\sum\limits_{n_{2}}{}^{^{\prime }}v_{0n_{2}}^{\alpha
1}(O^{(0)}-\kappa
^{2}zn_{2})v_{n_{2}0}^{\dag 1\alpha } \\
&\color{blue}+&\color{blue}\sum\limits_{n_{1}}v_{n_{1},-1}^{\alpha
1}\Bigl(O^{(1)}+\kappa ^{2}z(n_{1}+1)+\kappa
^{2}zce_{0}^{2}\Bigl(2|e_{1}|^{2}-\frac{1}{c}\Bigr)\Bigr)v_{-1,n_{1}}^{\dag 1%
\alpha } \\
&\color{blue}+&\color{blue}\sum\limits_{n_{2}}v_{0n_{2}}^{1\alpha
}\Bigl(O^{(1)}-\kappa ^{2}zn_{2}+\kappa ^{2}zce_{0}^{2}\Bigl(2|e_{1}|^{2}-%
\frac{1}{c}\Bigr)\Bigr)v_{n_{2}0}^{\dag \alpha 1}\Biggl \} \\
&-&\frac{\sigma _{xx}}{4}\kappa ^{2}zc\int d\eta d\theta
\sum\limits_{\alpha =2}^{N_{r}}\Biggl \{\sum\limits_{n_{1},n_{2}}{}^{^{%
\prime \prime }}e_{0}\Bigl[\bar{e}_{1}v_{n_{1}+1,n_{2}}^{1\alpha
}v_{n_{2}n_{1}}^{\dag\alpha 1}+e_{1}v_{n_{1}n_{2}}^{1\alpha
}v_{n_{2},n_{1}+1}^{\dag\alpha
1}-\bar{e}_{1}v_{n_{1},n_{2}-1}^{\alpha 1}v_{n_{2}n_{1}}^{\dag
1\alpha }-e_{1}v_{n_{1}n_{2}}^{\alpha
1}v_{n_{2}-1,n_{1}}^{\dag 1\alpha }\Bigr] \\
&+&\sum\limits_{n_{1}}{}^{^{\prime }}\Bigl[e_{0}\bar{e}_{1}
v_{n_{1}+1,-1}^{1\alpha }v_{-1,n_{1}}^{\dag \alpha
1}+e_{0}e_{1}v_{n_{1},-1}^{1\alpha }v_{-1,n_{1}+1}^{\dag \alpha 1}\Bigr]%
-\sum\limits_{n_{2}}{}^{^{\prime }}\Bigl[e_{0}\bar{e}_{1}
v_{0,n_{2}-1}^{\alpha 1}v_{n_{2},0}^{\dag 1\alpha
}+e_{0}e_{1}v_{0,n_{2}}^{\alpha 1}v_{n_{2}-1,0}^{\dag 1\alpha }\Bigr] \\
&+&\sum\limits_{n_{2}}{}^{^{\prime }}\Bigl[e_{0}e_{1}^{2}v_{1,n_{2}}^{1%
\alpha }v_{n_{2},0}^{\dag\alpha
1}+e_{0}\bar{e}_{1}^{2}v_{0,n_{2}}^{1\alpha
}v_{n_{2},1}^{\dag\alpha 1}\Bigr]-\sum\limits_{n_{1}}{}^{^{\prime }}\Bigl[%
e_{0}e_{1}^{2}v_{n_{1},-2}^{\alpha 1}v_{-1,n_{1}}^{\dag 1\alpha
}+e_{0}\bar{e}_{1}^{2}v_{n_{1},-1}^{\alpha 1}v_{-2,n_{1}}^{\dag 1\alpha }\Bigr%
]\Biggr \} \\
&& \\
\hline
&& \\
\color{red}\delta S^{(2)}=\color{blue}- &&\color{blue}\frac{\sigma _{xx}}{4}%
\int d\eta d\theta \Biggl \{\sum\limits_{n_{1}\cdots
n_{4}}{}^{^{\prime \prime \prime \prime
}}v_{n_{1}n_{2}}^{11}\Bigl((O^{(0)}+\kappa ^{2}zn_{12})\delta
_{n_{1}n_{3}}\delta _{n_{2}n_{4}}-\kappa ^{2}zc\delta
_{n_{12},n_{34}}\Bigr)v_{n_{4}n_{3}}^{\dag 11} \\
&\color{blue}+&\color{blue}\sum\limits_{n_{1}}{}^{^{\prime
}}v_{n_{1},-1}^{11}\Bigl(O^{(1)}+\kappa ^{2}z(n_{1}+1)-\kappa
^{2}zc+\kappa
^{2}zce_{0}^{2}\Bigl(2|e_{1}|^{2}-\frac{1}{c}\Bigr)\Bigr)v_{-1,n_{1}}^{\dag 11} \\
&\color{blue}+&\color{blue}\sum\limits_{n_{2}}{}^{^{\prime
}}v_{0n_{2}}^{11}\Bigl(O^{(1)}-\kappa ^{2}zn_{2}-\kappa
^{2}zc+\kappa
^{2}zce_{0}^{2}\Bigl(2|e_{1}|^{2}-\frac{1}{c}\Bigr)\Bigr)v_{n_{2}0}^{\dag 11} \\
&\color{blue}+&\color{blue}v_{0,-1}^{11}\Bigl(O^{(2)}+\kappa
^{2}z(1-c)+2\kappa ^{2}zce_{0}^{2}\Bigl(3|e_{1}|^{2}-\frac{1}{c}%
\Bigr)\Bigr)v_{-1,0}^{\dag 11}\Biggr \} \\
&-&\frac{\sigma _{xx}}{4}\kappa ^{2}zc\int d\eta d\theta \Biggl \{%
\sum\limits_{n_{1},n_{2}}{}^{^{\prime \prime
}}\Bigl[e_{0}\bar{e}_{1}v_{n_{1}+1,n_{2}}^{11}v_{n_{2}n_{1}}^{\dag
11}+e_{0}e_{1}v_{n_{1}n_{2}}^{11}v_{n_{2},n_{1}+1}^{\dag 11}-e_{0}
\bar{e}_{1}v_{n_{1},n_{2}-1}^{11}v_{n_{2}n_{1}}^{\dag 11}-e_{0}e_{1}v_{n_{1}n_{2}}^{11}
v_{n_{2}-1,n_{1}}^{\dag 11}\Bigr%
] \\
&& \\
&-&\sum\limits_{n_{1}\cdots n_{3}}{}^{^{\prime \prime \prime
}}v_{n_{1}n_{2}}^{11}\Bigl[\bar{e}_{1}\delta
_{n_{12},n_{3}+1}-e_{0}\delta
_{n_{12},n_{3}}\Bigr]v_{-1,n_{3}}^{\dag
11}-\sum\limits_{n_{1}\cdots n_{3}}{}^{^{\prime \prime \prime
}}v_{n_{3},-1}^{11}\Bigl[e_{1}\delta
_{n_{12},n_{3}+1}-e_{0}\delta _{n_{12},n_{3}}\Bigr]v_{n_{2},n_{1}}^{\dag 11} \\
&-&\sum\limits_{n_{2}\cdots n_{4}}{}^{^{\prime \prime \prime
}}v_{n_{3}n_{2}}^{11}\Bigl[\bar{e}_{1}\delta
_{n_{32},-n_{4}}+e_{0}\delta
_{n_{32},1-n_{4}}\Bigr]v_{n_{4},0}^{\dag
11}-\sum\limits_{n_{2}\cdots n_{4}}{}^{^{\prime \prime \prime
}}v_{0,n_{4}}^{11}\Bigl[e_{1}\delta
_{n_{32},-n_{4}}+e_{0}\delta _{n_{32},1-n_{4}}\Bigr]v_{n_{2}n_{3}}^{\dag 11} \\
&-&e_{0}\sum\limits_{n_{1}}{}^{^{\prime }}\Bigl[\bar{e}_{1}^{2}v_{n_{1},-2}^{11}v_{-1,n_{1}}^{\dag 11}
+e_{1}^{2}v_{n_{1},-1}^{11}v_{-2,n_{1}}^{\dag 11}\Bigr%
]+e_{0}\sum\limits_{n_{2}}{}^{^{\prime }}\Bigr[\bar{e}_{1}^{2}v_{1,n_{2}}^{11}v_{n_{2},0}^{\dag 11}
+e_{1}^{2}v_{0,n_{2}}^{11}v_{n_{2},1}^{\dag 11}\Bigr%
] \\
&+&2\sum\limits_{n_{1}}{}^{^{\prime }}\Bigl[e_{0}\bar{e}_{1}v_{n_{1}+1,-1}^{11}v_{-1,n_{1}}^{\dag 11}
+e_{0}e_{1}v_{n_{1},-1}^{11}v_{-1,n_{1}+1}^{\dag 11}\Bigr%
]-2\sum\limits_{n_{2}}{}^{^{\prime }}\Bigl[e_{0}\bar{e}_{1}v_{0,n_{2}-1}^{11}v_{n_{2},0}^{\dag 11}
+e_{0}e_{1}v_{0,n_{2}}^{11}v_{n_{2}-1,0}^{\dag 11}\Bigr%
] \\
&-&(1-2e_{0}^{2})\sum\limits_{n_{1}}{}^{^{\prime }}\Bigl[%
v_{n_{1},-1}^{11}v_{-n_{1}-1,0}^{\dag 11}+v_{0,-n_{1}-1}^{11}v_{-1,n_{1}}^{\dag 11}%
\Bigr]+e_{0}(\bar{e}_{1}-e_{1})\sum\limits_{n_{1}}{}^{^{\prime }}\Bigl[%
v_{n_{1},-1}^{11}v_{-n_{1},0}^{\dag
11}-v_{0,-n_{1}}^{11}v_{-1,n_{1}}^{\dag 11}\Bigr]
\\
&+&e_{0}(\bar{e}_{1}^{2}v_{1,-1}^{11}v_{0,-1}^{11}+e_{1}^{2}v_{-1,1}^{\dag 11}v_{-1,0}^{\dag 11}-
\bar{e}_{1}^{2}v_{0,-2}^{11}v_{0,-1}^{11}-e_{1}^{2}v_{-2,0}^{\dag 11}v_{-1,0}^{\dag 11})+%
e_0^2\Bigl[e_{1}^{2}v_{-1,0}^{\dag 11}v_{-1,0}^{\dag
11}+\bar{e}_{1}^{2}v_{0,-1}^{11}
v_{0,-1}^{11}\Bigr]%
\Biggr \} \\
&& \\
\hline
&& \\
\color{red}\delta S_\textrm{linear}^{(2)} &=&\frac{\sigma
_{xx}}{2}\kappa ^{2}zc\int d\eta d\theta \Biggl
\{e_{0}^{2}(\bar{e}_{1}v_{0,-2}^{11}+e_{1}v_{-2,0}^{\dag
11}-\bar{e}_{1}v_{1,-1}^{11}- e_{1}v_{-1,1}^{\dag
11})+e_{0}(1-2e_{0}^{2})(\bar{e}_{1}v_{0,-1}^{11}+e_{1}v_{-1,0}^{\dag
11})\Biggr \}
\end{eqnarray*}
\begin{eqnarray*}
\hline\hline
\end{eqnarray*}
\end{ruledtabular}
\end{table*}
%
\subsubsection{\label{Slin} $\delta S_\textrm{linear}^{(2)}$}
%
The contributions linear in $v$ and $v^\dag$ originate from the
singlet interaction term $S_F$ and are written in the bottom line
of Table ~\ref{Tf}. They can be written in terms of the
eigenfunctions $\Phi^{(a)}_{JM}$ as follows
\begin{gather}
\int d\eta d\theta \left(
e_{0}^2\bar{e}_{1}v_{0,-2}^{11}+e_{0}^2e_{1}v_{-2,0}^{\dag 11} \right )\hspace{3cm}{}\notag \\
\hspace{1.7cm}{}  \propto  \int d\eta d\theta \Bigl (
\bar{\Phi}^{(1)}_{2,1}
v_{0,-2}^{11}+ \Phi^{(1)}_{2,1} v^{\dag 11}_{-2,0} \Bigr)  \\
\int d\eta d\theta \left(
e_{0}^2\bar{e}_{1}v_{1,-1}^{11}+e_{0}^2e_{1}v_{-1,1}^{11} \right ) \hspace{3cm}{}\notag \\
\hspace{1.7cm} \propto \int d\eta d\theta \Bigl (
\bar{\Phi}^{(1)}_{2,1} v_{1,-1}^{11}  + \Phi^{(1)}_{2,1} v^{\dag
11}_{-1,1} \Bigr) \\
\int d\eta d\theta \left(
e_{0}^3\bar{e}_{1}v_{0,-1}^{11}+e_{0}^3e_{1}v_{-1,0}^{\dag 11}
\right ) \hspace{3cm}{}\notag
\\ \hspace{1.7cm} \propto \int d\eta d\theta\Bigl ( \bar{\Phi}^{(2)}_{2,1} v_{0,-1}^{11} +
\Phi^{(2)}_{2,1} v^{\dag 11}_{-1,0} \Bigr).
\end{gather}
Since the $\Phi^{(1)}_{2,1}$ and $\Phi^{(2)}_{2,1}$ do not
correspond to the zero modes of the operators $O^{(1)}$ and
$O^{(2)}$ one can eliminate these terms by performing a simple
shift in $v$, $v^\dag$. This leads to an insignificant
contribution to the classical action of the order $O(T^2)$. Next,
\begin{gather}
\int d\eta d\theta \Bigl(
e_{0}\bar{e}_{1}v_{0,-1}^{11}+e_{0}e_{1}v_{-1,0}^{\dag 11} \Bigr
)\hspace{3cm}{} \notag \\
\hspace{1.7cm} \propto
 \int d\eta d\theta \Bigl ( \bar{\Phi}^{(2)}_{1,-1}
v_{0,-1}^{11} + \Phi^{(2)}_{1,-1} v^{\dag 11}_{-1,0} \Bigr)\notag
\\ \hspace{1.7cm}\propto \frac{\delta
\lambda}{\lambda}.\hspace{3.9cm}{}
\end{gather}
This means that the fluctuations tangential to the instanton
parameter $\lambda$ are the only unstable fluctuations in the
problem. As will be discussed further below, these fluctuations
will be treated separately and we will proceed by formally
evaluating the quantum theory to first order in the temperature
$T$ only.
%
\subsubsection{\label{S0} Trivial vacuum}
%
For completeness we give the expression for the quantum
fluctuations about the trivial vacuum. The result can be written
as follows
\begin{equation}\label{Sfluc10}
S_0 = S_F[\Lambda] + \delta S_0
\end{equation}
where
\begin{eqnarray}\notag
\delta S_0= &-& \frac{\sigma _{xx}}{4} \int d\eta d\theta
\sum\limits_{\alpha ,\beta =1}^{N_{r}}\sum\limits_{n_{1}\cdots
n_{4}}\delta _{n_{12},n_{34}}v_{n_{1}n_{2}}^{\alpha \beta }\\
&\times & \Bigl[(O^{(0)} +\kappa
^{2}zn_{12})\delta _{n_{1}n_{3}}-\kappa ^{2}zc\delta ^{\alpha \beta }\Bigr]%
v_{n_{4}n_{3}}^{\dag\beta \alpha }.\hspace{0.7cm}{}\label{Sfh0}
\end{eqnarray}
%
%
%
\section{\label{EC}Details of computations}
%
%
In this Section we present the detailed computations of the
harmonic oscillator problem. In the first part we address the
thermodynamic potential which is in many ways standard. The
complications primarily arise from the infinite sums over
Matsubara frequencies which fundamentally alter the ultraviolet
singularity structure of the theory. We set up a systematic series
expansion of the thermodynamic potential in powers of the
temperature $T$. To perform the algebra we make use of the
complete set of eigenvalues and eigenfunctions obtained in the
previous Section as well as certain mathematical identities that
are all listed in Appendix~\ref{ME}. In the second part of this
Section we show (Section~\ref{Ptheory}) that the ultraviolet
singularity structure of the small oscillator problem is
identically the same as the one computed on the basis of the
theory of observable parameters. These important computations and
results, which are briefly summarized in Appendix~\ref{AppC},
permit one to proceed in an unambiguous manner and develop - in
the remaining part of this paper - a non-perturbative analysis of
the observable quantities of the theory.
%
\subsection{\label{QFAI.SZMRT.PV}Pauli-Villars regulators}
%
%
\subsubsection{\label{EC.E}Introduction}
%
Recall that after integration over the quantum fluctuations one is
in general left with two sources of divergences. First, there are
the ultraviolet singularities which eventually result in a
renormalization of the coupling constant or $\sigma_{xx}$. At
present we wish to extend the analysis to include the
renormalization of the $z$ and $zc$ fields. The ultraviolet of the
theory can be dealt with in a standard manner by employing
Pauli-Villars regulator fields with masses ${\mathcal M}_{f}$
($f=0,1,\dots, K$) and an alternating metric
$e_{f}$.~\cite{PauliVillars,tHooft,t'Hooft1,Pruisken2} We assume
$e_{0} =1$, ${\mathcal M}_{0}=0$ and large masses $\mathcal{M}_{f}
\gg 1$ for $f>1$. The following constraints are imposed
\begin{gather}\label{ConscM}
\sum \limits_{f=0}^{K} e_{f} \mathcal{M}^{k}_{f} = 0, \qquad
0 \leq k < K\\
  \sum \limits_{f=1}^{K} e_{f} \ln {\mathcal
M}_{f} = - \ln {\mathcal M}.
\end{gather}
The regularized theory is then defined as
\begin{equation}\label{SReg}
\delta S_{\rm reg} = \delta S_0 + \sum \limits_{f=1}^{K} e_{f}
\delta S_{f}.
\end{equation}
Here, action $\delta S_{f}$ is the same as $\delta S$ except that
the operators $O^{(a)}$ are all replaced by $O^{(a)}+ {\mathcal
M}_{f}^{2}$. Our task is to evaluate Eq.~\eqref{SReg} to lowest
orders in a series expansion in powers of $T$. This expansion
still formally diverges due to the zero modes of the operators
$O^{(a)}$. These zero modes, however, shall be treated separately
by employing the collective coordinate formalism introduced in
Ref.~\cite{Pruisken2}.

To simplify the notation we will next present the results while
omitting the alternating metric and the Pauli-Villars masses. This
can be done since in each case we consider one easily recognizes
how the metric and masses should be included. Consider the ratio
\begin{eqnarray}
\frac{Z_{\rm inst}}{Z_0} = \frac{\int \mathcal{D} [v,v^\dag] \exp
S}{ \int \mathcal{D} [v,v^\dag] \exp S_0} &=&\exp \Bigl [- 2 \pi
\sigma_{xx} + i \theta + S_F^{\rm
inst}\notag\\&&\hspace{0.5cm}+\Delta S_{\sigma} +\Delta S_F \Bigr
].\label{AD1}
\end{eqnarray}
Here, the quantum corrections denoted by $\Delta S_{\sigma}$ and
$\Delta S_F$ can be expressed in terms of the propagators
\begin{eqnarray}
\mathcal{G}_a(\omega) = \frac{1}{O^{(a)}+\omega} &=& \sum
\limits_{J M} \frac{|J M\rangle_{(a)} {}_{(a)}\langle J
M|}{E_J^{(a)}+\omega} \label{Propa} \\
\mathcal{G}_a^c (\omega)= \frac{1}{O^{(a)}+\alpha\omega} &=& \sum
\limits_{J M} \frac{|J M\rangle_{(a)} {}_{(a)}\langle J
M|}{E_J^{(a)}+\alpha\omega}\label{Ga}
\end{eqnarray}
where $a=0,1,2$. These expressions are directly analogous to those
that appear in flat space (see Ref.~[\onlinecite{Unify2}]). It is
important to emphasize that even at a Gaussian level the
integration over the field variables $v,v^\dag$ in Eq.~\eqref{AD1}
is not simple and straight forward. The main reason is that some
of the frequency sums can be written as an integral in the limit
$T \rightarrow 0$ and, along with that, they absorb a factor of
$T$. It is therefore not always obvious how the series expansion
in powers of $T$ should be evaluated. The simplest way to proceed
is to expand the functional integrals of Eq.~\eqref{AD1} in
non-diagonal elements which are proportional to $\kappa ^{2}\sim
T$. By inspection one can then convince oneself that in the
replica limit $N_{r}\rightarrow 0$, the expansion in the
non-diagonal terms can be truncated beyond third order only. We
shall next summarize the various different contributions to
$\Delta S_\sigma$ as well as $\Delta S_F$.

\begin{widetext}
%
\subsubsection{\label{EC.E.S}  $\Delta S_\sigma$}
%
The quantum correction $\Delta S_{\sigma }$ is obtained by
expanding the non-diagonal terms of Table~\ref{Tf} up to second
order. The results in the limit $T\rightarrow 0$,
$N_{r}\rightarrow 0$ and $N_{m}\rightarrow \infty $ can be written
as follows
\begin{eqnarray}
\Delta S_{\sigma } &=&2\tr [\ln \mathcal{G}_{1}(0)-\ln
\mathcal{G}_{0}(0)]- \tr[\ln \mathcal{G}_{2}(0)-\ln
\mathcal{G}_{0}(0)]  + 2c\int\limits_{0}^{\infty }d\omega
\tr[\mathcal{G}_{1}(\omega ) -\mathcal{G}_{0}(\omega )]
\label{dss0} \\ &+&2c^{2}\int \limits_{0}^{\infty }d\omega
\,\omega
\tr[\bar{e}_{1}\mathcal{G}_{0}^{c}(\omega )e_{1}%
\mathcal{G}_{1}(\omega )  + e_{0}\mathcal{G}_{0}^{c}(\omega
)e_{0}\mathcal{G}_{1}(\omega )-\mathcal{G}_{0}^{c}(\omega
)\mathcal{G}_{0}(\omega )]. \label{dss1}
\end{eqnarray}
\end{widetext}
In these expressions the trace is taken with respect to the
complete set of eigenfunctions of the operators $O^{(a)}$. To
evaluate these expressions we need the help of the identities
\eqref{MatEls} and \eqref{MatEls2} of Appendix~\ref{ME}. After
elementary algebra we obtain
\begin{equation}
\Delta S_{\sigma}=2\alpha D^{(1)}-D^{(2)}-2 c \Bigl( H^{(1)} \ln
\alpha  - H^{(2)}-c H^{(3)}\Bigr).  \label{Amass1}
\end{equation}
Here the quantities
\begin{equation}
D^{(r)}=\sum\limits_{J=1}^{\infty }(2J+r-1)\ln
E_{J}^{(r)}-\sum\limits_{J=0}^{\infty }(2J+1)\ln E_{J}^{(0)}
\label{D12}
\end{equation}
with $r=1,2$ originate from Eq.~\eqref{dss0}. The quantities
$H^{(i)}$ originate from Eq.~\eqref{dss1} and are defined by
\begin{equation}
H^{(1)}=\sum\limits_{J=0}^{\infty }\sum\limits_{J_1=J}^{J+1} J_1
\frac{E_{J}^{(0)}-E_{J_1}^{(1)}}{E_{J}^{(0)}-\alpha E_{J_1}^{(1)}}
\label{H1}
\end{equation}
\begin{equation}
H^{(2)}=\sum\limits_{J=0}^{\infty }\ln E_{J}^{(0)}
\sum\limits_{J_1=J}^{J+1}J_1\frac{E_{J}^{(0)}-E_{J_1}^{(1)}}{
E_{J}^{(0)}-\alpha E_{J_1}^{(1)}} \label{H2}
\end{equation}
\begin{equation}
H^{(3)}=\sum\limits_{J=0}^{\infty }\sum\limits_{J_1=J}^{J+1}J_1
\frac{E_{J_1}^{(1)}\ln E_{J_1}^{(1)} }{E_{J}^{(0)}-\alpha
E_{J_1}^{(1)}}. \label{H3}
\end{equation}

%
\subsubsection{\label{EC.E.SF} $\Delta S_F$}
%
To obtain the quantum correction $\Delta S_{F}$ we need to carry
out the expansion in the non-diagonal terms of Table~\ref{Tf} up
to the third order. By taking the appropriate limits as discussed
earlier we find the following results
\begin{widetext}
\setcounter{addeq}{\value{equation}}
\addtocounter{addeq}{1} \setcounter{equation}{0} \renewcommand{%
\theequation}{\arabic{addeq}.\arabic{equation}}
\begin{eqnarray}
\Delta S_{F} &=&2\kappa ^{2}z \Biggl \{\tr \Bigl [ (\alpha +c^{2})
(\mathcal{G}_{1}(0)-\mathcal{G}_{0}(0)) -\frac{\alpha}{2}(\mathcal{G}_{2}(0)-%
\mathcal{G}_{0}(0)) \Bigr ] \hspace{2cm}{} \label{dsF1.0} \\
&&\hspace{0.68cm}+  \tr \Bigl[ \alpha(2c|e_{1}|^{2}-1) e_0^{2}\mathop{\cal G}%
\nolimits_{1}(0)]- ( 3c|e_{1}|^{2}-1 )e_{0}^{2} \mathcal{G}_{2}(0)
-2c^{2} e_{0}^{2}\mathcal{G}_{1}(0)\Bigr ]  \label{dsF1.1}\\
&&\hspace{0.68cm}+\hspace{0.16cm}c^{3}\int_{0}^{\infty
}d\omega\,\omega\tr \Bigl[ \bar{e}_{1}\mathcal{G}_{0}^{c}(\omega
)e_{1}
\mathcal{G}_{1}^{2}(\omega )  + e_{0}\mathcal{G}_{0}^{c}(\omega )e_{0}\mathcal{G}_{1}^{2}(\omega )-\mathcal{G}_{0}^{c}(\omega )%
\mathcal{G}_{0}^{2}(\omega )\Bigr]  \label{dsF1.2}  \\
&&\hspace{0.68cm}-\hspace{0.16cm}c^{2}\int_{0}^{\infty
}d\omega\,\omega\tr \Bigl[ \bar{e}_{1}\mathcal{G}_{0}^{c}(\omega
)e_{1} \mathcal{G}_{1}(\omega
)e_{0}^{2}(2c|e_{1}|^{2}-1)\mathcal{G}_{1}(\omega )
+e_{0}\mathcal{G}_{0}^{c}(\omega )e_{0}\mathcal{G}_{1}(\omega
)e_{0}^{2}(2c|e_{1}|^{2}-1)\mathcal{G}_{1}(\omega )\Bigr]\hspace{0.5cm}{}   \label{dsF1.3} \\
&&\hspace{0.68cm} +2 c^{2}\int_{0}^{\infty
}d\omega\hspace{0.27cm}\tr \Bigl [ e_{0}^{2}\mathcal{G}_{1}(\omega
)
e_{0}^{2}\mathcal{G}_{1}(\omega ) \Bigr ] \label{dsF1.4} \\
&&\hspace{0.68cm}+5 c^{2}\int_{0}^{\infty
}d\omega\hspace{0.27cm}\tr\Bigl [ e_{0}e_{1}\mathcal{G}_{1}(\omega
)e_{0}\bar{e}_{1}
\mathcal{G}_{1}(\omega ) \Bigr ] \label{dsF1.5} \\
&&\hspace{0.68cm}-\hspace{0.16cm}c^{2}\int_{0}^{\infty
}d\omega\hspace{0.27cm}\tr \Bigl [
e_{0}e_{1}\mathcal{G}_{0}(\omega )e_{0}\bar{e}_{1}
\mathcal{G}_{0}(\omega ) \Bigr ] \label{dsF1.6} \\
&&\hspace{0.68cm}-\hspace{0.16cm}c^{2}\int_{0}^{\infty
}d\omega\,\omega\tr \Bigl [
 e_{0}\mathcal{G}_{0}^{c}(\omega
)e_{0}\mathcal{G}_{1}^{2}(\omega )\Bigr ] \hspace{8cm}{} \label{dsF1.7} \\
&&\hspace{0.68cm} - 4 c^{3}\int_{0}^{\infty }d\omega\,\omega\tr \Bigl [e_{0}e_{1}\mathcal{G}_{1}(%
\omega )e_{0}\mathcal{G}_{0}^{c}(\omega )\bar{e}_{1}
\mathcal{G}_{1}(\omega )\Bigr ]\Biggr \}.   \label{dsF1.8}
\end{eqnarray}
\setcounter{equation}{\value{addeq}}
\renewcommand{\theequation}{\arabic{equation}}
\end{widetext}

To evaluate these expressions we use the identities
\eqref{MatEls5}-\eqref{MatEls14} (see Appendix~\ref{ME}). After
some algebra we find
\begin{equation}
\Delta S_{F} = 2\kappa^2 z
^{2}\sum\limits_{i=1}^9B^{(i)}.\label{dsF2}
\end{equation}
Here, the nine contributions $B^{(i)}$, $i=1,\dots ,9$ correspond
to the nine equations \eqref{dsF1.0}-\eqref{dsF1.8}. The first two
of them  are given by
\begin{eqnarray}
B^{(1)}&=&(\alpha +c^{2})(Y^{(1)}-Y^{(0)})-\frac{\alpha
}{2}(Y^{(2)}-Y^{(0)})\hspace{0.5cm}{}\label{b1}\\
B^{(2)}&=&\frac{\alpha }{2}\left( \frac{2c}{3}-1\right) Y^{(1)}+\frac{\alpha }{%
2}Y^{(2)}-c^{2}Y^{(1)}\label{b2}
\end{eqnarray}
where we have introduced
\begin{equation}\label{Y}
Y^{(s)}=\sum\limits_{J=1}^{\infty
}\frac{2J+(s-1)^2}{E_{J}^{(r)}},\qquad s=0,1,2.
\end{equation}
The next two terms can be written as
\begin{eqnarray}
B^{(3)} &=& c^{3}\sum\limits_{J=0}^{\infty
}\sum\limits_{J_1=J}^{J+1} J_1
K_{\alpha}(E_{J}^{(0)},E_{J_1}^{(1)})\notag   \\
&-&c^3 \sum\limits_{J=0}^{\infty } (2J+1)K_{\alpha
}(E_{J}^{(0)},E_{J}^{(0)}) \label{b3}
\end{eqnarray}
and
\begin{equation}
B^{(4)} =-\frac{c^{2}}{2}\left( \frac{2c}{3}-1\right)
\sum\limits_{J=0}^{\infty }\sum\limits_{J_1=J}^{J+1}J_1 K_{\alpha
}(E_{J}^{(0)},E_{J_1}^{(1)}).  \label{b4}
\end{equation}
The function $K_{\alpha }(x,y)$ is defined as
\begin{equation}
K_{\alpha }(x,y)=\frac{x}{(x-\alpha y)^{2}}\ln \frac{x%
}{\alpha y}-\frac{1}{x-\alpha y}.  \label{Kalpha}
\end{equation}%
Notice that $K_{\alpha }(x,x)=-\ln (\alpha +c)/(c^{2}x)$. The next
three terms are given by
\begin{eqnarray}
B^{(5)}&=&2c^{2}\sum\limits_{J=1}^{\infty }\Biggl [\frac{J(6J^{2}-1)}{%
3(4J^{2}-1)}\frac{1}{E_{J}^{(1)}}+\frac{J(J+1)}{3(2J+1)}\notag \\
&&\hspace{1.3cm} \times L(E_{J}^{(1)},E_{J+1}^{(1)})\Biggr ]
\label{b5}
\end{eqnarray}
\begin{eqnarray}
B^{(6)}&=&\frac{5c^{2}}{3}\sum\limits_{J=1}^{\infty }\Biggl [\frac{J}{4J^{2}-1}%
\frac{1}{E_{J}^{(1)}}+2\frac{J(J+1)}{2J+1}\notag \\ &&
\hspace{1.3cm} \times
L(E_{J}^{(1)},E_{J+1}^{(1)})\Biggr ]  \label{b6}\\
B^{(7)}&=&-\frac{c^{2}}{3}\sum\limits_{J=0}^{\infty
}(J+1)L(E_{J}^{(1)},E_{J+1}^{(1)}).  \label{b7}
\end{eqnarray}
We have introduced the function
\begin{equation}
L(x,y)=\frac{\ln x-\ln y}{x-y}  \label{L}
\end{equation}%
such that $L(x,x)=1/x$. Finally, the last to terms are
\begin{equation}
B^{(8)}=-\frac{c^{2}}{2}\sum\limits_{J=0}^{\infty
}\sum\limits_{J_1=J}^{J+1}J_1 K_{\alpha
}(E_{J}^{(0)},E_{J_1}^{(1)})  \label{b8}
\end{equation}
and
\begin{equation}  \label{b9}
B^{(9)}=\frac{c^{3}}{3}\sum\limits_{J=0}^{\infty }
\frac{(2J+1)^{2}+2}{2J+1} F_{\alpha
}(E_{J}^{(0)},E_{J}^{(1)},E_{J+1}^{1)}).
\end{equation}
Here the function $F_{\alpha }(x,y,z)$ is defined as follows
\begin{equation}
 F_{\alpha }(x,y,z)=\frac{1}{y-z}\left[ \frac{y}{%
 x-\alpha y}\ln \frac{x}{\alpha y}-\frac{z}{x-\alpha z}
 \ln \frac{x}{\alpha z}%
 \right].  \label{Falpha}
\end{equation}%
such that $F_{\alpha }(x,y,y)=K_{\alpha }(x,y)$.

%
\subsection{\label{EC.PVR}Regularized expressions}
%

To obtain the regularized theory one has to include the
alternating metric $e_{f}$ and add the masses $\mathcal{M}_{f}$ to
the energies $E_J^{(a)}$ in the expressions for $D^{(r)}$,
$H^{(i)}$, $Y^{(s)}$ and $B^{(i)}$ respectively. We will proceed
by discussing the regularization of $\Delta S_\sigma$ and $\Delta
S_F$ separately.
%
\subsubsection{\label{EC.E.S} $\Delta S_\sigma$}
%
To start let us define the function
\begin{equation}
\Phi ^{(\Lambda )}(p)=\sum\limits_{J=p}^{\Lambda }2J\ln
(J^{2}-p^{2}). \label{Phi}
\end{equation}
According to Eq.~\eqref{SReg} the regularized function $\Phi
_{\mathrm{reg} }^{(\Lambda )}(p)$ is
\begin{eqnarray}
\Phi _{\mathrm{reg}}^{(\Lambda )}(p)
&=&\sum\limits_{f=1}^{K}e_{f}\sum\limits_{J=p}^{\Lambda }2J\ln
(J^{2}-p^{2}+
\mathcal{M}_{f}^{2})  \notag  \label{RPhi} \\
&&\hspace{0.5cm}+\sum\limits_{J=p+1}^{\Lambda }2J\ln
(J^{2}-p^{2}).
\end{eqnarray}
where we assume that the cut-off $\Lambda$ is much larger than
$\mathcal{M}_{f}$. In the presence of a large mass $\mathcal{M}_f$
we may consider the logarithm to be a slowly varying function of
the discrete variable $J$. We may therefore approximate the
summation by means of the Euler-Maclaurin formula
\begin{equation}
\sum\limits_{J=p+1}^{\Lambda }g(J)=\int \limits_{p}^{\Lambda
}g(x)dx+\frac{1}{2}g(x)\Bigr|_{p}^{\Lambda }+\frac{1}{12}g^{\prime
}(x)\Bigr|_{p}^{\Lambda }. \label{EMF}
\end{equation}
After some algebra we find that Eq.~\eqref{RPhi} can be written as
follows~\cite{Pruisken2}
\begin{eqnarray}
\Phi _{\mathrm{reg}}^{(\Lambda )}(p) &=&\!\!-2\Lambda (\Lambda
+1)\ln \Lambda +\Lambda ^{2}-\frac{\ln e\Lambda
}{3}+4\sum\limits_{J=1}^{\Lambda
}\!\!J\ln J  \notag  \label{RPhi1} \\
&+&\frac{1-6p}{3}\ln
\mathcal{M}+2p^{2}-2\sum\limits_{J=1}^{2p}(J-p)\ln J.
\end{eqnarray}%
The regularized expression for $D^{(r)}$ can now be obtained as
\begin{equation}
D_{\mathrm{reg}}^{(r)}=\lim\limits_{\Lambda \rightarrow \infty
}\left[ \Phi
_{\mathrm{reg}}^{(\Lambda )}\left( \frac{1+r}{2}\right) -\Phi _{\mathrm{reg}%
}^{(\Lambda )}\left( \frac{1}{2}\right) \right].  \label{Dr}
\end{equation}%
The final results are obtained as follows
\begin{eqnarray}
D_{\mathrm{reg}}^{(1)} &=&-\ln \mathcal{M}+\frac{3}{2}-2\ln 2
\label{D12reg1} \\
D_{\mathrm{reg}}^{(2)} &=&-2\ln \mathcal{M}+4-3\ln 3-\ln 2.
\label{D12reg2}
\end{eqnarray}%
The evaluation of $H^{(i)}_{\mathrm{reg}}$ is somewhat more subtle
but proceeds along similar lines. The results can be written as
follows
\begin{equation}
H_{\mathrm{reg}}^{(1)}=-\frac{\alpha }{c^{2}}\left[ 2\ln \mathcal{M}
+1-\psi \left( \frac{3c-1}{c}\right) -\psi \left( \frac{1}{c}%
\right) \right] .  \label{H1.2}
\end{equation}%
The Euler digamma function $\psi (z)$ appears as a result of the
following summation $\sum_{J=0}^{\infty }\left[
(J+1)^{-1}-(J+z)^{-1}\right] =\psi (z)-\psi (1)$.
Similarly we
find
\begin{eqnarray}
H_{\mathrm{reg}}^{(2)} &=&-\lim\limits_{\Lambda \rightarrow \infty }\Phi _{%
\mathrm{reg}}^{(\Lambda )}\left( \frac{1}{2}\right) -\frac{\alpha }{c^{2}}%
\Bigl[2\ln \mathcal{M}+1+2\ln ^{2}\mathcal{M}
\notag \\
&+&4\gamma _{S}+f\left( \frac{\alpha }{c}\right) +f\left( 1-\frac{\alpha }{c}%
\right) \Bigr]  \label{H2.1} \\
H_{\mathrm{reg}}^{(3)} &=&\frac{1}{c}\lim\limits_{\Lambda
\rightarrow \infty
}\Phi _{\mathrm{reg}}^{(\Lambda )}(1)+\frac{1}{c^{3}}\left( 2\ln %
\mathcal{M}+1\right) +\frac{\alpha }{c^{3}}\Bigl[2\ln ^{2}%
\mathcal{M}  \notag \\
&+&4\gamma_{S}+f\left( \frac{1}{c}\right) +f\left( 1-\frac{\alpha }{c}%
\right) -2c^{2}\frac{\ln 2}{1-2\alpha }\Bigr].  \label{H3.1}
\end{eqnarray}%
where $\gamma_{S}\approx -0.0728$ is the Stieltjes constant and
\begin{equation}
f(z)=2z^{2}\sum\limits_{J=2}^{\infty }\frac{\ln
J}{J(J^{2}-z^{2})}. \label{fz}
\end{equation}%
\begin{figure}
\includegraphics[width=80mm]{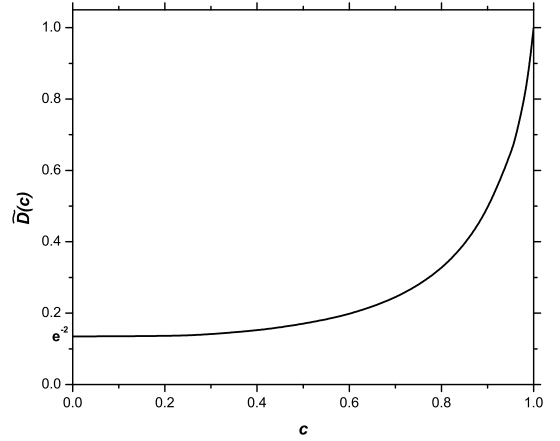}
\caption{ $\tilde{\mathcal{D}}(c)$ versus $c$, see text}
\label{Fig.Dc}
\end{figure}
We finally have the following total result for the quantum
correction $\Delta S_\sigma$
\begin{equation}
\exp \Delta S^\textrm{reg}_{\sigma
}=\frac{27}{8}\tilde{\mathcal{D}}(c)\exp \Biggl [4\left(
1+\frac{\alpha \ln \alpha }{c}\right) \ln \mathcal{M}+1\Biggr]
\label{Amassfin}
\end{equation}%
where
\begin{eqnarray}
\ln \tilde{\mathcal{D}}(c) &=&-2\frac{\alpha }{c}\Biggl\{ \left[
\psi \left( \frac{3c-1}{c}\right) +\psi \left( \frac{1}{c}\right)
-1\right] \ln \alpha
\notag   \\
&-&f\left( \frac{1-c}{c}\right) +f\left( \frac{1}{c}\right) -2\ln 2\frac{%
c^{2}}{2c-1}\Biggr\}.\hspace{1cm}{}\label{Dc}
\end{eqnarray}%
Notice that according to Eq.~\eqref{Dc} the quantity
$\tilde{\mathcal{D}}(c)$ depends on the crossover parameter $c$ in
a highly non-trivial fashion. Some of contributions diverge at the
points $c_{k}=1/k$ with $k=1,2,3,\dots $ but the final total
answer remains finite for all values of $c$ in the interval $0\leq
c\leq 1$ ranging from $\tilde{\mathcal{D}}(0)=e^{-2}$ to
$\tilde{\mathcal{D}}(1)=1$. A plot of the function
$\tilde{\mathcal{D}}(c)$ with varying $c$ is shown in
Fig.~\ref{Fig.Dc}.
%
\subsubsection{\label{EC.E.SF} $\Delta S_F$}
%
Notice that in contrast to the expression for $\Delta
S_\sigma^{\rm reg}$ where the numerical constants play an
important role, the expression for $\Delta S_F^{\rm reg}$ can only
be determined up to the logarithmic singularity in the
Pauli-Villars mass $\mathcal{M}$. In the latter case the constant
terms should actually be considered to be of order $1/\sigma_{xx}$
which is beyond the level of approximation as considered in this
paper. Keeping this in mind we proceed and define the following
function
\begin{equation}
Y^{(\Lambda )}(p)=\sum\limits_{J=p}^{\Lambda
}\frac{2J}{J^{2}-p^{2}}.  \label{cYL}
\end{equation}
According to Eq.~\eqref{SReg} the regularized function
$Y_{\mathrm{reg}}^{(\Lambda )}(p)$ is given by
\begin{equation}
Y_{\mathrm{reg}}^{(\Lambda
)}(p)=\sum\limits_{f=1}^{K}e_{f}\sum\limits_{J=p}^{\Lambda }\frac{2J}{%
J^{2}-p^{2}+\mathcal{M}_{f}^{2}}+\sum\limits_{J=p+1}^{\Lambda }\frac{2J}{%
J^{2}-p^{2}}  \label{RcYL}
\end{equation}
where as before we assume that $\Lambda \gg \mathcal{M}_{f}$.
Proceeding along similar lines as discussed earlier we now find
\begin{equation}\label{Yres}
Y^{(\Lambda)}_{\rm reg}(p) = 2 \ln \mathcal{M} + 2\gamma_E - \sum
\limits_{J=1}^{2 p} \frac{1}{J}
\end{equation}
where $\gamma_E\approx 0.577$ is the Euler constant. The
regularized expressions for $Y^{(s)}$ can be written as
\begin{equation}\label{Yr}
Y^{(s)}_{\rm reg} = \lim \limits_{\Lambda\to \infty}
Y^{(\Lambda)}_{\rm reg}\left (\frac{1+s}{2}\right ) .
\end{equation}
We finally obtain
\begin{equation}\label{Y01}
Y^{(s)}_{\rm reg}= 2 \ln \mathcal{M} + 2 \gamma_E - 1 -
\sum\limits_{J=2}^{s+1} \frac{1}{J} .
\end{equation}
Within the same logarithmic accuracy we can substitute $K_{\alpha
}(x,x)$ for the functions $K_{\alpha }(x,y)$ and $F_{\alpha
}(x,y,z)$ in Eqs.~\eqref{b3}-\eqref{b5} and \eqref{b9}. Similarly
we write $L(x,x)$ for $L(x,y)$ in Eqs.~\eqref{b6}-\eqref{b8}. With
the help of Eq.~\eqref{Y01}  we then find
\begin{eqnarray}
B^{(1)} &=&0\ln \mathcal{M}  \label{BiReg1} \\
B^{(2)} &=&\left( \frac{2c(1-c)}{3}-2c^{2}\right)\ln \mathcal{M}   \\
B^{(3)} &=&0\ln \mathcal{M}  \\
B^{(4)} &=&\left( \frac{2c}{3}-1\right) (\ln \alpha +c)\ln \mathcal{M}   \\
B^{(5)} &=&\frac{4c^{2}}{3}\ln \mathcal{M} \\
B^{(6)} &=&\frac{5c^{2}}{3}\ln \mathcal{M}   \\
B^{(7)} &=&-\frac{c^{2}}{3}\ln \mathcal{M}   \\
B^{(8)} &=&(\ln \alpha +c)\ln \mathcal{M}   \\
B^{(9)} &=&-\frac{2c}{3}(\ln \alpha +c)\ln \mathcal{M}.
\label{BiReg9}
\end{eqnarray}
The final total result for $\Delta S_{F}^\textrm{reg}$ can now be
written as follows
\begin{equation}
\Delta S_{F}^\textrm{reg}=\frac{32}{3\sigma _{xx}}\pi T z c( \ln
\mathcal{M}e^{\gamma _{E}-1/2} +\mathrm{const}). \label{dsF2}
\end{equation}
%
%
\subsubsection{\label{EC.REGZ} Regularized $Z_{\textrm{inst}}/Z_0$}
\begin{figure}
\includegraphics[width=80mm]{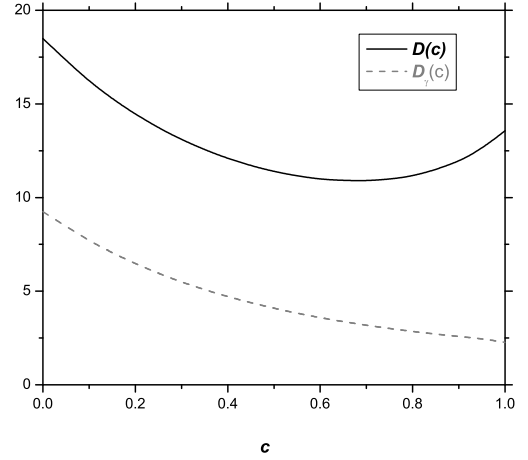}
\caption{ $D(c)$ and $D_{\gamma }(c)$ versus $c$, see text
.}\label{FIG.DDc1}
\end{figure}

%
%
We next collect the various different contributions together and
obtain the following result for the instanton contribution to the
thermodynamic potential
\begin{eqnarray}
\ln &&\hspace{-0.7cm} \left [\frac{Z_{\rm inst}}{Z_0}\right
]^\textrm{reg}  \notag
\\  &=& 3 \ln 3 - 7\ln 2 -\ln \pi +\ln \mathcal{D}(c) + i \theta
\label{const-reg}\hspace{2cm}\,\\ &-& \hspace{0.45cm} 2 \pi
\sigma_{xx} \Bigl [1 - \frac{2}{\pi \sigma_{xx}}\Bigl(1+
 \frac{\alpha }{c}\ln \alpha \Bigr )\ln \mathcal{M}
e^{\gamma _{E}}\Bigr ]\label{sigma-reg}\\
&+& \frac{16 \pi ^{2}}{3} T z c \Bigl [ 1 - \frac{1}{\pi \sigma
_{xx}}\ln \mathcal{M} e^{\gamma _{E}-1/2}\Bigr ] \label{zc-reg}\\
&-& \hspace{0.21cm} 16 \pi ^{2} T z \Bigl [ 1 - \frac{c}{\pi
\sigma _{xx}}\ln \mathcal{M} e^{\gamma _{E}-1/2}\Bigr
].\label{z-reg}
\end{eqnarray}
We have introduced new function ${\mathcal{D}}(c)$ which is
defined as
\begin{equation}
\mathcal{D}(c)=16\pi\tilde{\mathcal{D}}(c)\exp \Bigl[
1-4\Bigl(1+\frac{\alpha }{c}\ln \alpha \Bigr)\gamma _{E}\Bigr] .
\label{Dchat}
\end{equation}
A plot of ${\mathcal{D}}(c)$ with varying $c$ is shown in
Fig.~\ref{FIG.DDc1}.
%
\subsection{\label{Ptheory} Observable theory in Pauli-Villars regularization}
%
The most important result next is that the quantum corrections to
the parameters $\sigma_{xx}$, $zc$, and $z$ in
Eqs.~\eqref{sigma-reg}-\eqref{z-reg} are all identically the same
as those obtained from a perturbative expansion of the observable
parameters $\sigma_{xx}$, $z^\prime c^\prime$, and $z^\prime$
introduced in Section~\ref{PO}. In Appendix~\ref{AppC} we give the
details of the computation. Denoting the results for
$\sigma_{xx}^\prime$, $z^\prime$ and $c^\prime$ by
$\sigma_{xx}(\mathcal{M})$, $z(\mathcal{M})$ and $c(\mathcal{M})$
respectively then we have
\begin{eqnarray}
\sigma_{xx}(\mathcal{M}) &=& \sigma_{xx} \left[1 -
\frac{\beta_0(c)}{
\sigma_{xx} } \ln \mathcal{M} e^{\gamma_E} \right]\label{pr0}\\
z(\mathcal{M})c(\mathcal{M})  &=&\hspace{0.23cm} z c \left[1
+\frac{\gamma_0}{\sigma_{xx}} \ln \mathcal{M} e^{\gamma_E
-1/2}\right ]\hspace{1cm}\, \label{pr1} \\
z(\mathcal{M})  &=&\hspace{0.41cm}z \left[1 +\frac{c
\gamma_0}{\sigma_{xx}} \ln \mathcal{M} e^{\gamma_E -1/2}\right
].\label{pr2}
\end{eqnarray}
The results of Eqs.~\eqref{const-reg}-\eqref{z-reg} can therefore
be written as follows
\begin{equation}
\left [\frac{Z_{\rm inst}}{Z_0}\right ]^\textrm{reg}=
\frac{27\mathcal{D}(c)}{128\pi} \exp\left (-2 \pi
\sigma_{xx}(\mathcal{M}) + i \theta + {\check
S}_F^\textrm{inst}[\rho]\right )
 \label{exp1a}
\end{equation}
where
\begin{eqnarray}\label{SRrho}
{\check S}_F^\textrm{inst}[\rho] &=& \pi T z(\mathcal{M})\int d
\mathbf{r} \mu^2(\textbf{r}) \Bigl (
c(\mathcal{M})\sum\limits_{\alpha n}\mathop{\rm tr} \nolimits
I_{n}^{\alpha }\rho \mathop{\rm tr}\nolimits I_{-n}^{\alpha }\rho
\notag \\ && \hspace{2.8cm} + 4 \mathop{\rm tr}\nolimits\eta
\rho\Bigr ).
\end{eqnarray}
Notice that the expression in the exponent is similar to the
classical action with the rotation matrix $\mathcal{T}_0$ put
equal to unity. The main difference is in the expressions for
$\sigma_{xx}(\mathcal{M})$, $z(\mathcal{M})c(\mathcal{M})$ as well
as $z(\mathcal{M})$ which are all precisely the radiative
corrections as obtained from the observable theory.

At this stage of the analysis several remarks are in order. First
of all, it is important to stress that our result for the
observable theory, Eq.~\eqref{pr0}, uniquely fixes the amplitude
$\mathcal{D}(c)$ of the thermodynamic potential which is left
unresolved otherwise. This aspect of the problem is going to play
a significant role when extracting the non-perturbative
renormalization behavior of the theory. In fact, we shall see
later on, in Section~\ref{Dlambda}, that the most important
features of the theory, notably the values of $\mathcal{D}(c)$ at
$c=0$ and $c=1$ respectively, are universal in the sense that they
are independent of the specific regularization scheme that one
uses to define the renormalized theory. Secondly, our results
demonstrate that the idea of spatially varying masses does not
alter the ultraviolet singularity structure of the instanton
theory. In particular, Eqs.~\eqref{const-reg}-\eqref{SRrho}
display exactly the same logarithms as found previously in flat
space and by employing dimensional regularization.~\cite{Unify2}
The detailed computations of Appendix~\ref{AppC} provide a deeper
understanding of this aspect of the problem, especially where it
says that the Pauli-Villars regularization scheme retains the
translational invariance of the electron gas.
%
%
\section{\label{TCFS} Transformation from curved space to flat
space}
%
%
%
%
\subsection{\label{Imani} Instanton manifold}
%
%
%
%
\subsubsection{Integration over zero frequency modes}
%
%
We are now in a position to extend the results for the
thermodynamic potential to include the integration over the zero
modes. The complete expression for ${Z_{\rm inst}}/{Z_0}$ can be
written as follows~\cite{Pruisken2}
\begin{equation}\label{Z.lambda.r0.T0}
 \left[ \frac{Z_{\rm inst}}{Z_0} \right]^\textrm{reg} \rightarrow
 \frac{A_\textrm{inst}}{A_0} \frac{\int \mathcal{D} [Q_{\rm inst} ]
 }{\int \mathcal{D} [Q_{0}]}\left [\frac{Z_{\rm inst}
 [Q_\textrm{inst}]}{Z_0 [Q_0]} \right]^\textrm{reg} .
\end{equation}
The meaning of the symbols is as follows.
\begin{widetext}
\begin{equation}
 \left [\frac{Z_{\rm inst} [Q_\textrm{inst}]}{Z_0 [Q_0]}
 \right]^\textrm{reg} = \frac{27\mathcal{D}(c)}{128\pi} \frac{
 \exp \left \{ -2 \pi \sigma_{xx}(\mathcal{M}) + i \theta +
 z(\mathcal{M}) \int d\eta d\theta \left(
 c(\mathcal{M})\sum\limits_{\alpha n}\mathop{\rm tr} \nolimits
 I_{n}^{\alpha } Q_\textrm{inst} \mathop{\rm tr}\nolimits I_{-n}^{\alpha }
 Q_\textrm{inst}
 + 4 \mathop{\rm tr}\nolimits\eta Q_\textrm{inst}
 \right ) \right \} }{\exp \left \{ z(\mathcal{M}) \int d\eta d\theta \left (
 c(\mathcal{M})\sum\limits_{\alpha n}\mathop{\rm tr} \nolimits
 I_{n}^{\alpha } Q_0 \mathop{\rm tr}\nolimits I_{-n}^{\alpha } Q_0
 + 4 \mathop{\rm tr}\nolimits\eta Q_0
 \right) \right\} }.
 \label{exp2a}
\end{equation}
\end{widetext}
Here, $Q_{\rm inst}$ denotes the manifold of the instanton
parameters as is illustrated in Fig.~\ref{FIG7}
\begin{eqnarray}\label{mani1}
 \int \mathcal{D} [Q_{\rm inst} ] &=& \int d \textbf{r}_0 \int
 \frac{d \lambda}{\lambda^{3}} \int {\mathcal D}[{\mathcal T}_0 ]
 .
\end{eqnarray}
Here, the integral over ${\mathcal T}_0$ can be decomposed
according to
\begin{equation}
 \int {\mathcal D}[{\mathcal T}_0 ] = \int {\mathcal D}[t_0 ] \int
 {\mathcal D}[ U ]
\end{equation}
where schematically we can write
\begin{eqnarray}
\int {\mathcal D}[t_0 ] &=&  ~~\int {\mathcal D} \left[
\frac{U(2N)}{U(N)\times U(N)} \right]\\
 \int
 {\mathcal D}[ U ] &=& ~~\int {\mathcal D}\left[ \frac{U(N)}{U(1)\times U(N-1)}
 \right] \notag \\
 && \times \int {\mathcal D}\left[ \frac{U(N)}{U(1)\times U(N-1)}
 \right]
 \notag \\
 && \times \int {\mathcal D} ~ \left[ U(1) \right] .
\end{eqnarray}
On the other hand, the $Q_{0}$ are the zero modes associated with
the trivial vacuum
\begin{equation}
 \int \mathcal{D} [Q_{0} ] = \int {\mathcal D}
 \left[ \frac{U(2N)}{U(N)\times U(N)} \right].
 \label{mani0}
\end{equation}
The numerical factors $A_{\rm inst}$ and $A_0$ are given by
\begin{eqnarray}
A_{\rm inst} &=&  \langle e^{4}_{0} \rangle \langle |e_{1}|^{4}
\rangle \langle e^{2}_{0} |e_{1}|^{2}\rangle \left( \langle
e^{2}_{0} \rangle \langle |e_{1}|^{2}
\rangle \right)^{2N-2} \notag \\
&\times & \langle 1 \rangle^{(N-1)(N-1)} \pi^{-N^2-2N}\label{Ainst}\\
A_{0} &=&  \langle 1 \rangle^{N^2} \pi^{-N^2}\label{A0}
\end{eqnarray}
where the average $\langle \dots \rangle$ is with respect to the
surface of a sphere
\begin{equation}\label{<>}
\langle f \rangle = \sigma_{xx} \int \limits_{-1}^{1} d \eta \int
\limits_{0}^{2 \pi} d \theta f(\eta, \theta).
\end{equation}
%

\subsubsection{\label{EC.ZMR} $U$ rotation}

%
We have already mentioned earlier that the fluctuations in the
Goldstone modes $t_0 , Q_0 \in U(2N)/U(N)\times U(N)$ have an
infinite action in flat space and eventually drop out. We can
therefore write the result of Eq.~\eqref{Z.lambda.r0.T0} as
follows
\begin{equation}\label{OmegaInstRes}
 \left[ \frac{Z_{\rm inst}}{Z_0} \right]^\textrm{reg} =
 \frac{27}{128\pi} \int d\textbf{r}_0 \int \frac{d
 \lambda}{\lambda^{3}} \int \mathcal{D} [U]
 \frac{A_\textrm{inst}\mathcal{D}(c)}{A_0} e^{S^{\prime}_{\rm inst}} .
\end{equation}
Here,
\begin{equation}
 S^{\prime}_{\rm inst} = - 2\pi  \sigma_{xx}({\mathcal M}) \pm i
 \theta +\check{S}_F^\textrm{inst}[U^{-1}\rho U] \label{S-prime}
\end{equation}
with $\check{S}_F$ defined by Eq. ~\eqref{SRrho}. Next, by making
use of the identity~\cite{Pruisken2}
\begin{equation}\label{vol}
\int {\mathcal D} \left[ \frac{U(k)}{U(1)\times U(k-1)} \right] =
\frac{\pi^{k-1}} {\Gamma(k)}
\end{equation}
we can write the result for the thermodynamic potential in the
limit $N_r\to 0$ in a more compact fashion as follows
\begin{equation}\label{OmegaInstRes1}
 \left[ \frac{Z_{\rm inst}}{Z_0} \right]^\textrm{reg} =
 \frac{N^2}{8\pi^2} \int d\textbf{r}_0 \int \frac{d
 \lambda}{\lambda^{3}}
 \mathcal{D}(c) \langle e^{S^{\prime}_{\rm inst}} \rangle_U
\end{equation}
where the average is defined according to
\begin{equation}
\langle X \rangle_{U} =\frac{\int \mathcal{D} [U] X }{\int
\mathcal{D}[U]}.
\end{equation}
%

\subsubsection{\label{Flat1} Curved space versus flat space}

%
Our final result of Eq.~\eqref{OmegaInstRes1} still involves a
spatially varying momentum scale $\mu (\textbf{r})$ and our task
next is to express the final answer in quantities that are defined
in {\em flat} space, rather than {\em curved} space. The first
step is to rewrite the integral $\int d\eta d\theta$ in
$\tilde{S}_F^\textrm{inst}$ as an integral over {\em flat} space
following the substitution
\begin{equation}
\int d\eta d\theta = \int d \textbf{r} \mu^2 (\textbf{r}) \quad
\rightarrow \quad \int d \textbf{r} .
\end{equation}
The expression for $\tilde{S}_F^\textrm{inst}$ now reads
\begin{eqnarray}
\tilde{S}_F^\textrm{inst}[U^{-1} \rho U] &=& \pi T \int^{\prime} d
\textbf{r} z(\mathcal{M})\Bigl [ c(\mathcal{M})
\sum\limits_{\alpha n}\tr I_{n}^{\alpha }U^{-1}\rho U \notag \\
&&\times \tr I_{-n}^{\alpha }U^{-1} \rho U + 4 \tr\eta U^{-1}\rho
U \Bigr ] \label{SFinsttilde}
\end{eqnarray}
where the ``prime'' on the integral sign reminds us of the fact
that the expression for $\tilde{S}_F^\textrm{inst}$, as it now
stands, still diverges logarithmically in the sample size. What
remains, however, is to perform the next step which is to express
the Pauli-Villars masses $\mathcal M$ in terms of the appropriate
quantities that are defined in {\em flat} space. Notice hereto
that $\mathcal M$ actually describes a {\em spatially varying}
momentum scale $\mu (r) \mathcal M$. In
Section~\ref{Transformation} below as well as in the remainder of
this paper we will embark on the general problem of how to
translate a momentum scale in {\em curved space} into a quantity
$\mu_0$ that is defined in {\em flat space}. As an extremely
important consequence of this procedure we shall show in what
follows that the final expression for the interaction term
$\tilde{S}_F^\textrm{inst}$ is finite in the infrared. This
remarkable result is the primary reason as to why one can proceed
and obtain the non-perturbative corrections to the renormalization
of the quantities $z$ and $c$.
%

\subsection{\label{PO1}Physical observables}
%
\subsubsection{\label{PO1.LR} Linear response}
%
Our results for the thermodynamic potential are easily extended to
include the quantities $\sigma^\prime_{xx}$ and $\theta^\prime$
defined by Eqs.~\eqref{resp1} and \eqref{resp2}. To leading order
in $\sigma_{xx}$ we obtain the following result (see also
Ref.~[\onlinecite{Pruisken3}])
\begin{gather}
 \sigma_{xx}^{\prime} = \sigma_{xx}(\mathcal{M}) + \int
 \frac{d \lambda}{\lambda} \mathcal{D}(c) \Bigl \langle \left (J_{xx}[Q_\textrm{inst}]e^{i\theta} + c.c.\right
 )\notag\\
\hspace{2cm}\times  \exp \left (-2\pi\sigma_{xx}(\mathcal{M}) +\tilde{S}^\textrm{inst}_F\right )\Bigr
\rangle_{U} \label{ss1_0}\\
 \frac{\theta^{\prime}}{2\pi} = \frac{\theta}{2\pi}\hspace{0.8cm}
 + \int
 \frac{d \lambda}{\lambda} \mathcal{D}(c) \Bigl \langle \left (J_{xy}[Q_\textrm{inst}]e^{i\theta} + c.c.\right
 )\notag\\
\hspace{2cm} \times  \exp \left (-2\pi\sigma_{xx}(\mathcal{M})
+\tilde{S}^\textrm{inst}_F\right )
 \Bigr \rangle_{U} . \label{ss2_0}
\end{gather}
Here, we have introduced the quantity $J_{ab}[Q_\textrm{inst}]$
which is given as
\begin{eqnarray}
J_{jk}[Q_\textrm{inst}] &=& N^2 \frac{\sigma _{xx}^{2}}{32 \pi^2 n
\lambda^{2}} \int d \mathbf{r}\tr I_{n}^{\alpha }U\rho \nabla
_{j}\rho U^{-1} \notag \\ &&\hspace{1.6cm}\times \int d
\mathbf{r}^\prime\tr I_{-n}^{\alpha }U\rho \nabla _{k}\rho
U^{-1}.\hspace{0.6cm}{} \label{J2ab2}
\end{eqnarray}%
The interaction term $\tilde{S}^\textrm{inst}_F$ in
Eqs.~\eqref{ss1_0} and \eqref{ss2_0} does not contribute in the
limit $T \to 0$ and can be dropped. By using the normalization
conditions
\begin{eqnarray}
\sum\limits_{n_{1},\alpha }(U^{-1})_{0,n_{1}}^{1\alpha
}(U)_{n_{1},0}^{\alpha 1}&=&1\label{norm1}
\\
\sum\limits_{n_{2},\alpha }(U^{-1})_{-1,n_{2}}^{1\alpha
}(U)_{n_{2},-1}^{\alpha 1}&=&1 \label{norm2}
\end{eqnarray}
we find the following results for the expressions bilinear in the
$U$
\begin{eqnarray}
\left \langle (U)_{n_{1},0}^{\alpha 1}(U^{-1})_{0,n_{3}}^{1\beta
}\right \rangle &=& \frac{1}{N} \delta
_{n_{1}n_{3}}\delta^{\alpha \beta } \label{avW1}\\
\left \langle (U)_{n_{2},-1}^{\alpha 1}(U^{-1})_{-1,n_{4}}^{1\beta
}\right \rangle &=& \frac{1}{N}\delta _{n_{3}n_{4}}\delta^{\alpha
\beta }.\label{avW2}
\end{eqnarray}
For the quartic combinations we find
\begin{gather}\label{avW3}
\left \langle (U)^{\alpha 1}_{n_{1},0} (U^{-1})^{1\beta}_{0,n_{3}}
(U)^{\gamma 1}_{n_{5},0} (U^{-1})^{1 \delta}_{0,n_{7}} \right
\rangle \hspace{1.9cm}\,\notag\\
\hspace{3cm}= \frac{\delta_{n_{1}n_{3}}^{\alpha \beta}
\delta_{n_{5}n_{7}}^{\gamma \delta} + \delta_{n_{1}n_{7}}^{\alpha
\delta} \delta_{n_{5}n_{3}}^{\gamma \beta}}{N(1+N)}  \\
\left \langle (U)^{\alpha 1}_{n_{2},-1}
(U^{-1})^{1\beta}_{-1,n_{4}} (U)^{\gamma 1}_{n_{6},-1} (U^{-1})^{1
\delta}_{-1,n_{8}} \right
\rangle \hspace{1cm} \, \notag\\
\hspace{3cm} = \frac{\delta_{n_{2}n_{4}}^{\alpha \beta}
\delta_{n_{6}n_{8}}^{\gamma \delta} + \delta_{n_{2}n_{8}}^{\alpha
\delta} \delta_{n_{4}n_{6}}^{\gamma \beta}}{N(1+N)}.\label{avW4}
\end{gather}
We have used the shorthand notation
$\delta_{n_1n_3}^{\alpha\beta}\equiv
\delta_{n_1n_3}\delta^{\alpha\beta}$. In the limit where $N_r \to
0$ we obtain
\begin{eqnarray}
\langle J_{jk}[Q_\textrm{inst}]\rangle_{U}&=&\frac{\sigma
_{xx}^{2}}{32\pi^2\lambda^2}\int d \mathbf{r}(\rho \nabla
_{j}\rho)^{11}_{-1,0}\notag \\
&\times & \int d\mathbf{r}^\prime(\rho \nabla
_{k}\rho)^{11}_{0,-1} \notag
\\ &=& \frac{\sigma_{xx}^2}{2} \left (\delta_{jk}- i
\varepsilon_{jk}\right ).\label{PiW}
\end{eqnarray}
The expressions for $\sigma_{xx}^\prime$ and $\theta^\prime$ can
now be written as follows
\begin{equation}
 \sigma_{xx}^{\prime} = \sigma_{xx}(\mathcal{M})- \int^{\prime}
 \frac{d \lambda}{\lambda} \mathcal{D}(c)\sigma_{xx}^2
 e^{-2\pi\sigma_{xx}(\mathcal{M})}\cos\theta \label{ss1}
\end{equation}
\begin{equation}
 \frac{\theta^{\prime}}{2\pi} =
 \frac{\theta}{2\pi}\hspace{0.8cm}
 -\int^{\prime}
 \frac{d \lambda}{\lambda} \mathcal{D}(c)\sigma_{xx}^2
 e^{-2\pi\sigma_{xx}(\mathcal{M})}\sin\theta. \label{ss2}
\end{equation}
%
\subsubsection{\label{PO1.SH} Specific heat}
%
The simplest way of obtaining the parameters $z^\prime$ and
$z^\prime c^\prime$ is by using the definitions in
Section~\ref{PO} and expand the instanton result in powers of
$\tilde{S}_F[U]$. This leads to the expression
\begin{eqnarray}
 z^{\prime} = z(\mathcal{M}) &-&  \int^{\prime}
 \frac{d \lambda}{\lambda} \mathcal{D}(c) e^{-2\pi\sigma_{xx}(\mathcal{M})} \cos \theta \notag \\
&\times& \frac{N^2}{8\pi^3 \lambda^2 T \tr\eta\Lambda}
  \langle \tilde{S}_F[U]\rangle_{U}
 \label{zi2}\\
z^{\prime}c^\prime = z(\mathcal{M})c(\mathcal{M}) &-&
\int^{\prime}
 \frac{d \lambda}{\lambda} \mathcal{D}(c) e^{-2\pi\sigma_{xx}(\mathcal{M})} \cos \theta \notag\\
&\times& \frac{N^2}{8\pi^3 \lambda^2 T \tr\eta\Lambda}
  \langle \tilde{S}_F[U]\rangle_{U}.
 \label{zi1}
\end{eqnarray}
\noindent{The} expectations can be evaluated along the same lines
as was done in the previous Section. It should be mentioned,
however, that there are certain subtleties associated with the
limit $T=0$ in this case and these will be addressed in detail in
Section ~\ref{Omega}. Here we just state the result
\begin{equation}\label{OFAvW}
\left \langle
 \tilde{S}_F[U]\right \rangle_{U} = \frac{2\pi T}{N^2}
 \int d \mathbf{r} z(\mathcal{M})c(\mathcal{M})
 |\rho_{00}^{11}(r)| \tr \eta \Lambda.
\end{equation}
Eqs.~\eqref{zi1} and \eqref{zi2} therefore greatly simplify and we
obtain
\begin{eqnarray}
 z^{\prime} = z(\mathcal{M}) &+& \frac{\gamma_0}{4\pi} \int^{\prime}
 \frac{d \lambda}{\lambda} \mathcal{D}(c) e^{-2\pi\sigma_{xx}(\mathcal{M})} \cos \theta \notag\\
 &\times& \int^{\prime} \frac{d
 \textbf{r}}{\lambda} z(\mathcal{M})c(\mathcal{M}) \mu(\textbf{r}) \label{zi3}\\
 z^{\prime}c^\prime = z(\mathcal{M})c(\mathcal{M}) &+&
 \frac{\gamma_0}{4\pi} \int^{\prime}
 \frac{d \lambda}{\lambda} \mathcal{D}(c) e^{-2\pi\sigma_{xx}(\mathcal{M})} \cos \theta \notag\\
 &\times& \int^{\prime} \frac{d
 \textbf{r}}{\lambda} z(\mathcal{M})c(\mathcal{M}) \mu(\textbf{r}).\label{zi}
\end{eqnarray}
\noindent{The} most important feature of these results is that the
non-perturbative contributions to the observable parameters
$\sigma_{xx}^\prime$, $\theta^\prime$, $c^\prime$ and $z^\prime$
are all unambiguously expressed in terms of the perturbative
quantities $\sigma_{xx} ({\mathcal M})$, $\theta (\nu)$,
$c({\mathcal M})$ and $z({\mathcal M})$.
%
%
\subsection{\label{Transformation}Transformation ~$\mu^{2}(\mathbf{r})
{\mathcal M} \rightarrow \mu_0$}
%
%
As a last step in the development of a quantum theory we next wish
to express the Pauli-Villars masses which carry the metric of a
sphere $\mu^2(\textbf{r}) \mathcal{M}^2$ in terms of a mass or
momentum scale in flat space, say $\mu_0^2$. By changing the
momentum scale from $\mu (\textbf{r}) \mathcal{M}$ to $\mu_0$ one
changes the renormalized theory according to
\begin{gather}
 \sigma_{xx}({\mathcal M}) \rightarrow \sigma_{xx}({\mathcal M}) \left[ 1
 + \frac{\beta_0(c)}{\sigma_{xx}} \ln \frac{\mu(\textbf{r})\mathcal{M}}{\mu_{0}}
 \right]\hspace{4cm}{}\notag\\
 = \sigma_{xx} \left[ 1
 - \frac{\beta_0(c)}{\sigma_{xx}} \ln \frac{\mu_{0}}{\mu(\textbf{r})}
 e^{\gamma_E} \right] = \sigma_{xx} (\mu(\textbf{r}))\label{countersigma} \\
c({\mathcal M})\rightarrow c({\mathcal M})
 \left[ 1 + \alpha\frac{\gamma_0}{\sigma_{xx}} \ln
 \frac{\mu(\textbf{r}){\mathcal M}}{\mu_{0}} \right]\hspace{4cm}{}\notag\\
 =c \left[ 1 - \alpha \frac{\gamma_0}{\sigma_{xx}}
 \ln \frac{\mu_{0}}{\mu(\textbf{r})} e^{\gamma_E -1/2} \right] = c
 (\mu(\textbf{r}))\label{counterzc}\\
z({\mathcal M})\rightarrow z({\mathcal M})
 \left[ 1 + c \frac{\gamma_0}{\sigma_{xx}} \ln
 \frac{\mu(\textbf{r}){\mathcal M}}{\mu_{0}} \right]\hspace{4cm}{}\notag\\
 =z \left[ 1 - c \frac{\gamma_0}{\sigma_{xx}}
 \ln \frac{\mu_{0}}{\mu(\textbf{r})} e^{\gamma_E -1/2} \right] = z
 (\mu(\textbf{r})).\label{counterz}
\end{gather}
\noindent{The} introduction of spatially varying parameters
$\sigma_{xx} (\mu(\textbf{r}))$, $c (\mu(\textbf{r}))$ and $z
(\mu(\textbf{r}))$ means that the action $S^{\prime}_{\rm inst}$
gets modified according to the prescription
\begin{equation}\label{S-prime1a}
S^\prime_{\rm inst}\rightarrow - \int d\textbf{r}\,
 \sigma_{xx} (\mu(\textbf{r}))
 \tr (\nabla Q_{\rm inst} (\textbf{r}))^2
 \pm i \theta + \hat{S}_F[W]
\end{equation}
where
\begin{eqnarray}
\hat{S}_F[U] &=& \pi T \int^{\prime} d \mathbf{r}
z(\mu(\textbf{r}))\Bigl [ c(\mu(\textbf{r}))
\sum\limits_{\alpha n}\tr I_{n}^{\alpha }U^{-1}\rho U \notag \\
&&\times \tr I_{-n}^{\alpha }U^{-1} \rho U + 4 \tr\eta U^{-1}\rho
U \Bigr ].\label{S-prime}
\end{eqnarray}
Notice that in these expressions the instanton quantity $\rho$
depends explicitly on $\textbf r$ and should be read as $\rho =
\rho (\textbf{r})$.

%

\subsection{\label{BF} The quantities $\sigma_{xx}$ and
$\sigma_{xx}^\prime$ in flat space}

%
\subsubsection{Transformation}
Let us first evaluate the first spatial integral in Eq.
\eqref{S-prime1a} which can be written as
\begin{eqnarray}
 \int d\mathbf{r} \,\sigma_{xx} (\mu(\mathbf{r})) \tr (\nabla
 Q_\textrm{inst}(\mathbf{r}))^2
 &=&
 \int d \mathbf{r} \mu^2 (\mathbf{r}) \sigma_{xx}
 (\mu(\mathbf{r}))\notag \\
&=& 2\pi \sigma_{xx} (\zeta \lambda)  \label{newsigma}
\end{eqnarray}
where
\begin{equation}\label{sxxzl}
\sigma_{xx} (\zeta \lambda) = \sigma_{xx}
 - \beta_0(c) \ln {\zeta \lambda \mu_{0}}
 e^{\gamma_E}, \qquad \zeta =e^2 /4.
\end{equation}
We have introduced the quantity $\zeta$ that from now onward
denotes the different numerical factors that one in general can
associate with each of the different regularization schemes that
one uses. Notice that the expression for $\sigma_{xx}^{\prime}$,
Eq.~\eqref{ss1} now becomes
\begin{equation}
\sigma_{xx}^{\prime} =
 \sigma_{xx}({\mathcal M}) - \int^{\prime}
 \frac{d \lambda}{\lambda} \mathcal{D}(c)\sigma_{xx}^2
 e^{-2\pi\sigma_{xx}(\zeta \lambda)} \cos \theta .
\end{equation}
To complete the transformation from curved space to flat space we
still have to perform similar operations on the {\em observable}
theory. Write
\begin{equation}
\sigma_{xx}^{\prime} ({\mathcal M})  \rightarrow
\sigma_{xx}^{\prime} (\mu^\prime ({\bf r})) \notag
\end{equation}
then completely analogous to the definition of
Eq.~\eqref{newsigma} we obtain the observable parameter
$\sigma_{xx}^\prime$ in flat space according to the prescription
\begin{equation}
\label{sxxprimezl} \sigma_{xx}^{\prime} (\zeta\lambda^\prime) =
\frac{1}{2\pi} \int d {\bf r} (\mu^\prime ({\bf r}))^2
\sigma_{xx}^{\prime} (\mu^\prime ({\bf r})) .
\end{equation}
One can think of the quantity $\mu^\prime ({\bf r})
=2\lambda^\prime / (r^2 +\lambda^{\prime 2})$ as being the result
of a background instanton with a large scale size
$\lambda^\prime$. The result for $\sigma_{xx}^{\prime}$ and
$\theta^{\prime}$ in flat space can now be written as follows
\begin{eqnarray}
 \sigma_{xx}^{\prime} (\zeta \lambda^{\prime})&=&
 \sigma_{xx}(\zeta \lambda^{\prime}) - \int^{\prime}
 \frac{d[\zeta  \lambda]}{\zeta \lambda} \mathcal{D}(c)\sigma_{xx}^2
 e^{-2\pi\sigma_{xx}(\zeta \lambda)}\notag \\
 & &\hspace{3cm}\times  \cos \theta \label{ss1a}\\
 {\theta^{\prime} (\zeta \lambda^{\prime})} &=&
 {\theta} \hspace{0.5cm}- 2\pi \int^{\prime}
 \frac{d[\zeta \lambda]}{\zeta \lambda} \mathcal{D}(c)\sigma_{xx}^2
 e^{-2\pi\sigma_{xx}(\zeta \lambda)}\notag \\
 && \hspace{3cm}\times \sin \theta .\label{ss2a}
\end{eqnarray}
\subsubsection{\label{Dlambda} Integration over scale sizes $\lambda$}
Notice that the expression for $\sigma_{xx}(\zeta
\lambda^{\prime})$ has precisely the same meaning as
Eq.~\eqref{sxxzl} with $\lambda$ replaced by $\lambda^\prime$. In
Table ~\ref{REGULAR} we compare this expression with the result
obtained in dimensional regularization. To discuss the effect of
the arbitrary factor $\zeta$ it is convenient to write the
quantity $\sigma_{xx}(\zeta \lambda^{\prime})$ as an integral over
scale sizes
\begin{equation}
\sigma_{xx} (\zeta \lambda^{\prime}) =
 \sigma_{xx}^0 - \int^{\zeta \lambda^\prime}_{1/\mu_0 e^{\gamma_E}}
 \frac{d[\zeta  \lambda]}{\zeta \lambda} \beta_0 (c)
 \label{flatsigmaprime}
\end{equation}
where $\sigma_{xx}^0 = \sigma_{xx} (1/\mu_0 e^{\gamma_E})$. We
thus obtain the following natural expression for the observable
theory
\begin{widetext}
\begin{eqnarray}
 \sigma_{xx}^{\prime} (\zeta \lambda^{\prime})&=&
 \sigma_{xx}^0 - \int^{\zeta \lambda^\prime}_{1/\mu_0 e^{\gamma_E}}
 \frac{d[\zeta  \lambda]}{\zeta \lambda}
 \left[ \beta_0 (c) + \mathcal{D}(c)\sigma_{xx}^2
 e^{-2\pi\sigma_{xx}(\zeta \lambda)} \cos \theta \right] \label{ss1aa} \\
 {\theta^{\prime} (\zeta \lambda^{\prime})} &=&
 {\theta} - 2\pi \int^{\zeta \lambda^\prime}_{1/\mu_0 e^{\gamma_E}}
 \frac{d[\zeta \lambda]}{\zeta \lambda}
 \left[ 0 ~~~~~+ \mathcal{D}(c)\sigma_{xx}^2
 e^{-2\pi\sigma_{xx}(\zeta \lambda)} \sin \theta \right] .\label{ss2aa}
\end{eqnarray}
\end{widetext}
Notice that the contributions from instantons are finite in the
ultraviolet and the limit $\mu_0 \rightarrow \infty$ was taken
implicitly in the computation of the original expressions of
Eqs.~\eqref{ss1a} and \eqref{ss2a}. By comparing
Eqs.~\eqref{flatsigmaprime}-\eqref{ss2aa} with the result of
Eq.~\eqref{rgsigma} obtained from the theory in dimensional
regularization one clearly sees that the integral over scale sizes
$\lambda$ should be interpreted in terms of the integral over
momentum scales that generally defines the relation between the
observable and renormalized theories. Hence, we have found the
natural meaning for the instanton parameter $\lambda$. This
meaning obviously does not emerge from free energy considerations
alone. The results of this paper therefore fundamentally resolve
the infrared controversies that historically were associated with
the problem of instantons and instanton gases in scale invariant
theories.~\cite{BergLusher1,BergLusher2,Rajaraman}

Eqs~\eqref{ss1aa} and ~\eqref{ss2aa} show furthermore that the
factor $\zeta$ can be absorbed in a redefinition of $\lambda$.
Different values of $\zeta$ simply amount to different values of
the momentum scale that one associates with the bare parameters
$\sigma_{xx}^0$ and $\theta$. These differences, however, do not
affect the expressions $[ \dots ]$ of the integrand which are
therefore independent of the specific regularization scheme that
has been used to define the renormalized theory. This aspect of
universality has recently been exploited for the purpose of making
detailed comparisons~\cite{PruiskenBurmistrov} between the quantum
critical predictions of the free electron theory ($c=0$) and the
results known from numerical experiments.

Before embarking on the renormalization of the theory with
interactions we shall first address the various difficulties
associated with the observable parameters $z^\prime$ and $z^\prime
c^\prime$. This will be done in the Sections below and we will
come back to the $\beta$ and $\gamma$ functions of the theory in
Section~\ref{bandgprime}.

\subsection{\label{secAAA} The quantities $z$, $zc$ and $z^\prime$,
$z^\prime c^\prime$ in flat space}

%
In this Section we extend the various steps of
Eqs.~\eqref{newsigma}-\eqref{sxxprimezl} and translate the
parameters $z({\mathcal M})$ and $z^\prime ({\mathcal M})$ as well
as $z({\mathcal M}) c({\mathcal M})$ and $z^\prime ({\mathcal M})
c^\prime ({\mathcal M})$ into the appropriate quantities that are
defined in {\em flat} space. As an important check upon the
procedure we make sure that the relation $z^\prime \alpha^\prime =
z \alpha$ is satisfied at different stages of the analysis. For
the main part, however, the present Section proceeds along the
similar lines as those presented in the study of the ordinary
Grassmannian theory.~\cite{PruiskenBurmistrov}

\subsubsection{Transformation}
Let us first introduce the spatially varying momentum scales $\mu
({\textbf r})$ and $\mu^\prime ({\textbf r})$ according to
Eqs.~\eqref{counterzc} and \eqref{counterz}
\begin{eqnarray}
 z^{\prime} &=& z({\mu^\prime ({\textbf r})}) \notag \\
 &+& \frac{\gamma_0}{4\pi} \int^{\prime}
 \frac{d \lambda}{\lambda} \mathcal{D}(c) \mathcal{A}_1
 e^{-2\pi\sigma_{xx}(\zeta \lambda^\prime)}
 \cos \theta \\
 z^{\prime}c^\prime &=& z(\mu^\prime ({\textbf r})) c(\mu^\prime ({\textbf r}))
 \notag \\
 &+& \frac{\gamma_0}{4\pi} \int^{\prime}
 \frac{d \lambda}{\lambda} \mathcal{D}(c) \mathcal{A}_1
 e^{-2\pi\sigma_{xx}(\zeta \lambda^\prime)}
 \cos \theta .
\end{eqnarray}
The amplitude $\mathcal{A}_1$ is given as
\begin{equation}\label{A}
 \mathcal{A}_1 = \int^{\prime} d\textbf{r}
 \frac{\mu(\textbf{r})}{\lambda} z(\mu ({\bf r}))c(\mu ({\bf r}))
 .
\end{equation}
By using exactly the same procedure as in Eqs.~\eqref{newsigma}
and \eqref{sxxprimezl} we next define the quantities in flat space
$z({\zeta \lambda})$ and $z({\zeta \lambda}) c({\zeta \lambda})$
according to
\begin{eqnarray}\label{zzl}
z({\zeta \lambda})&=& \frac{1}{2\pi}\int d {\bf r} \mu^2 ({\bf r})
z(\mu ({\bf r})) \label{zprimezl0}\\
z({\zeta \lambda})c({\zeta \lambda})&=& \frac{1}{2\pi}\int d {\bf
r} \mu^2 ({\bf r}) z(\mu ({\bf r})) c(\mu ({\bf r})) .
\label{zprimezl1}
\end{eqnarray}
From this one obtains the explicit results
\begin{eqnarray}\label{flatz1}
z(\zeta\lambda) &=&\hspace{0.2cm} z\left[ 1
 - \frac{c \gamma_0}{\sigma_{xx}} \ln {\zeta \lambda \mu_{0}}
 e^{\gamma_E -1/2} \right] \\\label{flatz2}
 z(\zeta\lambda) c(\zeta\lambda) &=& zc\left[ 1
 - \frac{\gamma_0}{\sigma_{xx}} \ln {\zeta \lambda \mu_{0}}
 e^{\gamma_E -1/2} \right] .
\end{eqnarray}
In Table~\ref{REGULAR} we show that these results are precisely
consistent with those of the theory in dimensional regularization.
To proceed let us first apply the transformations to obtain the
observable parameters in flat space. Completely analogous to
Eq.~\eqref{sxxprimezl} we have
\begin{eqnarray}
 z^{\prime}({\zeta \lambda^\prime}) &=& z({\zeta \lambda^\prime})
 \notag\\
 &+& \frac{\gamma_0}{4\pi} \int^{\prime}
 \frac{d \lambda}{\lambda} \mathcal{D}(c) \mathcal{A}_1
 e^{-2\pi\sigma_{xx}(\zeta \lambda^\prime)}
 \cos \theta \\
 z^{\prime}c^\prime({\zeta \lambda^\prime}) &=& z({\zeta \lambda^\prime})
 c({\zeta \lambda^\prime}) \notag\\
 &+& \frac{\gamma_0}{4\pi} \int^{\prime}
 \frac{d \lambda}{\lambda} \mathcal{D}(c) \mathcal{A}_1
 e^{-2\pi\sigma_{xx}(\zeta \lambda^\prime)}
 \cos \theta .\hspace{0.5cm}\,{}
\end{eqnarray}
Here, the $z({\zeta \lambda^\prime})$ and $z({\zeta
\lambda^\prime}) c({\zeta \lambda^\prime})$ are defined by Eqs
~\eqref{flatz1} and \eqref{flatz2} with $\lambda$ replaced by
$\lambda^\prime$.

The problem that clearly remains is how to express the amplitude
$\mathcal{A}_1$, Eq.~\eqref{A}, in terms of the spatially flat
quantities defined in Eqs ~\eqref{flatz1} and \eqref{flatz2}.
%
\subsubsection{Amplitude $\mathcal{A}_1$}
%
To evaluate $\mathcal{A}_1$ further it is convenient to introduce
the quantity $M_1(\textbf{r})$ according to
\begin{eqnarray}
 \mathcal{A}_1 &=& z(\mu(0)) c(\mu(0)) \mathcal{M}_1 \\
 \mathcal{M}_1 &=& - 2\pi \int_{\mu(0)}^{\mu(L^{'})} d [\ln \mu
 (\textbf{r}) ] M_1 (\textbf{r})\\
 M_0 (\textbf{r}) &=& \frac{z(\mu(\textbf{r})) c(\mu(\textbf{r}))}
 {z(\mu(0)) c(\mu(0))}.
\end{eqnarray}
Since the anomalous dimension $\gamma_{zc}=\gamma_z/c$ is negative
the quantity $M_1 (\textbf{r})$ is in all respects like a
spatially varying {\em spontaneous magnetization} in the classical
Heisenberg ferromagnet. The associated momentum scale
$\mu(\textbf{r})$ strongly varies from {\em large} values
$O(\lambda^{-1})$ at short distances ($|\textbf{r}| \ll \lambda$)
to {\em small} values $O(\lambda/(L^\prime)^2)$ at very large
distances ($|\textbf{r}| \approx L^\prime \gg \lambda$). This
means that at distances sufficiently far from the center of the
instanton the system is effectively in the {\em symmetric} or {\em
strong coupling} phase where $M_1 (\textbf{r})$ vanishes. Hence we
expect the amplitude $\mathcal{M}_1$ to remain finite as $L'
\rightarrow \infty$. This is in spite of the fact that the
amplitude $\mathcal{A}_1$ diverges at a classical level.
%
\subsubsection{Details of computation}
%
The expression for $\mathcal{M}_1$ can be written in terms of the
$\gamma_{zc}$ function as follows
\begin{eqnarray}
 \mathcal{M}_1 = &-& 2\pi \int_{\ln \mu(0)}^{\ln \mu(L^\prime)}
 d [\ln \mu(\textbf{r})] \notag\\
 &\times & \exp\Bigl \{ -\int_{\ln \mu(0)}^{\ln
 \mu(\textbf{r})} d [\ln \mu ]\gamma_{zc} \Bigr \}.\label{hAi10_1}
\end{eqnarray}
Taking the derivative with respect to $\ln\lambda$ we find that
$\mathcal{M}_1$ obeys the following differential equation
\begin{equation}\label{Dif1}
 \left ( -\frac{d}{d \ln\lambda} +\gamma_{zc}\right
 )\mathcal{M}_1 = 2\pi \left (1 + M_1 (L^\prime)\right ).
\end{equation}
We can safely take the limit $L^\prime = \infty$ and put
$M_1(L^\prime) = 0$ from now onward. At the same time one can
solve Eq.~\eqref{Dif1} in the weak coupling limit where $\lambda
\to 0$, $\mu(0)\to\infty$ and $\sigma_{xx}(\mu(0)) \to \infty$.
Under these circumstances it suffices to insert the perturbative
expressions of Eqs.~\eqref{pb1}, \eqref{pb2} and \eqref{bc1} for
the $\gamma_{zc}$, $\beta_{\sigma}$ and $\beta_c$ functions such
that the quantity $\mathcal{M}_1 = \mathcal{M}_1(\sigma_{xx}
(\mu(0)), c(\mu(0)))$ is obtained as the solution of the
differential equation
\begin{equation}\label{Dif2}
 \left ( \beta_\sigma \frac{\partial }{\partial \sigma_{xx}}
 +\beta_c \frac{\partial }{\partial c} +\gamma_{zc} \right )
 \mathcal{M}_1
 = 2\pi,
\end{equation}
where to leading order $\beta_\sigma = \beta_0(c)$, $\beta_c =
c(1-c)\gamma_0/\sigma_{xx}$ and $\gamma_{zc} = -
\gamma_0/\sigma_{xx}$. The result for $\mathcal{M}_1$ can
generally be expanded in powers of $\sigma_{xx}^{-1} (\mu(0))$
\begin{equation}\label{Dif4}
\mathcal{M}_1 = 2\pi^2 \sigma_{xx} m_1^{(1)} (c) + m_0^{(1)} (c) +
\sigma_{xx}^{-1} m_{-1}^{(1)}(c) + \dots
\end{equation}
We are interested in the leading order quantity $m_1^{(1)}(c)$
which obeys the following differential equation
\begin{equation}\label{DifR}
\left( - \gamma_0 c (1-c) \frac{d }{d c} + (\beta_0(c)-\gamma_0)
\right) m_1^{(1)}(c) = \frac{1}{\pi}.
\end{equation}
The solution can be written as
\begin{equation}\label{solR}
 m_1^{(1)}(c) =\frac{\alpha }{c}\exp \left[ \frac{2}{c}\ln \alpha %
 \right] \int_{0}^{c}\mathrm{d}s (1-s)^{-2-2/s}.
\end{equation}
The quantity $m_1^{(1)}(c)$ varies between the Fermi liquid value
$m_1^{(1)}(0)$ and the Coulomb interaction value $m_1^{(1)}(1)$
which are obtained as
\begin{equation}\label{BCR}
 m_1^{(1)}(0) = 1,\qquad m_1^{(1)}(1) = 1/3.
\end{equation}
The result for $\mathcal{A}_1$ becomes
\begin{equation}
\mathcal{A}_1 = - 2 \pi^2 z(\mu(0)) c(\mu(0)) \sigma_{xx} (\mu(0))
m_1^{(1)}(c(\mu(0))).
\end{equation}
As a final step we wish to express $\sigma_{xx} (\mu(0))$,
$c(\mu(0))$ and $z(\mu(0))$ in terms of the spatially flat
quantities $\sigma_{xx}(\zeta\lambda)$, $c(\zeta\lambda)$ and
$z(\zeta\lambda)$ respectively. The following relations are
obtained
\begin{eqnarray}
\sigma_{xx}(\mu(0)) &=& \sigma_{xx} (\zeta\lambda) \left[
1+ \frac{\beta_0(c)}{\sigma_{xx} (\zeta\lambda)} \ln 2\zeta \right] \\
c(\mu(0)) &=& \hspace{0.4cm}c(\zeta\lambda) \left[ 1+ \frac{\alpha
\gamma_0}{\sigma_{xx} (\zeta\lambda)} \ln 2\zeta \right] \\
z(\mu(0)) &=& \hspace{0.4cm}z(\zeta\lambda) \left[ 1+ \frac{c
\gamma_0}{\sigma_{xx} (\zeta\lambda)} \ln 2\zeta \right] .
\end{eqnarray}
For our purposes the correction terms $O(\sigma_{xx}^{-1})$ are
unimportant. Hence we obtain the final result for the amplitude
$\mathcal{A}_1$ which can be written as follows
\begin{equation}\label{Aifinal}
 \mathcal{A}_1 = - 2\pi^2 z(\zeta\lambda)c(\zeta\lambda) \sigma_{xx}(\zeta\lambda)
 m_1^{(1)}(c(\zeta\lambda)) .
\end{equation}
The function $m_1^{(1)}(c)$ is given by Eq. \eqref{solR}. The
complete expressions for the quantities $z^\prime$ and $z^\prime
c^\prime$ now become
\begin{eqnarray}
 z^{\prime} (\zeta\lambda^\prime) &=& z(\zeta\lambda^\prime)
 \notag \\
 &-& \int^{\prime} \frac{d [\zeta \lambda]}{\zeta \lambda} z c
 {\mathcal D}_\gamma (c) \sigma_{xx}
 e^{-2\pi\sigma_{xx}} \cos \theta \label{ziprimezlprime} \\
 z^{\prime} (\zeta\lambda^\prime) c^\prime (\zeta\lambda^\prime)
 &=& z(\zeta\lambda^\prime) c(\zeta\lambda^\prime) \notag \\
 &-& \int^{\prime} \frac{d [\zeta \lambda]}{\zeta \lambda} z c
 {\mathcal D}_\gamma (c) \sigma_{xx}
 e^{-2\pi\sigma_{xx}} \cos \theta ~~~~~~\label{zciprimezlprime}
\end{eqnarray}
where
\begin{equation}
\mathcal{D}_\gamma(c) = -\frac{\gamma_0 \pi}{2}\mathcal{D}(c)
m_1^{(1)}(c) \label{Dgc}.
\end{equation}
In Fig.~\ref{FIG.DDc1} we plot the function
$\mathcal{D}_\gamma(c)$ with varying $c$. It has the Fermi liquid
value $\mathcal{D}_\gamma(0)=1/2$ and the Coulomb interaction
value $\mathcal{D}_\gamma(1)=1/6$.

\subsubsection{ Integration over scale sizes $\lambda$}
As before we can write the renormalized parameters $z(
\zeta\lambda^\prime)$ and $z( \zeta\lambda^\prime) c(
\zeta\lambda^\prime)$ as an integral over scale sizes. This leads
to the more general expression for the observable theory
\begin{widetext}
\begin{eqnarray}
 z^{\prime} (\zeta\lambda^\prime) &=& z_0 ~~
 - \int^{\zeta\lambda^\prime}_{1/\mu_0 e^{\gamma_E}} \frac{d [\zeta \lambda]}{\zeta \lambda}
 z c \left( \frac{\gamma_0}{\sigma_{xx}} + {\mathcal D}_\gamma (c) \sigma_{xx}
 e^{-2\pi\sigma_{xx} } \cos \theta \right) \\
 z^{\prime} (\zeta\lambda^\prime) c^\prime (\zeta\lambda^\prime) &=& z_0
 c_0
 - \int^{\zeta\lambda^\prime}_{1/\mu_0 e^{\gamma_E}} \frac{d [\zeta \lambda]}{\zeta \lambda}
 z c \left( \frac{\gamma_0}{\sigma_{xx}} + {\mathcal D}_\gamma (c)
 \sigma_{xx} e^{-2\pi\sigma_{xx}} \cos \theta \right)
\end{eqnarray}
\end{widetext}
where the parameters $z_0$ and $z_0 c_0$ are defined for a fixed
microscopic length scale $1/\mu_0 e^{\gamma_E}$. Again we compare
the results with those obtained from the theory in dimensional
regularization, Eqs.~\eqref{rgz1} and \eqref{rgz2}. This
comparison further demonstrates the validity of the statement made
earlier which says that the significance of the instanton
parameter $\lambda$ is primarily found in the fundamental relation
that exists between the observable and renormalized theories. At
the same time we conclude that the numerical factor $\zeta$ has
exactly the same meaning as discussed earlier and is immaterial.
%
%
\begin{table*}[tbp]
\begin{ruledtabular}
\caption{Observable theory using different regularization schemes,
see text}
\begin{tabular}{||c||c|c|c||}\hline
 &
 Pauli-Villars regularization ~~~~~~~&
 Pauli-Villars regularization ~~~~~~~&
 Dimensional regularization~\cite{Unify2,Unify5}
 ~~~~\\
 & ({\em curved space}) & ({\em flat space}) & \\
 & & & \\
 \hline\hline
 & & & \\
 $\sigma_{xx}^\prime$ & $\sigma_{xx}^\prime (\mathcal{M}) =
 \sigma_{xx} - \beta_0 (c)
 \ln \mathcal{M} e^{\gamma_E}$ & $\sigma_{xx}^\prime (\zeta \lambda) =
 \sigma_{xx} -
 \beta_0 (c) \ln  \zeta \lambda \mu_0 e^{\gamma_E}$ &
 $\sigma_{xx}^\prime (\mu) = \sigma_{xx} - \beta_0 (c) \ln \frac{\mu_0}{\mu}$ \\
 & & & \\
 \hline
 & & & \\
 $z^\prime$ & $z^\prime (\mathcal{M}) = z \left( 1 - \frac{c\gamma_0}{\sigma_{xx}}
 \ln \mathcal{M}e^{\gamma_E-\frac{1}{2}}\right)$ &
 $z^\prime (\zeta \lambda) = z \left( 1 - \frac{c\gamma_0}{\sigma_{xx}}
 \ln \zeta \lambda \mu_0 e^{\gamma_E-\frac{1}{2}}\right)$ &
 $z^\prime (\mu) = z \left( 1 - \frac{c\gamma_0}{\sigma_{xx}}
 \ln\frac{\mu_0}{\mu}e^{-\frac{1}{2}}\right)$ \\
 & & & \\
 \hline
 & & & \\
 $z^\prime c^\prime$ & $z^\prime (\mathcal{M}) c^\prime (\mathcal{M})
 =zc \left( 1 - \frac{\gamma_0}{\sigma_{xx}}
 \ln \mathcal{M}e^{\gamma_E-\frac{1}{2}} \right)$ &
 $z^\prime (\zeta \lambda) c^\prime (\zeta \lambda) = zc
 \left( 1 - \frac{\gamma_0}{\sigma_{xx}}
 \ln  \zeta \lambda \mu_0 e^{\gamma_E-\frac{1}{2}} \right)$ &
 $z^\prime (\mu) c^\prime (\mu) = zc \left( 1 - \frac{\gamma_0}{\sigma_{xx}}
 \ln \frac{\mu_0}{\mu}e^{-\frac{1}{2}}\right)$ \\
 & & &\\
\hline
\end{tabular}
\label{REGULAR}
\end{ruledtabular}
\end{table*}
%
%
\subsection{\label{Omega} Thermodynamic potential}
As an important general check on the consistency of the procedure
we next reconstruct the thermodynamic potential $\Omega$ of the
electron gas in the limit where $T$ goes to zero. It turns out
that the integration over the zero modes $U \in U(N) \times U(N)$
is not always as trivial as one might expect on the basis of the
previous Sections. For example, there is an ambiguity in
evaluating the expectation of $S_F [U]$ as given by
Eq.~\eqref{OFAvW} and the answer depends on cut-off procedure that
one uses in the summation over the $I_n$ matrices in the
definition of $S_F$. To obtain an unambiguous result we must take
the limit $T=0$ in a more careful fashion. As we next shall see,
this aspect of the problem has direct consequences for the
statement of $\mathcal F$ invariance as well as the statement made
in the beginning which says that the quantity $z\alpha$ is
unrenormalized.
\subsubsection{$t_0 = {\bf 1}$}
We start from the expression for the singlet interaction term
$\hat S_F[U]$ as given by Eq.~\eqref{S-prime} which still contains
the spatially varying momentum scale ${\mu(\mathbf{r})}$. In order
to separate the spatial integrals from the global matrices $U$ we
introduce the matrices $\hat{\Lambda}$ and $\hat{\bf 1}$
\begin{eqnarray}
\hat{\Lambda}_{nm}^{\alpha \beta } &=& \delta ^{\alpha 1}\delta
^{\beta 1}\delta _{nm}[\delta _{n0}-\delta _{n,-1}] \label{Aneta}
\\
\hat{\bf 1}_{nm}^{\alpha \beta } &=& \delta ^{\alpha 1}\delta
^{\beta 1}\delta _{nm}[\delta _{n0} + \delta _{n,-1}] .
\label{Aneta1}
\end{eqnarray}
Eq. ~\eqref{S-prime} can then be written as follows
\begin{equation}\label{SFprime1}
\hat S_F (U) = \hat S_i (U) + \hat S_\eta (U)
\end{equation}
where
\begin{eqnarray}
\hat S_i &=& - \frac{\pi}{2} T\lambda^2 \left( \mathcal{A}_1 -
\frac{5}{2}\mathcal{A}_2\right ) \sum \limits_{\alpha n}\tr
I^\alpha_n U^{-1}
\hat\Lambda U I^\alpha_{-n} U^{-1}\hat\Lambda U \notag\\
 && +\frac{\pi}{2} T \lambda^2 \left (\mathcal{A}_1-\frac{1}{2}\mathcal{A}_2\right )
 \sum \limits_{\alpha n}\tr
I^\alpha_n U^{-1}
\hat{\bf 1} U I^\alpha_{-n} U^{-1} \hat{\bf 1} U \notag\\
\hat S_\eta &=& -4 \pi T \lambda^2 \mathcal{A}_3 \tr \eta
U^{-1}\hat\Lambda U.
\end{eqnarray}
Here, the spatial integrals are all contained in the quantities
$\mathcal{A}_i$. $\mathcal{A}_1$ is defined in Eq.~\eqref{A}
whereas $\mathcal{A}_i$ are given as
\begin{eqnarray}
\mathcal{A}_{2}&=& \int^\prime d\mathbf{r}
\mu^2(\mathbf{r})z(\mu(\mathbf{r})) c(\mu(\mathbf{r})) \label{Oi2}  \\
\mathcal{A}_{3} &=& \int^\prime d\mathbf{r}
\frac{\mu(\mathbf{r})}{\lambda}z(\mu(\mathbf{r})). \label{Oi3}
\end{eqnarray}
Notice that ${\hat S}_F[U]$ has the same form as the classical
expression $S_F^\textrm{inst}[U]$ except that the amplitudes
$\mathcal{A}_i$ are replaced by $\mathcal{A}_i^\textrm{inst}$
according to
\begin{eqnarray}
\mathcal{A}_{1}^\textrm{inst} &=& z c \int^\prime d\mathbf{r}
\frac{\mu(\mathbf{r})}{\lambda} \label{Oi1class}\\
\mathcal{A}_{2}^\textrm{inst} &=& z c \int^\prime d\mathbf{r}
\mu^2(\mathbf{r})
\label{Oi2class}  \\
\mathcal{A}_{3}^\textrm{inst} &=& z ~~\int^\prime d\mathbf{r}
\frac{\mu(\mathbf{r})}{\lambda}. \label{Oi3class}
\end{eqnarray}
We have already mentioned earlier that the classical expression
$S_F^\textrm{inst}[U]$, in particular the amplitudes
$\mathcal{A}_{1}^\textrm{inst}$ and
$\mathcal{A}_{3}^\textrm{inst}$, diverge logarithmically in the
sample size. By following the same procedure as discussed in
Section~\ref{secAAA}, however, we find that the final expressions
for the amplitudes $\mathcal{A}_2$ and $\mathcal{A}_3$ are finite
\begin{eqnarray}
\mathcal{A}_2 &=& -2 \pi z(\zeta\lambda)  \label{Oif2}\\
\mathcal{A}_3 &=& -2 \pi^2 z(\zeta\lambda)
\sigma_{xx}(\zeta\lambda)m^{(3)}_{1}(c(\zeta\lambda)).
\label{Oif3}
\end{eqnarray}
Here, the quantity $m^{(3)}_1$ is given by
\begin{eqnarray}
m^{(3)}_1(c) &=& \alpha \exp \Bigl [ \frac{2}{c}\ln\alpha\Bigr ]
\int \limits_{0}^c \frac{d
s}{s(1-s)^2}\notag \\
&&\hspace{1.8cm} \times \exp \Bigl [ - \frac{2}{s}\ln(1-s)\Bigr
].\label{m3}
\end{eqnarray}
\begin{figure}[tbp]
\includegraphics[width=80mm]{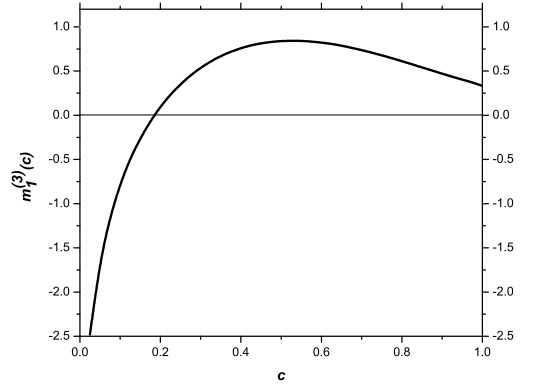}
\caption{ $m^{(3)}(c)$ versus $c$, see text.} \label{FIGm3}
\end{figure}
Notice that we can neglect the amplitude $\mathcal{A}_2$ relative
to the quantities $\mathcal{A}_1$ and $\mathcal{A}_3$ which are of
order $\sigma_{xx} (\zeta\lambda)$. On the other hand, in
Fig.~\ref{FIGm3} we plot of the function $m_1^{(3)}(c)$ that
defines the quantity $\mathcal{A}_3$. We see that $m^{(3)}_{1}(c)$
diverges as $c$ tends to $0$. This means that for $c=0$ the
leading term in $\mathcal{A}_3$ is proportional to $\sigma
_{xx}^{2}$ rather than $\sigma _{xx}$. Keeping these remarks in
mind we finally obtain the instanton contribution to the
thermodynamic potential as follows
\begin{eqnarray} \label{TDPU}
\Omega_\textrm{inst} &=& \frac{N^2}{4\pi^2} \int d\mathbf{r}_0
\int^\prime \frac{d\lambda}{\lambda^3}
\mathcal{D}(c(\zeta\lambda)) e^{-2 \pi \sigma_{xx}(\zeta\lambda)}
\cos\theta\notag \\
&& \hspace{1.5cm} \times \langle e^{ \hat S_i (U) + \hat S_\eta
(U) } \rangle_U .
\end{eqnarray}

\subsubsection{Expansion in $T$}
Next, in a naive expansion of the thermodynamic potential $\Omega$
in powers of the temperature $T$ one would proceed by replacing
the quantity $ \hat S_F (U) $ by its expectation with respect to
the matrix $U$. In the limit where $N=N_r N_m \rightarrow 0$ this
expectation is given by
\begin{eqnarray}\label{Si}
\langle\hat S_i (U) \rangle_U &=& ~~\frac{2 \pi \lambda^2 T
\mathcal{A}_1}{N^2} \tr \eta \Lambda
\\ \label{Seta}  \langle\hat S_\eta (U) \rangle_U
&=& -\frac{4 \pi \lambda^2 T \mathcal{A}_3}{N} \tr
\eta \Lambda.
\end{eqnarray}
To the lowest order in $T$ only the quantity $\langle\hat
S_i(U)\rangle_U$ survives in Eq.~\eqref{TDPU} whereas the term
$\langle\hat S_\eta (U) \rangle_U$ vanishes in the limit where $N
\rightarrow 0$. We have already mentioned, however, that the
expectation of $S_F$, in particular Eq.~\eqref{Si}, is complicated
and cut-off dependent. These as well as other complications
disappear once it is recognized that the frequency term $\hat
S_\eta (U)$ in the action is actually not a perturbative quantity
at all and should generally be retained in the exponential of
$\Omega_\textrm{inst}$. The correct series expansion in powers of
$T$ therefore has the following general form
\begin{eqnarray}
\Omega_\textrm{inst} &=& \frac{N^2}{4\pi^2} \int d\mathbf{r}_0
\int^\prime \frac{d\lambda}{\lambda^3}
\mathcal{D}(c(\zeta\lambda)) e^{-2 \pi \sigma_{xx}(\zeta\lambda)}
\cos\theta \notag \\ && \hspace{1.5cm} \times \langle e^{\hat
S_\eta (U)}(1 + \hat S_i (U) + \dots)
\rangle_U.\hspace{0.5cm}\,{}\label{Omegainst}
\end{eqnarray}
The problem that remains is to evaluate expectations of the type
\begin{equation}
\langle X \rangle_\epsilon = \langle X e^{-\epsilon \tr \eta
U^{-1} \hat{\Lambda} U} \rangle_U
\end{equation}
where we have written $\epsilon = -4\pi \lambda^2 T
\mathcal{A}_3$. For our purposes the only expectations that we
shall need are the following results which are valid in the limit
$N \rightarrow 0$
\begin{eqnarray}
\left\langle ( U^{-1} \hat{\Lambda} U )^{\alpha\beta}_{nm}
\right\rangle_\epsilon &=& \frac{1}{N}
\Lambda^{\alpha\beta}_{nm} e^{-\epsilon |n|} \label{AvLn}\\
\left\langle ( U^{-1} \hat{\bf 1} U )^{\alpha\beta}_{nm}
\right\rangle_\epsilon &=& \frac{1}{N} {\bf{1}}^{\alpha\beta}_{nm}
e^{-\epsilon |n|}.\label{AvEn}
\end{eqnarray}
We see that the main effect of $\epsilon$ is to exponentially
suppress the large Matsubara frequency components. To justify
Eqs.~\eqref{AvLn} and \eqref{AvEn} we proceed as follows. Since
the averaging over positive and negative frequency blocks is
independent of one another we first introduce for brevity the
symbol $P^\alpha_n = (U^{-1}\mathbf{\hat 1}
U)^{\alpha\alpha}_{nn}$ where $n$ is limited to, say, positive
frequency indices only. Equations~\eqref{AvLn} and \eqref{AvEn}
can then be expressed in terms of an infinite series expansion in
powers of $\epsilon$ with coefficients of the type
\begin{equation}
\langle P^{\beta_1}_{m_1}\dots P^{\beta_k}_{m_k}\rangle_U .
\end{equation}
The lowest order coefficients we already have, in particular
\begin{equation}
\langle P^\alpha_{n_1}\rangle_U = \frac{1}{N}, \qquad \langle
P^\alpha_{n_1} P^\beta_{n_3}\rangle_U = \frac{1}{N}\frac{1+
\delta^{\alpha\beta}_{n_1n_3}}{1+N}.
\end{equation}
The second of these equations simplifies in the limit $N \to 0$
and can be replaced by the following expression
\begin{equation}
\langle P^\alpha_{n_1} P^\beta_{n_3}\rangle_U = \frac{1}{N}
\delta^{\alpha\beta}_{n_1n_3}.
\end{equation}
It is clear that the terms that have been left out are all of
higher order in $N_r$ and therefore insignificant. Proceeding
along the same lines one can prove by induction that the general
expression can be written as
\begin{equation}
\langle P^{\beta_1}_{m_1}\dots P^{\beta_k}_{m_k}\rangle_U =
\frac{1}{N} \delta^{\beta_1\dots\beta_k}\delta_{m_1\dots m_k}.
\end{equation}
Using this result one can re-exponentiate the series in powers of
$\epsilon$ and the result can be written as follows
\begin{equation}
\Bigl \langle P^\alpha_{n} \exp\Bigl (-\epsilon
\sum\limits_{\beta,m>0} m P^\beta_{m} \Bigr )\Bigr \rangle =
\frac{1}{N} \exp (-\epsilon |n|).
\end{equation}
\begin{widetext}
This, then, directly leads to the result of Eqs.~\eqref{AvLn} and
\eqref{AvEn}.

\subsubsection{Effective action for $t_0$}

On the basis of Eqs.~\eqref{AvLn} and \eqref{AvEn} one can write
the expectation $\langle \hat S_i (U) \rangle_\epsilon $ as
follows
\begin{equation}\label{SFprime1a}
 \langle \hat S_i (U) \rangle_\epsilon = - \frac{\pi}{2}T \lambda^2 \mathcal{A}_1
 \sum \limits_{\alpha n} \tr
 \left (  \left \langle [ I^\alpha_n
 , U^{-1} \hat\Lambda U ] \right \rangle_\epsilon
 \left \langle [ I^\alpha_{-n} , U^{-1}\hat\Lambda U
 ] \right \rangle_\epsilon
 - \left\langle [ I^\alpha_n , U^{-1}
 \hat{\bf 1} U ] \right\rangle_\epsilon \left\langle [ I^\alpha_{-n} , U^{-1} \hat{\bf 1} U
 ] \right\rangle_\epsilon \right ) .
\end{equation}
\end{widetext}
This expression is important for a variety of reasons. First of
all, a finite value of $\epsilon$ permits us to take the frequency
cut-off $N_m$ appearing in the {\em size} of the matrices $[
I^\alpha_n , U^{-1} \hat\Lambda U ]$ and $[ I^\alpha_n , U^{-1}
\hat{\bf 1} U ]$ to infinity first. An explicit computation leads
to
\begin{eqnarray}
 \langle \hat S_i (U) \rangle_\epsilon &=&
\frac{2\pi \lambda^2 T \mathcal{A}_1}{N^2} \sum \limits_{\alpha n}
|n| e^{-\epsilon |n|} \notag \\ &=& \frac{2\pi \lambda^2 T
\mathcal{A}_1}{N^2}
 \tr \left (\eta \Lambda e^{-\epsilon \eta\Lambda}\right ).
\end{eqnarray}
This indicates that the large frequency components $n \gtrsim
\epsilon^{-1}$ are being suppressed by the theory itself and the
results are clearly independent of the arbitrary cut-off $N_m$ as
they should be.

Secondly, we can now proceed and extent the result of
Eq.~\eqref{SFprime1a} and, hence, the thermodynamic potential
$\Omega_\textrm{inst}$ to include the zero modes $t_0$ or $q_0 \in
U(2N) /U(N) \times U(N)$. Although we have seen that these zero
modes do not appear in the final answer, they can nevertheless be
used as an important check on the general statement which says
that the quantity $z\alpha$ is unrenormalized. Notice that as a
general prescription for inserting the rotation $t_0$ we can use
the procedure of Appendix~\ref{AppA} which shows how to deal with
the electrodynamic $U(1)$ gauge invariance of the theory.
Replacing in Eq. ~\eqref{SFprime1a}
\begin{equation}\label{prescr}
U^{-1}\hat\Lambda U  \to t^{-1}_0 U^{-1}\hat\Lambda U t_0
\end{equation}
then one should think of the matrix $t_0$ as being a ``small''
$U(2n)/U(n) \times U(n)$ rotation with $n=N_r n_m$ much smaller
than $N=N_r N_m$. Indeed, according to the rules of $\mathcal F$
algebra one considers the different cut-offs $n_m << \epsilon^{-1}
<< N_m$ as a general prescription that should be followed before
taking the limit to infinite frequency space, i.e. $n_m <<
\epsilon^{-1} << N_m \to \infty$.~\cite{Unify2} Evaluating the
theory of Eq.~\eqref{SFprime1a} in the presence of the matrix
field $t_0$ we can write
\begin{equation}
 \langle \hat S_i (Ut_0 ) \rangle_\epsilon = -\frac{\pi
 \lambda^2 T \mathcal{A}_1}{2 N^2}
 \Gamma [t_0 ]
\end{equation}
where
\begin{equation}\label{subtle}
 \Gamma [t_0 ] = \sum\limits_{\alpha n} \tr [I^\alpha_n, t^{-1}_0
 \Lambda e^{-\epsilon\eta\Lambda} t_0 ][I^\alpha_{-n}, t^{-1}_0 \Lambda
 e^{-\epsilon\eta\Lambda} t_0 ].
\end{equation}
To appreciate the subtleties that are associated with the ``{\em
finiteness}'' of $\epsilon$ as well as the ``{\em smallness}'' of
the background field $t_0$ we next analyze the result of
Eq.~\eqref{subtle} in some detail. First, $n_m \ll \epsilon^{-1}$
means that we can write
\begin{equation}
[e^{-\epsilon\eta\Lambda}, t^{-1}_0 ] \approx 0, \qquad
[e^{-\epsilon\eta\Lambda}, t_0 ]\approx 0.
\end{equation}
Hence,
\begin{equation}\label{Gammaq}
\Gamma [q_0 ] = \sum\limits_{\alpha n} \tr [I^\alpha_n, q_0
][e^{-\epsilon\eta\Lambda} I^\alpha_{-n} e^{-\epsilon\eta\Lambda}
, q_0 ]
\end{equation}
where $q_0 =t^{-1}_0 \Lambda t_0$. Eq.~\eqref{Gammaq} shows that
the results correctly display $U(N) \times U(N)$ invariance as it
should be.  Next we split the matrix $q_0$ into ``small''
components $q_0 -\Lambda$ and ``large'' components $\Lambda$ and
write
\begin{eqnarray}
\Gamma [q_0 ] &=&~~~ \sum\limits_{\alpha n} \tr [I^\alpha_n,
\Lambda]e^{-\epsilon\eta\Lambda} [I^\alpha_{-n},\Lambda]
e^{-\epsilon\eta\Lambda}\notag \\
&&+2\sum \limits_{\alpha n} \tr [I^\alpha_n, (q_0 -\Lambda)
]e^{-\epsilon\eta\Lambda} [I^\alpha_{-n},\Lambda]
e^{-\epsilon\eta\Lambda}\notag \\
&&+~\sum \limits_{\alpha, n} \tr [I^\alpha_n, (q_0 -\Lambda)
][I^\alpha_{-n},(q_0 -\Lambda)].
\end{eqnarray}
By using the following identity
\begin{equation}
e^{-\epsilon\eta\Lambda} [I^\alpha_{-n},\Lambda]
e^{-\epsilon\eta\Lambda} = e^{-\epsilon|n|}[
I^\alpha_{-n},\Lambda]
\end{equation}
we finally obtain two equivalent expressions for the quantity
$\Gamma$
\begin{eqnarray}
\Gamma [q_0 ] &=& \sum\limits_{\alpha n} e^{-\epsilon|n|}\tr
[I^\alpha_n, q_0 ][I^\alpha_{-n}, q_0 ] \\
\Gamma[q_0 ] &=& 2 \Bigl ( \sum\limits_{\alpha n} \tr I^\alpha_n
q_0 \tr I^\alpha_{-n} q_0 + 4 \tr \eta (q_0 -\Lambda) \notag \\
&&\hspace{0.5cm} -2 \tr \eta \Lambda e^{-\epsilon \eta \Lambda}
\Bigr ).
\end{eqnarray}
Notice that for all practical purposes we can represent the
results in terms of a reduced matrix space of size $\tilde{N}_m
\times \tilde{N}_m$ as follows
\begin{eqnarray}
\Gamma [q_0 ] &=& {\sum\limits_{\alpha n}}^\prime \tr
[I^\alpha_n, q_0 ][I^\alpha_{-n}, q_0 ] \label{oldform1}\\
\Gamma [q_0 ] &=& 2 \left ( \sum\limits_{\alpha n} \tr I^\alpha_n
q_0 \tr I^\alpha_{-n} q_0 + 4 \tr \eta q_0  - 6 \tr \eta \Lambda
\right ).\notag\\
\label{oldform2}
\end{eqnarray}
Here, the size $\tilde{N}_m$ of the matrices $I_n$ and $\lambda$
is such that $n_m \ll \tilde{N}_m \ll \epsilon^{-1}$. The prime on
the summation sign in Eq.~\eqref{oldform1} denotes the restriction
$-\tilde{N}_m < n < \tilde{N}_m$. In summary we can say that the
background field quantity $\Gamma [q_0 ]$ has the familiar
$\mathcal F$ invariant form.~\cite{Unify2}

The result for the thermodynamic potential, Eq.~\eqref{Omegainst},
in the presence of a global background field $t$ becomes
\begin{eqnarray}
\Omega_\textrm{inst} &=& \frac{N^2}{4\pi^2} \int d\mathbf{r}_0
\int^\prime \frac{d\lambda}{\lambda^3}
\mathcal{D}(c(\zeta\lambda)) e^{-2 \pi \sigma_{xx}(\zeta\lambda)}
\cos\theta \notag \\ && ~~~~~~ \times (1 + \frac{2\pi \lambda^2 T
\mathcal{A}_1}{N^2}
 \Gamma [q_0] + {\mathcal O} (T^2)
) .\label{Omegainst1}
\end{eqnarray}
To obtain the final total expression for the effective action
$S_\textrm{eff}[q_0]$ we have to add the results obtained for the
trivial vacuum. Splitting the thermodynamic potential in $T=0$ and
$T \neq 0$ parts
\begin{equation}
 \Omega = \Omega (T=0) + \Omega (T)
\end{equation}
then in the limit where $N \to 0$ we obtain
\begin{eqnarray}\label{Omegat0}
\Omega (T) &=& \ln \int \mathcal{D} [q_0] ~e^{S_\textrm{eff}[q_0]} \notag \\
S_\textrm{eff}[q_0] &=& ~~(L^2 T) z^\prime c^\prime
{\sum\limits_{\alpha n}}^\prime \tr [ I_{n}^{\alpha } , q_0 ] [
I_{-n}^{\alpha
} , q_0 ] \notag \\
&& + (L^2 T) z \alpha  \left( 4 \tr\eta q_0 - 6 \tr\eta \Lambda
\right ) .
\end{eqnarray}
These expressions are all well defined with $L$ denoting the
linear dimensions of the system. The most important result is that
quantity $z^\prime c^\prime$ is given precisely by
Eq.~\eqref{zciprimezlprime} whereas $z\alpha$ is unrenormalized.
Notice that in the limit where $L \to \infty$ only the classical
value $q_0 = \Lambda$ contributes as mentioned before. $\Omega
(T)$ therefore reduces to $S_\textrm{eff}[\Lambda]= -2 (L^2 T)
z^\prime \tr\eta \Lambda $ which by itself does not determine the
renormalization of $z$ and/or $c$. On the basis of
Eq.~\eqref{Omegat0} we conclude, however, that the observable
parameters $z^\prime$ and $z^\prime \alpha^\prime$ are correctly
given by the definitions of Eqs.~\eqref{zcren} and \eqref{zalpha}.
At the same time we have explicitly verified the $T$ dependent
part of the effective action as presented in Appendix~\ref{AppA},
Eq.~\eqref{SeffA7}.

%
%
\section{\label{bandgprime} The $\beta^\prime$ and $\gamma^\prime$ functions}
%
We next summarize the results obtained for the observable
parameters and derive expressions for the renormalization group
$\beta^\prime$ and $\gamma^\prime$ functions of the interacting
electron gas. The final expressions that we obtain in this Section
are amongst the most important results of the present paper.
%
\subsection{Observable and renormalized theories}
%
Introducing an arbitrary scale size $\lambda_0$ we can rewrite
Eqs.~\eqref{ss1a}, \eqref{ss2a}, \eqref{ziprimezlprime} and
\eqref{zciprimezlprime} in the following manner
\begin{eqnarray}
 \sigma_{xx}^{\prime} (\zeta \lambda^\prime) &=&
 {\sigma}_{xx}^\prime (\zeta \lambda_0)  \notag \\
 &&  - \int^{\zeta \lambda^\prime}_{\zeta \lambda_0} \frac{d [\zeta \lambda]}{\zeta
 \lambda} \beta_\sigma^\prime (\sigma_{xx} , \theta, c)
 \label{ss2a0}\\
 {\theta^{\prime}} (\zeta \lambda^\prime) &=&
 ~{{\theta}^\prime (\zeta \lambda_0 )} \notag \\
 &&  - \int^{\zeta \lambda^\prime}_{\zeta \lambda_0} \frac{d [\zeta
 \lambda]}{\zeta \lambda} \beta_\theta^\prime (\sigma_{xx} , \theta,
 c) \label{ss2a1} \\
 z^\prime (\zeta \lambda^\prime) &=& z^\prime (\zeta\lambda_0)
 \notag \\
 && + \int_{\zeta\lambda_0}^{\zeta \lambda^\prime}
 \frac{d[\zeta\lambda]}{\zeta\lambda}
 z^\prime \gamma_z^\prime (\sigma_{xx} , \theta, c, c^\prime)
 \label{ssss}\\
 z^\prime (\zeta \lambda^\prime) c^\prime (\zeta \lambda^\prime)
 &=& z^\prime (\zeta\lambda_0) c^\prime (\zeta\lambda_0) \notag \\
 &&  + \int_{\zeta\lambda_0}^{\zeta \lambda^\prime}
 \frac{d[\zeta\lambda]}{\zeta\lambda}
 z^\prime c^\prime \gamma_{zc}^\prime (\sigma_{xx} , \theta, c,
 c^\prime)~~~~
\end{eqnarray}
where
\begin{eqnarray}
 \beta_{\sigma}^\prime (\sigma_{xx} , \theta, c) &=&
 -\frac{d\sigma_{xx}^\prime}{d\ln \lambda} \notag \\
 &=& \beta_0 (c) + \mathcal{D}(c) \sigma_{xx}^2
 e^{-2\pi\sigma_{xx}}\cos \theta\label{betafunctionsa} \\
 \beta_\theta^\prime (\sigma_{xx} , \theta, c) &=&
 -\frac{d \theta^\prime}{d\ln
 \lambda} \notag \\
 &=& ~~~~~~~ 2\pi \mathcal{D}(c) \sigma_{xx}^2
 e^{-2\pi\sigma_{xx}}\sin \theta
 \hspace{1cm}{}\label{betafunctionsb}\\
 \gamma_z^\prime (\sigma_{xx} , \theta, c, c^\prime) &=&
 \frac{d\ln z^\prime}{d\ln\lambda} = \frac{(1-c^\prime)c}{1-c} \Bigl[ \frac{ \gamma_0}{\sigma_{xx}}
 \notag \\
 &+&  \mathcal{D}_\gamma
 (c) \sigma_{xx} e^{-2\pi\sigma_{xx}} \cos\theta \Bigr]\hspace{1cm}{}\\
 \gamma_{zc}^\prime (\sigma_{xx} , \theta, c, c^\prime) &=&
 \frac{d\ln z^\prime c^\prime}{d\ln\lambda} = \frac{(1-c^\prime)
 c}{(1-c~)c^\prime}\Bigl[
 \frac{\gamma_0}{\sigma_{xx}}
 \notag \\
 &+& \mathcal{D}_\gamma
 (c) \sigma_{xx} e^{-2\pi\sigma_{xx}} \cos\theta \Bigr].
\end{eqnarray}
The difference between the {\em observable} theory
$\sigma_{xx}^\prime$, $\theta^\prime$, $c^\prime$ and $z^\prime$
and the {\em renormalized} theory $\sigma_{xx}$, $\theta$, $c$ and
$z$ can be expressed in terms of the renormalization group
functions as follows
\begin{eqnarray}
 \beta_{\sigma} (\sigma_{xx} , c) &\Leftrightarrow&
 \beta_{\sigma}^\prime (\sigma_{xx} , \theta, c) \\
 \beta_{\theta} =0 ~~~~&\Leftrightarrow&
 \beta_{\theta}^\prime (\sigma_{xx} , \theta, c) \\
 \gamma_z (\sigma_{xx} , c) &\Leftrightarrow&
 \gamma_z^\prime (\sigma_{xx} , \theta, c, c^\prime) \\
 \gamma_{zc} (\sigma_{xx} , c) &\Leftrightarrow&
 \gamma_{zc}^\prime (\sigma_{xx} , \theta, c, c^\prime).
\end{eqnarray}
Our final task is to express the $\beta^\prime$ and
$\gamma^\prime$ functions of the {\em observable} theory in terms
of the {\em observable} parameters $\sigma_{xx}^\prime$,
$\theta^\prime$ and $c^\prime$ alone, rather than the {\em
renormalized} quantities $\sigma_{xx}$, $\theta$ and $c$. To
ensure that this can be done in a legitimate fashion we proceed as
follows. First of all it is important to notice that the following
general relations hold
\begin{eqnarray}
 \gamma_z (\sigma_{xx} , c) &=& c \gamma_z (\sigma_{xx} , c) \\
 \label{gammazprime}
 \gamma_z^\prime (\sigma_{xx} , \theta, c, c^\prime) &=&
 c^\prime \gamma_{zc}^\prime (\sigma_{xx} , \theta, c, c^\prime) .
\end{eqnarray}
This means that both quantities $z\alpha$ and $z^\prime
\alpha^\prime$ are unrenormalized as they should be. Next, we
compare the renormalization behavior of the quantities $c$ and
$c^\prime$
\begin{eqnarray}
 \beta_c (\sigma_{xx}, c) &=&
 -\frac{dc}{d \ln \lambda} \notag \\
 &=& (1-c){c}
 \gamma_{zc} (\sigma_{xx}, c) \\
 \beta_c^\prime (\sigma_{xx}, \theta , c, c^\prime) &=&
 -\frac{d c^\prime}{d \ln \lambda} \notag \\ \label{betacprime}
 &=& (1-c^\prime) c^\prime
 \gamma_{zc}^\prime (\sigma_{xx}, \theta , c, c^\prime) .~~~~~
\end{eqnarray}
We see that the Fermi liquid plane $c=c^\prime = 0$ and the
Coulomb interaction plane $c=c^\prime = 1$ correspond to zero's of
both the $\beta_c$ and $\beta_c^\prime$ functions provided the
$\gamma_{zc}^\prime$ is well behaved.
\subsection{The $\beta^\prime$ and $\gamma^\prime$ functions}
%
The relation between the observable and renormalized theories can
be obtained by solving the following differential equations
\begin{eqnarray}
 \beta_\sigma (\sigma_{xx} , c) \frac{\partial\sigma_{xx}^\prime}
 {\partial \sigma_{xx}} + \beta_c (\sigma_{xx} , c)
 \frac{\partial \sigma_{xx}^\prime}{\partial c}&=&
 \beta_\sigma^\prime (\sigma_{xx}, \theta , c) \label{dea} \\
 {\beta_\sigma (\sigma_{xx} , c)}
 \frac{\partial \theta^\prime}{\partial \sigma_{xx}} +
 ~~ {\beta_c (\sigma_{xx} , c)}
 \frac{\partial \theta^\prime }{\partial c}&=&
 \beta_\theta^\prime (\sigma_{xx}, \theta , c)
 \label{deb} \\
 \beta_\sigma (\sigma_{xx} , c) \frac{\partial c^\prime}{\partial\sigma_{xx}} +
 ~~ \beta_c (\sigma_{xx} , c)
 \frac{\partial c^\prime }{\partial c}&=&
 \beta_c^\prime (\sigma_{xx}, \theta , c, c^\prime) .~~~~ \notag \\
\end{eqnarray}
To obtain solutions that are meaningful in the entire range $0
\leq c \leq 1$ we must work with the two loop results for the
$\beta_\sigma$ function as in Eq.~\eqref{pb1}. It is next a matter
of simple algebra to show that the results can be expressed in
terms of an infinite double series in powers of $\exp
(-2\pi\sigma_{xx}^\prime)$ and the trigonometric functions of
$\theta^\prime$. The first few terms in the series can be written
as follows
\begin{widetext}
\begin{eqnarray}
 \beta_{\sigma}^\prime (\sigma_{xx}^\prime , \theta^\prime ,
 c^\prime) &=& \left\{ \beta_\sigma (\sigma_{xx}^\prime, c^\prime)
 + F_0^\prime e^{-4\pi\sigma_{xx}^\prime } ~ \right\} +
 \left\{
 {\mathcal{D}(c^\prime)} \left( \sigma_{xx}^\prime \right)^2
 e^{-2\pi\sigma_{xx}^\prime } \right\} \cos \theta^\prime
 + \left\{ F_2^\prime e^{-4\pi\sigma_{xx}^\prime } \right\}
 \cos2\theta^\prime + \dots \label{finalbetaprime}
 \\
 \beta_\theta^\prime (\sigma_{xx}^\prime , \theta^\prime ,
 c^\prime) &=& ~~~~~~~~~~~~~~~~~~~~~~~~~~~~~~~~~~~~~~~
 \left\{
 {\mathcal{D}(c^\prime)} \left( \sigma_{xx}^\prime \right)^2
 e^{-2\pi\sigma_{xx}^\prime }\right\} \sin \theta^\prime +
 \left\{ F_2^\prime e^{-4\pi\sigma_{xx}^\prime } \right\} \sin 2\theta^\prime
 + \dots  \\
 \gamma_{zc}^\prime (\sigma_{xx}^\prime , \theta^\prime ,
 c^\prime) &=& \left\{ \gamma_{zc} (\sigma_{xx}^\prime, c^\prime) +
 H_0^\prime e^{-4\pi\sigma_{xx}^\prime } \right\}
 +  \left\{
 {\mathcal{D}_\gamma (c^\prime)} \sigma_{xx}^\prime
 e^{-2\pi\sigma_{xx}^\prime }~~~\right\} \cos
 \theta^\prime + \left\{ H_2^\prime e^{-4\pi\sigma_{xx}^\prime }
 \right\} \cos 2\theta^\prime + \dots \hspace{0.5cm}\,{}
 \label{betafunctionsprimea}
\end{eqnarray}
\end{widetext}
where the $F^\prime$, and $H^\prime$ are rational
functions in $\sigma_{xx}^\prime$ and to leading order are given
by
\begin{eqnarray}
 F_0^\prime &=&
 -\frac{\left( \sigma_{xx}^\prime \right)^4}
 {\beta_\sigma (\sigma_{xx}^\prime, c^\prime)} \mathcal{D}^2
 (c^\prime)
 \label{f0} \\
 F_2^\prime &=&
 \frac{\left( \sigma_{xx}^\prime
 \right)^3 }
 {2\pi \beta_\sigma (\sigma_{xx}^\prime, c^\prime)} \mathcal{D} (c^\prime)
 \times \notag \\
 && \times \left[ \mathcal{D} (c^\prime)
 +\frac{1}{2}c^\prime (1-c^\prime)\mathcal{D}_\gamma(c^\prime)
 \partial_{c^\prime} \mathcal{D}_\gamma (c^\prime) \right]\qquad {} \label{f2}\\
  H_0^\prime &=&
 -\frac{\left( \sigma_{xx}^\prime \right)^3}
 {\beta_\sigma (\sigma_{xx}^\prime, c^\prime)}
 \mathcal{D}_\gamma (c^\prime) \mathcal{D} (c^\prime)
 \label{f0} \\
 H_2^\prime &=& \frac{\left( \sigma_{xx}^\prime \right)^2}
 {4\pi \beta_\sigma (\sigma_{xx}^\prime , c^\prime)}  \mathcal{D}_\gamma (c^\prime) \times \notag \\
 && \times \left[ \mathcal{D} (c^\prime) + \mathcal{D}_\gamma (c^\prime)
 + c^\prime (1-c^\prime) \partial_{c^\prime}
 \mathcal{D}_\gamma (c^\prime) \right].\qquad {} \label{h2}
\end{eqnarray}
We see that the renormalization group $\beta^\prime$ and
$\gamma^\prime$ functions are formally given as a sum over all
topological sectors of the theory. This is in spite of the fact
that we started out the computation with single instanton only.

It is interesting to digress on the higher order terms in the
series that are actually beyond the scope of the present analysis.
For example, it is clear that all the exponential terms
proportional to $\exp(-4\pi\sigma_{xx}^\prime)$ in
Eqs.~\eqref{finalbetaprime} - \eqref{betafunctionsprimea}
generally become important when {\em multi instanton}
configurations are taken into account. In particular, the terms
with $H_0^\prime$ and $F_0^\prime$ indicate that the trivial
vacuum is affected by {\em instanton} and {\em anti instanton}
combinations. Similarly, the terms proportional to $F^\prime_2$
and $H^\prime_2$ are recognized as the {\em disconnected} pieces
that appear in the contributions from instantons of topological
charge $\pm 2$. It is not difficult to see that a consistent
procedure for multi instantons is likely to involve the effects of
merons.~\cite{Gross}

To summarize the main results of this paper we can say that the
theory of observable parameters can be expressed as follows
\begin{eqnarray}
 \sigma_{xx}^{\prime} (\zeta \lambda^\prime) &=&
 \sigma_{xx}^\prime (\zeta \lambda_0) -
 \int^{\zeta \lambda^\prime}_{\zeta \lambda_0} \frac{d [\zeta \lambda]}{\zeta
 \lambda}  \beta_{\sigma}^\prime (\sigma_{xx}^\prime ,
 \theta^\prime , c^\prime ) \notag \\ \label{ss1a10} \\
 \frac{\theta^{\prime} (\zeta \lambda^\prime)}{2\pi} &=&
 ~\frac{\theta^\prime (\zeta \lambda_0 )}{2\pi}
 \hspace{0.18cm}-
 \int^{\zeta \lambda^\prime}_{\zeta \lambda_0} \frac{d [\zeta \lambda]}{\zeta \lambda}
 \beta_{\theta}^\prime (\sigma_{xx}^\prime ,
 \theta^\prime , c^\prime ) \notag \\ \label{ss2a10} \\
 z^{\prime} (\zeta \lambda^\prime) &=& ~z^\prime (\zeta \lambda_0 )
 \hspace{0.18cm}-
 \int^{\zeta \lambda^\prime}_{\zeta \lambda_0}
 \frac{d [\zeta \lambda]}{\zeta \lambda}
 z^\prime c^\prime \gamma^\prime_{zc} (\sigma_{xx}^\prime ,
 \theta^\prime , c^\prime ) \notag \\ \label{ss3a10}
\end{eqnarray}
\begin{equation}
 z^\prime (\zeta \lambda^\prime) \alpha^\prime (\zeta \lambda^\prime)
 = z^\prime (\zeta \lambda_0 )
 \alpha^\prime (\zeta \lambda_0 ). \label{ss4a10}
\end{equation}
Here, $\beta_\sigma^\prime$, $\beta_\theta^\prime$ and
$\gamma^\prime_{zc}$ are given to the appropriate order by
Eqs.~\eqref{finalbetaprime}-\eqref{betafunctionsprimea}. These
final results generalize the perturbative expressions of
Eqs.~\eqref{rgsigma}-\eqref{rgz2}.
%
%
\section{\label{SUM} Discussion}
%
%
In this paper we have extended the perturbative theory of
localization and interaction effects to include the highly
non-trivial effects of the $\theta$ term. The analysis that we
have presented is an important technical as well as conceptual
advance since it directly relates to some of the most fundamental
and the long standing problems of the interacting electron gas on
the strong coupling side.

We have seen, first of all, that the appearance of {\em massless
chiral edge excitations} has important consequences for the low
energy dynamics of the instanton vacuum and can be used, amongst
many other things, to formulate a Thouless-like criterion for the
quantum Hall effect. Our introduction of an effective action for
the edge excitations resolves the previously encountered
ambiguities in the Kubo formulae and renormalization group, in
particular the general problem of boundary conditions as well as
the quantization of topological charge. The effective action
procedure for edge excitations uniquely defines the {\em response
parameters} or {\em physical observables} $\sigma_{xx}^\prime$ and
$\theta^\prime$. Moreover, by recognizing the differences between
the {\em edge} excitations and {\em bulk} excitations we have
fundamentally explained the various different aspects of symmetry
in the problem, notably {\em particle-hole} symmetry and {\em
periodicity} in $\sigma_{xy}^\prime$. Furthermore, the conditions
for the quantum Hall effect can now quite generally be expressed
by saying that $\sigma_{xx}^\prime = \theta^\prime =0$. This means
that the bulk of the system renders insensitive to changes in the
boundary conditions. This generally happens when the bulk
excitations of the system generate a mass gap.
\begin{figure}[tbp]
\includegraphics[width=80mm]{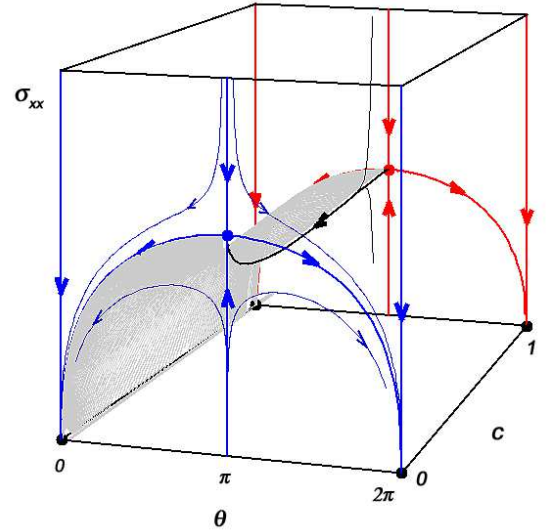}
\caption{Renormalization group flow diagram in the parameter space
$\sigma_{xx}$, $\theta$ and $c$. The arrows indicate the direction
toward the infrared} \label{RGall}
\end{figure}
These general statements have motivated us to develop a unified
microscopic theory for the physical observables
$\sigma_{xx}^\prime$ and $\theta^\prime$ of the electron gas in
the presence of electron-electron interactions. The complete list
of observable parameters includes also the parameter $c^\prime$
which distinguishes between {\em finite range} electron-electron
interactions ($0<c^\prime <1$) and {\em infinite range}
interactions ($c^\prime =1$), as well as the parameter $z^\prime$
which controls the temperature and frequency dependence of the
electron gas. The most important results of this work are given by
Eqs.~\eqref{ss1a10} -\eqref{ss4a10} expressing how the observable
parameters are related to the renormalization group $\beta^\prime$
and $\gamma^\prime$ functions of the theory. The closed set of
renormalization group functions $\beta_\sigma^\prime$,
$\beta_\theta^\prime$ and $\beta_c^\prime$ that we have obtained
(Eqs.~\eqref{finalbetaprime}-\eqref{betafunctionsprimea}) controls
the low energy dynamics of the electron gas at $T=0$ and zero
external frequency. The principal features of this theory are
encapsulated in the three dimensional renormalization group flow
diagram as sketched in Fig.~\ref{RGall}. The regime of finite
range electron-electron interactions $0<c<1$, like the theory in
$2+\epsilon$ dimensions, lies the domain of attraction of the
Fermi liquid plane $c=0$ which is stable in the infrared. These
results are in accordance with the principle of $\mathcal{F}$
invariance which states the distinctly different problems of the
Coulomb interaction $c=1$ and finite range electron-electrons
interactions $0 \leq c <1$ are preserved separately under the
action of the renormalization group.

\subsection{Robust quantization of Hall conductance}

\begin{figure}[tbp]
\includegraphics[width=80mm]{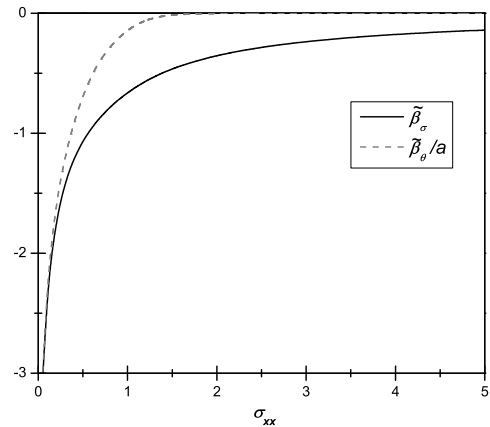}
\caption{Renormalization group functions $\tilde \beta_\sigma$ and
$\tilde \beta_\theta/a$ with varying $\sigma_{xx}$ and $\theta=0$,
see text.} \label{tbt}
\end{figure}

We are now in a position to elaborate on the quantum Hall effect
which is represented in Fig.~\ref{RGall} by the infrared fixed
points located at precise values of $\sigma_{xy}^\prime = k(\nu)$
or $\theta^\prime = 0$ and $\sigma_{xx}^\prime =0$. For this
purpose let us consider the renormalization group equations along
the lines $\sigma_{xy}^\prime \approx k(\nu)$ or $\theta^\prime
\approx 0$. Specializing to the most interesting case $c=1$ then
we can write
\begin{eqnarray}
\frac{d\ln \sigma_{xx}^\prime}{d\ln\lambda} &=&
\tilde{\beta}_\sigma (\sigma_{xx}^\prime) \notag  \\
&=& -\frac{2}{\pi \sigma_{xx}^\prime} -
\frac{\beta_1(1)}{(\sigma_{xx}^\prime)^2} -{\mathcal{D}(1)}
\sigma_{xx}^\prime e^{-2\pi\sigma_{xx}^\prime}
\label{betafunctiona}\\
\frac{d\ln |\theta^\prime |}{d\ln\lambda} &=&
\tilde{\beta}_\theta (\sigma_{xx}^\prime) \label{betafunctionb} \\
&=& ~~~~~~~~~~~~~~~~~~~~- 2\pi {\mathcal{D}(1)}
(\sigma_{xx}^\prime)^2 e^{-2\pi\sigma_{xx}^\prime}.\notag
\end{eqnarray}
These results are clearly consistent with the Thouless-like
criterion presented in Section~\ref{qHe} which tells us that along
the lines $\theta^\prime \approx 0$ both quantities
$\sigma_{xx}^\prime$ and $\theta^\prime$ should become
exponentially small for large scale sizes $\lambda$. Recall from
the discussion in Section~\ref{Flows} that the perturbative
$\tilde{\beta}_\sigma$ function usually indicates that the
response parameter $\sigma^\prime_{xx}$ scales from $-(2/\pi) \ln
(\lambda/\xi)$ for small values of $\lambda$ to $\exp (
-\lambda/\xi)$ for large values of $\lambda$. Here, $\xi$ is the
{\em dynamically generated} correlation length (localization
length), see Eqs.~\eqref{obs2} and \eqref{obs3}. From
Eq.~\eqref{betafunctiona} we see that the instanton contribution
generally enhances the tendency of the electron gas to localize at
large distances. In Fig.~\ref{tbt} we sketch the overall behavior
of the $\tilde{\beta}_\sigma$ function which is given by the weak
coupling result of Eq.~\eqref{betafunctiona} for large values of
$\sigma_{xx}^\prime$ and the strong coupling result
\begin{equation}
\tilde{\beta}_\sigma = \ln \sigma_{xx}^\prime \label{strongsigma}
\end{equation}
as $\sigma_{xx}^\prime$ goes to zero. These results give rise to
the well known scaling scenario of localization in two spatial
dimensions. ~\cite{gangoffour} However, Eq.~\eqref{betafunctionb}
shows that $|\theta^\prime |$ deceases at a much slower rate with
increasing values of $\lambda$ which means that the quantum Hall
regime is generally confined to the regime of ``bad conductors"
$\sigma_{xx}^\prime \lesssim 1$ only. Similar to
$\sigma_{xx}^\prime$ there seems to be something remarkably
universal about the exponential form with which $|\theta^\prime |$
vanishes in the strong coupling regime. The experiments on the
quantum Hall effect,~\cite{experimentsNew} for example, indicate
that $\theta^\prime \propto (\sigma^\prime_{xx})^{a}$ with some
positive value for the exponent $a$ which is presumably equal to
two. The same behavior has recently been found in strong coupling
studies of closely related two dimensional models of the instanton
vacuum.~\cite{o3} Analogous to Eq.~\eqref{strongsigma} one
therefore expects that
\begin{equation}
\tilde{\beta}_\theta = a \ln \sigma_{xx}^\prime
\label{strongtheta}
\end{equation}
in the limit where $\sigma_{xx}^\prime$ goes to zero. In
Fig.~\ref{tbt} we compare the scaling results for the Hall
conductance $\tilde{\beta}_\theta$ with those for the longitudinal
conductance $\tilde{\beta}_\sigma$. These scaling results indicate
that the quantization phenomenon is a {\em (super) universal
strong coupling} feature of the instanton vacuum concept,
independent of the specific application of this concept or, for
that matter, independent of the presence of electron-electron
interactions.

\subsection{Fermi liquid versus non-Fermi liquid theory}

The most important features next are the {\em quantum critical}
fixed points that are located at $\theta = \pi$ or half-integer
values of $\sigma_{xy}$. Fig.~\ref{RGall} shows that the {\em Fermi
liquid} fixed point located at $c=0$ is distinctly different from
the {\em Coulomb interaction} fixed point at $c=1$. Like the {\em
mobility edge} problem in $2+\epsilon$ dimensions, the quantum
critical behavior of the {\em transitions} between adjacent
quantum Hall plateaus is very different for finite range
electron-electron interactions and the Coulomb potential, each
involving different exponent values as well as a fundamentally
different dynamical behavior. The results of this paper therefore
completely invalidate any attempt to explain the experimentally
observed exponent values on the basis of Fermi liquid type of
ideas.~\cite{Exponents2,Exponents4,Exponents5,Exponents7,Exponents8,Exponents10}

To understand the fundamental differences between Fermi liquid
theory and the Coulomb interaction problem in the quantum Hall
regime, a detailed understanding of the quantum phase transition
in $2+\epsilon$ spatial dimensions plays an absolutely essential
role. The main reason is that the mobility edge problem in
$2+\epsilon$ dimension is the only place where the various
different aspects of dynamical scaling of the electron gas can be
established and evaluated explicitly. This includes not only the
theory of quantum transport but also fundamental aspects of the
problem that one usually does not probe in quantum Hall
experiments such as the {\em specific heat} of the electron gas,
the {\em multi fractal} singularity spectrum
~\cite{VoropaevBurmistrovPruisken} etc. In fact, the long standing
problems associated with the theory of electron-electron
interactions have in many ways turned out to be an outstanding
laboratory for advanced methods in quantum field theory that one
cannot study in any different manner.~\cite{Unify2,Unify5}

\subsection{Super universality}

The results of this paper explain, at the same time, why the
scaling behavior of the free electron gas and the Coulomb
interaction problem in strong magnetic fields look so similar. In
spite of the fact that the underlying theories are fundamentally
different they have nevertheless important features in common such
as asymptotic freedom, instantons, massless edge excitations etc.
Since in both cases the topological concepts are the same it is
natural to expect that the basic phenomena are the same, in
particular the existence of {\em robust topological quantum
numbers} that explain the observability and precision of the
quantum Hall effect, as well as {\em quantum criticality} at
$\theta = \pi$ that generally facilitates a {\em transition} to
take place between different quantum Hall plateaus. Finally, by
recognizing the fact that quantum Hall physics actually reveals
itself as a generic, {\em super universal} feature of the
instanton vacuum in asymptotically free field theory one has
essentially laid the foundation for a more ambitious {\em
unifying} theory that includes besides integral quantum Hall
regime also the scaling behavior of completely different physical
phenomena such as the {\em abelian} quantum Hall states.

\begin{acknowledgements}

This research was funded in part by the Dutch National Science
Foundations \textit{NWO} and \textit{FOM}. One of us (\textit{IB})
is indebted to the Russian Foundation for Basic Research
(\textit{RFBR}), the Russian Ministry of Science and Russian
Science Support Foundation for financial support.

\end{acknowledgements}

\appendix


\setcounter{addeq}{\value{equation}} \setcounter{equation}{0}
\renewcommand{\theequation}{A.\arabic{equation}}

\section{\label{AppA} Linear response versus background field procedure}

With the introduction of $\mathcal F$ {\em algebra} it has become
possible to show that observable quantities $\sigma_{xx}^\prime$,
$\sigma_{xy}^\prime$, $z^\prime$ and $c^\prime$ which are usually
obtained by means of {\em background field} procedures or {\em
momentum shell} procedures are, in fact, precisely the same as the
expressions for the conductances at zero temperature that one
derives from ordinary linear response theory in the external
vector potential.~\cite{Unify1} In this Appendix we briefly repeat
the argument for the special case where the infrared of the system
is regulated by a finite size $L \times L$. We show in particular
that the linear response formulae given by Eqs.~\eqref{resp1} and
\eqref{resp2} are the same expressions for $\sigma_{xx}^\prime$
and $\sigma_{xy}^\prime$ as those appearing in the effective
action for the edge modes, Eq.~\eqref{Response1}.
\subsection{Linear response theory}
Specializing to the theory of Eqs. ~\eqref{S}-\eqref{SF} with {\em
spherical} boundary conditions on the field variables $Q$ then the
response of the system to an external vector potential
$\mathbf{A}$ can generally be written in terms of an effective
action $S_\textrm{eff}[\mathbf{A}]$ according to
\begin{equation}\label{SeffA}
 \exp S_\textrm{eff}[\mathbf{A}] = \int_{\partial
 V}\mathcal{D}[Q_0] \exp \Bigl ( S_\sigma [Q_0,\mathbf{A}] +
 S_F[Q_0]\Bigr ) .
\end{equation}
The vector potential $\mathbf{A}$ couples the free electron part
of the action $S_\sigma$ only~\cite{Unify1}
\begin{eqnarray}
S_{\sigma} [Q_0, \mathbf{A}] =  &-&\frac{\sigma _{xx}}{8} \int
d\mathbf{r}
  \tr
 [D_j, Q_0][D_j, Q_0] \label{SsSs} \\
 &+& \frac{\sigma_{xy}}{8} \int d\mathbf{r}\tr \varepsilon_{j
 k}Q_0[D_j,Q_0][D_{k}, Q_0]\notag
\end{eqnarray}
with $D_j$ standing for the covariant derivative
\begin{equation}\label{cd}
 D_j = \nabla_j - i\,\hat A_j,\qquad  \hat A_j
 = \sum\limits_{\alpha n} A_j^\alpha (\nu_n) I^\alpha_n ~.
\end{equation}
Since we are interested in the global response at zero temperature
and frequency it suffices to take a spatially independent
$\mathbf{A}^\alpha (\nu_n)$ and consider a small range of values
$\nu_n =2\pi T n \approx 0$ only. The response parameters
$\sigma_{xx}^\prime$ and $\sigma_{xy}^\prime$ are then defined by
the following general form of the effective action
\begin{equation}\label{SeffA2}
 S_\textrm{eff}[\mathbf{A}] = -L^2 \sum\limits_{\alpha, n>0} n
 \Bigl [\sigma_{xx}^\prime \delta_{jk}+
 \sigma_{xy}^\prime \varepsilon_{jk}\Bigr ]A_j(\nu_n) A_k(-\nu_n) .
\end{equation}
By using this expression for the left hand side of
Eq.~\eqref{SeffA} it is easy to derive the results of
Eqs.~\eqref{resp1} and \eqref{resp2} for $\sigma_{xx}^\prime$ and
$\sigma_{xy}^\prime$ respectively which are the main objectives of
the present paper. These formulae are some of the most fundamental
quantities of the theory since they can generally be used for
studies at finite temperature and frequency rather than finite
sample sizes. Moreover, they facilitate an analysis of mesoscopic
fluctuations as well as important self-consistency checks in
practical computations such as the replica limit $N_r = 0$ and
$N_m \rightarrow \infty$.

However, the complications primarily arise if one wants to make
sure that the Finkelstein formalism preserves the fundamental
symmetries of the interacting electron gas, in particular the
electrodynamic $U(1)$ gauge invariance as well as $\mathcal F$
invariance which are properly defined in infinite Matsubara
frequency space only. As we shall see next, these complications
automatically arise in the attempt to lay the bridge between
linear response theory and the effective action for the edge
modes.

\subsection{$\mathcal F$ invariance}
To deal with electrodynamic gauge invariance in {\em finite}
frequency space we start out by embedding the matrix variables
$Q_0$ of size $2 N_r N_m \times 2 N_r N_m$ in a much larger matrix
space of size $2 N_r N_m^\prime \times 2 N_r N_m^\prime$ with $1
\ll N_m \ll N_m^\prime$. All matrix manipulations will be carried
out from now onward in the space of {\em large} matrices whereas
the unitary rotations $Q_0$ effectively retain their size $2 N_r
N_m \times 2 N_r N_m$ which we term {\em small}.

Let us next introduce the quantity $\varphi^\alpha_n ({\mathbf r})
= \mathbf{A}^\alpha (\nu_n) \cdot \mathbf{r}$. We can then express
the vector potential $\mathbf{\hat A}$ in terms of the {\em large}
unitary matrix $\hat{\varphi} =\hat{\varphi} (\mathbf{r})$
according to
\begin{equation}
 \mathbf{\hat A} = \nabla \hat \varphi = i W^{-1}\nabla W, \qquad W =
 \exp (-i \hat \varphi).
\end{equation}
Following the rules of $\mathcal F$ {\em algebra}~\cite{Unify1}
the unitary matrix $W$ just stands for an electrodynamic $U(1)$
gauge transformation in Matsubara frequency notation. The free
electron part of the action ~\eqref{SsSs} can be expressed in
terms of the $W$ rotation on the matrix field variable $Q_0$
according to
\begin{widetext}
 \begin{equation}\label{SsW}
 S_{\sigma} [Q_0, \mathbf{A}] = S_{\sigma} [ W^{-1} Q_0 W ] =
 -\frac{\sigma_{xx}}{8} \int d\mathbf{r} \tr [\nabla(W^{-1}Q_0
 W)]^2 +\frac{\sigma_{xy}}{8} \int d\mathbf{r}\tr \varepsilon_{j
k}Q_0\nabla_j (W^{-1}Q_0 W)\nabla_k (W^{-1}Q_0 W).
\end{equation}
\end{widetext}
Next we split the quantity $\mathcal{O}_F(Q_0)$ into an $\mathcal
F$ invariant part $\mathcal{O}_s (Q_0)$ and a symmetry breaking
part
\begin{equation}
\mathcal{O}_F(Q_0) = \mathcal{O}_s(Q_0) + \mathcal{O}_\eta (Q_0)
\end{equation}
where
\begin{eqnarray}
\mathcal{O}_s(Q_0) &=& z c \Bigl ( \sum\limits_{\alpha
n}\mathop{\rm tr} \nolimits I_{n}^{\alpha }Q_0\mathop{\rm
tr}\nolimits I_{-n}^{\alpha }Q_0 + 4 \mathop{\rm tr}\nolimits\eta
Q_0 \notag \\
&&\hspace{0.5cm}- 6 \mathop{\rm
tr}\nolimits\eta \Lambda \Bigr ) \notag \\
&=& z c {\sum\limits_{\alpha n}}^\prime \mathop{\rm tr} \nolimits
[
I_{n}^{\alpha } , Q_0 ] [ I_{-n}^{\alpha } , Q_0 ] \\
\mathcal{O}_\eta (Q_0) &=& z \alpha \left\{ 4 \mathop{\rm
tr}\nolimits\eta Q_0- 6 \mathop{\rm tr}\nolimits\eta \Lambda
\right\} .\label{OF1}
\end{eqnarray}
The statement of $\mathcal{F}$ {\em invariance} now says that
$\mathcal{O}_s (Q_0)$ is gauge invariant~\cite{Unify1}
\begin{eqnarray}\label{SsW}
 \mathcal{O}_s (Q_0)=\mathcal{O}_s (W^{-1}Q_0 W) .
\end{eqnarray}
On the other hand, as long as one evaluates the theory at zero
temperature and finite system sizes, the response parameters
$\sigma^\prime_{xx}$ and $\sigma^\prime_{xy}$ remain unchanged if
one inserts the $W$ rotation into the quantity $\mathcal{O}_\eta
(Q)$, i.e. the replacement
\begin{eqnarray}\label{SetaW}
 \mathcal{O}_\eta (Q_0) \rightarrow \mathcal{O}_\eta (W^{-1}Q_0 W)
\end{eqnarray}
does not affect the statement of Eq.~\eqref{SeffA2} where the
$\sigma^\prime_{xx}$ and $\sigma^\prime_{xy}$ depend on the system
size $L$. Linear response theory at zero temperature and finite
system sizes is therefore formally the same thing as evaluating
the theory in the presence of a gauge field $W$
\begin{equation}\label{SeffW}
 e^{\tilde{S}_\textrm{eff}[\mathbf{A}]} = \int_{\partial
 V}\mathcal{D}[Q_0] e^{S_\sigma [W^{-1} Q_0 W] +
 S_F[W^{-1} Q_0 W]}.
\end{equation}
The main reason for introducing the two different cut-offs $1 \ll
N_m \ll N_m^\prime$ in {\em finite} Matsubara frequency space is
to ensure that Eqs.~\eqref{SsW}, \eqref{SetaW} and \eqref{SeffW}
display the {\em exact} same symmetries that are known to exist in
the theory where $N_m$ and $N_m^\prime$ are being sent off to
infinity.

\subsection{Background field formalism}
It is clear the the statement of Eq.~\eqref{SeffW} is non-trivial
only due to the fact that that we work at zero temperature and
with fixed boundary conditions on the matrix field variable $Q_0$.
If on the other hand we were to work with {\em finite}
temperatures and {\em infinite} system sizes $L$ then
Eq.~\eqref{SeffW} is merely a statement of electrodynamic $U(1)$
gauge invariance which is clearly very different from
Eq.~\eqref{SeffA}.

Notice that Eq.~\eqref{SeffW} is not yet quite the same as the
{\em back ground field} methodology that previously has been
studied intensively for renormalization group purposes. This is
because the quantities $Q_0$ and $W^{-1} Q_0 W$ by construction
belong to different manifolds for any finite value of $N_m$ and
$N_m^\prime$. However, in order for the $W$ rotation in
Eqs.~\eqref{SsW}, \eqref{SetaW} and \eqref{SeffW} to represent an
{\em exact} electrodynamic $U(1)$ gauge transformation it is
imperative that the results do not fundamentally depend on the
details of how the frequency cut-offs $N_m$ and $N_m^\prime$ go to
infinity. Moreover, the statement of Eq.~\eqref{SeffW} renders
highly non-trivial if one recognizes that the unitary matrix $W$
can in general be written as the product of two distinctly
different matrices $t$ and $U_0$
\begin{equation}
W =\exp (-i \hat \varphi)=  U_0 ~t, \qquad U_0 \in
U(N^\prime)\times U(N^\prime) \label{WU0t}
\end{equation}
where $N^\prime = N_r N_m^\prime$. Here, $t$ is a ``small"
background matrix field in the true sense of the word
\begin{equation}\label{smallt}
t = \exp \left( \frac{i}{2} [\hat{\varphi} , \Lambda ] \Lambda
~+~\dots \right)
\end{equation}
whereas the ``large'' generators of $W$ are all collected together
in the $U(N^\prime)\times U(N^\prime)$ gauge $U_0$ which can be
written as
\begin{equation}
U_0 = \exp \left( \frac{i}{2} \{ \hat{\varphi} , \Lambda \}
\Lambda
 \right).
\end{equation}
Next we consider the change of variables
\begin{equation}\label{Ugauge}
U_0^{-1} Q_0 U_0 \rightarrow Q_0 .
\end{equation}
It is clear that this transformation preserves the spherical
boundary conditions and leaves the measure of the functional
integral unchanged. Equation~\eqref{SeffW} can therefore be
represented as follows
\begin{equation}\label{SeffA5}
\exp \tilde{S}_\textrm{eff}[\mathbf{A}] = \int_{\partial
V}\mathcal{D}[{Q}] \exp \Bigl ( S_\sigma [t^{-1} {Q_0} t] +
S_F[t^{-1} {Q_0} t] \Bigr )
\end{equation}
which precisely corresponds to the background field methodology
with the ``small'' matrix field $t$ given explicitly by Eq.
~\eqref{smallt}. This, then, leads to the principle result of this
Appendix  which says that Eq.~\eqref{SeffA5} in the limit where
$N_m , N_m^\prime \rightarrow \infty$ and $T=0$ is identically the
same as linear response theory Eqs.~\eqref{resp1} and
\eqref{resp2}.

Eq.~\eqref{smallt} together with Eq.~\eqref{SeffA5} can be used to
derive different or alternative expressions for the quantities
$\sigma_{xx}^\prime$ and $\sigma_{xy}^\prime$ which are completely
equivalent to those given by Eqs.~\eqref{resp1} and \eqref{resp2}.
Here we do not list these expressions but instead we simply verify
the correctness of the effective action of Eq.~\eqref{SeffA2}.
Since Eq.~\eqref{SeffA5} has the same form as the effective action
for the edge modes we can immediately write down the following
general result
\begin{equation}\label{SeffA6}
\tilde{S}_\textrm{eff}^0 [\mathbf{A}] =
-\frac{\sigma_{xx}^\prime}{8} \int d \mathbf{r} \tr (\nabla q)^2
+\frac{\sigma_{xy}^\prime}{8} \int d \mathbf{r}\tr \varepsilon_{j
k}q\nabla_j q\nabla_{k} q
\end{equation}
where the superscript ``$0$'' denotes the result at $T=0$.
Eq.~\eqref{SeffA6} can be obtained, as before, by expanding in the
gradients of the slowly varying matrix field $q=t^{-1} \Lambda t$.
By inserting the expression for $t$ in Eq.~\eqref{smallt} we
obtain
\begin{equation}\label{SeffA7}
\tilde{S}^{0}_\textrm{eff}[\mathbf{A}] =  - \int d\mathbf{r}
\sum\limits_{\alpha n>0} n \Bigl [\sigma_{xx}^\prime
\delta_{jk}+\sigma_{xy}^\prime \varepsilon_{jk}\Bigr ]\nabla_j
\varphi^\alpha_n \nabla_k \varphi^\alpha_{-n}.
\end{equation}
The following identities have been used
\begin{eqnarray}
\tr [\hat I^\alpha_n,\Lambda][\hat I^\alpha_{-n},\Lambda] &=& -4 n\\
\tr \Lambda [\hat I^\alpha_n,\hat I^\alpha_{-n}] &=& 2 n.
\end{eqnarray}
We see that we recover the same results as those in
Eq.~\eqref{SeffA2}.
\subsection{The quantities $z^\prime$ and $c^\prime$}
For completeness we next extend the results of the background
field methodology to include the terms obtained by expanding to
lowest order in $T$
\begin{eqnarray}
e^{\tilde{S}_\textrm{eff} [\mathbf{A}]} &=&
e^{\tilde{S}_\textrm{eff}^0 [\mathbf{A}]} \Bigl ( 1+ T z^\prime
c^\prime \int d \mathbf{r} {\sum\limits_{\alpha n}}^\prime
\mathop{\rm tr} \nolimits [ I_{n}^{\alpha } , q ] [ I_{-n}^{\alpha
} , q ] \notag \\
&& \hspace{0.8cm}+ T z \alpha \int d \mathbf{r} \left ( 4
\mathop{\rm tr}\nolimits\eta q- 6 \mathop{\rm tr}\nolimits\eta
\Lambda \right ) \Bigr ).\,{}\label{SeffA7}
\end{eqnarray}
These results indicate that the quantity $zc$ is renormalized
whereas the statement $z\alpha=z^\prime \alpha^\prime$ is a
physical constraint that should in general be imposed upon the
theory. Eq.~\eqref{SeffA7} has been verified in the theory of
perturbative expansions. In Section ~\ref{Omega} of this paper we
explicitly check the validity of this statement at a
non-perturbative level. As a final remark, it should be mentioned
that by taking $q=\Lambda$ in Eq.~\eqref{SeffA7} one immediately
obtains the expression for $z^\prime$, Eq.~\eqref{zcren}.
\setcounter{addeq}{\value{equation}} \setcounter{equation}{0}
\renewcommand{\theequation}{B.\arabic{equation}}
\section{\label{ME} Matrix elements}

The matrix elements of a function $f(\eta,\theta)$ are defined as
follows
\begin{widetext}
\begin{equation}\label{APP.ME1}
{}_{(a)}\left\langle J,M|f(\eta,\theta )|M^{^{\prime
}},J^{^{\prime }}\right\rangle _{(b)} = \int d\eta d\theta
\Phi_{J,M}^{(a)}(\eta,\theta )f(\eta,\theta)\bar{\Phi}
_{J^{^{\prime}},M^{^{\prime}}}^{(b)}(\eta,\theta)\notag
\end{equation}
where $a,b=0,1,2$. By using the following identity for the Jacobi
polynomials~\cite{GR}
\begin{equation}\label{rel1}
(2n+\alpha+\beta) P_n^{(\alpha-1,\beta)}(x)=(n+\alpha+\beta)
P_n^{(\alpha,\beta)}(x) - (n+\beta) P_n^{(\alpha,\beta)}(x)
\end{equation}
and the normalization condition
\begin{equation}\label{rel2}
\int\limits_{-1}^1 d x (1-x)^{\alpha}(1+x)^{\beta}
P_n^{(\alpha,\beta)}(x) P_m^{(\alpha,\beta)}(x) =
\delta_{n,m}2^{\alpha+\beta+1}
\frac{\Gamma(\alpha+n+1)\Gamma(\beta+n+1)}{(\alpha+\beta+2n+1)
\Gamma(n+1)\Gamma(\alpha+\beta+n+1)}
\end{equation}
we find that the matrix elements for $e_0$ and $e_1$ are given as
\begin{eqnarray}
{}_{(0)}\left\langle J,M|e_{1}^{\ast }|M-1,J\right\rangle
_{(1)}&=& \frac{1}{\sqrt{2}}\sqrt{\frac{J+M}{2J+1}}, \qquad
{}_{(0)}\left\langle J,M|e_{1}^{\ast }|M-1,J+1\right\rangle
_{(1)}= \frac{1}{\sqrt{2}}\sqrt{\frac{J-M+1}{2J+1}}
\label{MatEls}\\
{}_{(0)}\left\langle J,M|e_{0}|M,J\right\rangle _{(1)}&=&-\frac{1}{%
\sqrt{2}}\sqrt{\frac{J-M}{2J+1}}, \qquad\hspace{0.3cm}
{}_{(0)}\left\langle J,M|e_{0}|M,J+1\right\rangle _{(1)}=\frac{1}{%
\sqrt{2}}\sqrt{\frac{J+M+1}{2J+1}}. \label{MatEls2}
\end{eqnarray}
Next for $e_0^2$ we have
\begin{eqnarray}  \label{MatEls5}
{}_{(1)} \left \langle J,M | e_0^2| M,J-1\right \rangle_{(1)} &=&-%
\frac{\sqrt{(J-M-1)(J + M)}}{2(2J-1)},\qquad  {}_{(1)} \left
\langle J,M | e_0^2| M,J\right \rangle_{(1)} =\frac{1
}{2} \left [ 1+\frac{2M+1}{4J^2-1}\right ] \\
{}_{(1)} \left \langle J,M | e_0^2| M,J+1\right \rangle_{(1)} &=&-
\frac{\sqrt{(J+M+1)(J - M)}}{2(2J+1)},\qquad
{}_{(2)}\left \langle J,M | e_0^2| M,J\right \rangle_{(2)} =\frac{1%
}{2} \left [ \frac{M+1}{J(J+1)}\right ].\label{MatEls7}
\end{eqnarray}
The matrix elements of $e_0e_1$ are as follows
\begin{eqnarray}
{}_{(1)} \left \langle J,M-1 | e_0 e_1| M,J-1\right \rangle_{(1)}
&=&\frac{\sqrt{(J-M)(J-M-1)}}{2(2J-1)} \notag\\ {}_{(1)} \left
\langle J,M-1 | e_0 e_1| M,J\right \rangle_{(1)}
&=&\frac{\sqrt{(J-M)(J+M+1)}}{4J^2-1}  \notag\\
{}_{(1)} \left \langle J,M-1 | e_0 e_1| M,J+1\right \rangle_{(1)}
&=& \frac{\sqrt{(J+M)(J+M+1)}}{2(-2J-1)} \label{MatEls9}\\
{}_{(0)}\left \langle J,M | e_0 e_1^*| M-1,J-1\right \rangle_{(0)}
&=& \sqrt{\frac{(J+M-1)(J+M)}{4(2J-1)(2J+1)}}\notag \\ {}_{(0)}
\left \langle J,M+1 | e_0 e_1^*| M,J+1\right \rangle_{(0)}&=& -
\sqrt{\frac{(J-M)(J-M+1)}{4(2J+1)(2J+3)}}. \notag
\end{eqnarray}
Finally, the following summation theorems are of interest
\begin{equation}  \label{MatEls14}
\sum \limits_{M=-J}^{J-1}{}_{(1)} \left \langle J,M | e_0^2
|e_1|^2| M,J\right \rangle_{(1)} =\frac{J}{3},\qquad \sum
\limits_{M=-J-1}^{J-1}{}_{(2)} \left \langle J,M | e_0^2 |e_1|^2|
M,J\right \rangle_{(2)} =\frac{2J+1}{6}.
\end{equation}
\end{widetext}

\setcounter{addeq}{\value{equation}} \setcounter{equation}{0}
\renewcommand{\theequation}{C.\arabic{equation}}
\section{\label{AppC} Perturbative expansions of observable theory
using Pauli-Villars regularization}

\subsection{Renormalization of $\sigma_{xx}$}

For ordinary perturbation theory we use the expression for the
matrix field variable $Q$ as in Eqs.~\eqref{top1} - \eqref{vfield}
but with the matrices $\mathcal T$ and $R$ put equal to the unit
matrix. Evaluating Eq.~\eqref{resp1} to the second order in the
independent field variables $v$,$v^\dag$ we then obtain
\begin{eqnarray}
\sigma_{xx}^{\prime}=\sigma_{xx}&+&\frac{\sigma_{xx}^2}{2n} \int d
\textbf{r} \langle\tr I^\alpha_n v(\textbf{r})\nabla
v^{\dag}(\textbf{r})\notag \\ && \hspace{1cm}\times \tr
I^\alpha_{-n} v(\textbf{r}^\prime)\nabla
v^{\dag}(\textbf{r}^\prime)\rangle .\label{sren2}
\end{eqnarray}
Notice that in {\em flat space} one can choose the point
$\textbf{r}^{\prime}$ arbitrarily due to translational invariance.
In {\em curved space}, however, we must evaluate Eq.~\eqref{sren2}
in terms of the propagators of Eqs.~\eqref{Propa} and \eqref{Ga}
in which case translational invariance is no longer obvious. In
terms of the energies and eigenfunctions in curved space
Eq.~\eqref{sren2} reads as follows
\begin{eqnarray}
\sigma_{xx}^{\prime}=\sigma_{xx} &-& 4 c \int \limits_{0}^\infty
d\omega \sum \limits_{J} \frac{E_J^{(0)}}{(E_J^{(0)}+\omega)^2
(E_J^{(0)}+\alpha\omega)} \nonumber \\
&&\hspace{0.7cm}\times \sum \limits_{M=-J}^{J}
\Phi_{JM}^{(0)}(\eta^{\prime},\theta^{\prime})
\bar{\Phi}_{JM}^{(0)}(\eta^{\prime},\theta^{\prime}) \nonumber
\\\label{sren3}
\end{eqnarray}
where $\eta^{\prime},\theta^{\prime}$ denote the spherical
coordinates of the point $\textbf{r}^{\prime}$. Since the
eigenfunction $\Phi_{JM}^{(0)}$ is proportional to the Jacobi
polynomial $P_{J-M}^{M,M}(\eta)$ which itself is proportional to
the Gegenbauer polynomial $C_{J-M}^{M+1/2}(\eta)$ we can use the
well known summation theorem for Gegenbauer polynomials ~\cite{GR}
and obtain
\begin{eqnarray}\label{proj}
\sum \limits_{M=-J}^{J} \Phi_{JM}^{(0)}(\cos\phi,\theta)
\bar{\Phi}_{JM}^{(0)}(\cos\phi^{\prime},\theta) \hspace{2cm}\,&&
 \notag \\ & &\hspace{-5cm}= \frac{2J+1}{4\pi} C^{1/2}_{J} \left
(\cos(\phi-\phi^{\prime})\right ).
\end{eqnarray}
We recognize this identity as a projection operator statement
which means that Eq. ~\eqref{sren3} is in fact independent of
$\eta^{\prime},\theta^{\prime}$.

Next, introducing the Pauli-Villars masses as well as the
alternating metric, using $C^{1/2}_J(1)=1$ and after integrating
over $\omega$ we obtain
\begin{eqnarray}
\sigma_{xx}^{\prime}=\sigma_{xx}&-&\frac{\beta_0(c)}{2}\lim\limits_{\Lambda\to
\infty} \Biggl [ \sum\limits_{J=3/2}^\Lambda
\frac{2J(J^2-\frac{1}{4})}{(J^2-\frac{1}{4})^2} \notag \\ &+&
\sum\limits_{f=1}^K \hat e_f \sum\limits_{J=1/2}^\Lambda \frac{
2J(J^2-\frac{1}{4})}{(J^2-\frac{1}{4}+\mathcal{M}^2_f)^2}\Biggr
].\hspace{0.5cm}\,{}\label{sren5}
\end{eqnarray}
Evaluating the sums we finally have
\begin{eqnarray}
\sigma_{xx}^{\prime}&=&\sigma_{xx}-\frac{\beta_0(c)}{2}\left
(Y^{(0)}_{\rm reg} + 1 \right ) \notag \\ &=& \sigma_{xx}\left (1
-\frac{\beta_0(c)}{\sigma_{xx}} \ln \mathcal{M}e^{\gamma_E}\right
).\label{sreg5}
\end{eqnarray}

\subsection{Renormalization of $zc$}

The expression for $z^{\prime}c^\prime$ can be expanded in a
similar fashion. To lowest order in $v$, $v^\dag$ we can write the
contributing terms as follows
\begin{equation}
z^{\prime}c^\prime =  zc \left ( 1 - \frac{1}{\tr\eta\Lambda}
\sum\limits_{\alpha,n>0}\langle \tr I^\alpha_n v (\textbf{r}) \tr
I^\alpha_{-n} v^{\dag} (\textbf{r}) \rangle\right ).\label{SfhP1}
\end{equation}
In curved space this expression becomes
\begin{equation}
z^{\prime} c^\prime=  z c \Bigl [ 1 +
\frac{2\pi\gamma_0}{\sigma_{xx}} \sum \limits_{J}
\frac{1}{E_J^{(0)}} \sum \limits_{M=-J}^{J}
\Phi_{JM}^{(0)}(\eta,\theta) \bar{\Phi}_{JM}^{(0)}(\eta,\theta)
\Bigr ]
\end{equation}
with $\eta,\theta$ denoting the point $\textbf{r}$. Next, using
Eq.~\eqref{proj} as well as \eqref{Y01} we finally obtain
\begin{eqnarray}
z^{\prime}c^\prime &=& z c \left ( 1 +
\frac{\gamma_0}{2\sigma_{xx}} Y^{(0)}_\textrm{reg}\right
).\notag \\
&=&  z c  \left ( 1 + \frac{\gamma_0}{\sigma_{xx}} \ln
\mathcal{M}e^{\gamma_E-1/2}\right ).\label{SfhP3}
\end{eqnarray}


\end{document}